 \pdfoutput=1
\documentclass[11pt,a4paper]{article}
\usepackage{jheppub,bm,booktabs,multirow}

\usepackage{amsmath}
\allowdisplaybreaks[4]
\usepackage{amssymb}
\usepackage{graphicx}
\usepackage{booktabs}
\usepackage{bm}
\usepackage{psfrag}
\usepackage[normalem]{ulem}
\usepackage{color}
\usepackage{overpic}
\usepackage[utf8x]{inputenc}

\makeatletter
\def\@fpheader{~}
\makeatother

\def\as{\alpha_s}
\def\e{\epsilon}
\newcommand{\ep}{\epsilon}
\newcommand{\E}[1]{E_{\mathbf{#1}}}

\newcommand{\Dq}[1]{\int\hspace{-0.15cm}\frac{d^d{#1}}{(2\pi)^d}}

\newcommand{\Dqqq}[1]{ \int \hspace{-0.05cm} \left[d\Omega_{#1} \right]}
\newcommand{\Dqin}[1]{\int_\text{in}\hspace{-0.1cm}\frac{d^d{#1}}{(2\pi)^d}}

\newcommand{\DqqE}[1]{\int \hspace{-0.1cm}\frac{d^{d-1}\mathbf{#1}}{(2\pi)^{d-1} 2\E{q}}}
\newcommand{\Dqqqin}[1]{\int_\text{in}\hspace{-0.05cm} \left[d\Omega_{#1} \right]}
\newcommand{\Dqqqout}[1]{\int_\text{out}\hspace{-0.05cm} \left[d\Omega_{#1} \right]}
\newcommand{\DQQQ}[1]{\int \hspace{-0.05cm} \left[d^2\Omega_{#1} \right]}
\newcommand{\DQQQin}[1]{\int_\text{in}\hspace{-0.01cm}\left[d^2\Omega_{#1} \right]}
\newcommand{\DQQQout}[1]{\int_\text{out}\hspace{-0.01cm}\left[d^2\Omega_{#1} \right]}

\newcommand{\CDRflag}{c_R}

\title{Two-loop anomalous dimension for the resummation of non-global observables}
\author[a]{Thomas Becher,}
\author[a]{Thomas Rauh,}
\author[a]{and Xiaofeng Xu}
\affiliation[a]{Albert Einstein Center for Fundamental Physics, Institut f\"ur Theoretische Physik, Universit\"at Bern, Sidlerstrasse 5, CH-3012 Bern, Switzerland}

\emailAdd{becher@itp.unibe.ch}
\emailAdd{rauh@itp.unibe.ch}
\emailAdd{xuxiaofeng@itp.unibe.ch}

\date{December 3, 2021}

\preprint{\begin{flushright}
December 3, 2021
\end{flushright}}

\abstract
{The soft radiation emitted in jet cross sections can resolve the directions and colors of individual hard partons, leading to a complicated pattern of logarithmically enhanced terms in the perturbative series. Starting from a factorization theorem and solving the renormalization group equations for its ingredients, these large logarithms can be resummed. In this paper we extract the two-loop anomalous dimension governing the resummation of subleading logarithms in jet cross sections and other non-global observables. This anomalous dimension can be obtained by considering soft limits of hard amplitudes, but the presence of collinear singularities in intermediate expressions makes its extraction delicate. As a consistency check, we use our results to predict the known subleading non-global logarithms in the two-jet cross section.}

\begin{document}
\maketitle

\newpage

\section{Introduction}

Soft emissions off an energetic particle are remarkably simple in that they only depend on the direction and the charge of the emitting particle. Furthermore emissions off collinear energetic particles are only sensitive to their total charge, a property known as soft coherence. However, in QCD these properties often do not translate into simple all-order results for soft-emission contributions to cross sections. Specifically, it was shown in \cite{Dasgupta:2001sh} that for non-global observables, which constrain the energy of emissions only in certain phase-space regions, successive emissions inside the unconstrained region lead to a complicated pattern of logarithmically enhanced higher-order terms. Jet cross sections are in this category since emissions are unconstrained inside the jets, the simplest concrete example being the interjet energy flow. The leading non-global logarithms arising in this observable in the large-$N_c$ limit can be resummed either by generating the successive emissions through a dedicated parton shower \cite{Dasgupta:2001sh,Dasgupta:2002bw} or by solving a non-linear evolution equation, the Banfi-Marchesini-Smye (BMS) equation \cite{Banfi:2002hw}.  Both approaches were generalized beyond the large-$N_c$ limit \cite{Weigert:2003mm,AngelesMartinez:2018cfz} and first numerical results at leading-logarithmic accuracy are now available at $N_c =3$ \cite{Hatta:2013iba,Hagiwara:2015bia,Hatta:2020wre,DeAngelis:2020rvq}. While the subleading-color contributions are in general small at $e^+e^-$ colliders, the situation is different at hadron colliders since Glauber phases lead to the occurrence of double logarithms at higher orders, while only single logarithms are present in the large-$N_c$ limit. These so-called super-leading logarithms were discovered a long time ago \cite{Forshaw:2006fk,Forshaw:2008cq}, but their resummation was achieved only recently \cite{Becher:2021zkk}.

By now, the leading non-global logarithms have been resummed for a variety of observables, but unsurprisingly these resummations suffer from large uncertainties and are not sufficient for precision physics. Given the prevalence of non-global observables it is therefore important to develop methods to also resum subleading non-global logarithms. Factorization theorems provide the theoretical basis for resummations at higher accuracy and there has been a lot of progress in understanding the factorization properties of non-global observables. The crucial difference to global observables is that soft radiation in non-global observables resolves the directions of  individual hard partons. This translates into factorization theorems for non-global cross sections, which involve a product of hard functions, consisting of squared amplitudes for hard partons along fixed directions, and soft functions, given by matrix elements of Wilson lines along these directions \cite{Becher:2015hka,Becher:2016mmh}. The same structure is present in the formalism of \cite{Caron-Huot:2015bja}, which uses a ``color-density matrix'' to keep track of the individual hard partons and reconstructs the soft emissions by taking a suitable average over this matrix. To resum logarithms one evolves the hard function from its natural scale $\mu\sim Q$, where $Q$ is the center-of-mass energy, to the typical scale $\mu\sim Q_0$ of the soft emissions. This evolution is governed by a renormalization group (RG) equation. To resum the first tower of subleading non-global logarithms one needs the one-loop corrections to the hard and soft functions, together with the evolution driven by the two-loop anomalous dimension. The one-loop corrections to the hard and soft functions were implemented in \cite{Balsiger:2019tne} and the goal of the present paper will be the extraction of the two-loop anomalous dimension $\bm{\Gamma}^{(2)}$. Recently, an alternative approach was put forward \cite{Banfi:2021owj}, which formulates a generalization of the BMS equation to subleading logarithmic accuracy. This paper set up this equation in a way suitable for numerical implementation and tested that the framework captures all logarithms at two loops in the interjet energy flow. Very recently first resummed results were presented in this new framework \cite{Banfi:2021xzn}.

The paper \cite{Caron-Huot:2015bja} by Caron-Huot has presented the two-loop anomalous dimension in the color-density matrix formalism and even the three-loop result for planar $\mathcal{N} = 4$ super Yang-Mills theory is known \cite{Caron-Huot:2016tzz}. The determination of the anomalous dimension made use of a mapping between the evolution equation for non-global logarithms and the Balitsky-Kovchegov (BK) equation in $\mathcal{N} = 4$ super Yang-Mills theory  \cite{Weigert:2003mm,Caron-Huot:2015bja}. Given the close relationship between our two approaches, the anomalous dimension given in \cite{Caron-Huot:2015bja} should be relevant also for our formalism \cite{Becher:2016mmh}. However, the derivation presented in \cite{Caron-Huot:2015bja} is indirect and quite intricate. An independent, direct computation of the anomalous dimension is therefore desirable. The paper \cite{Caron-Huot:2015bja} also did not perform any checks on the result, or detail how one would translate the anomalous dimension into a result for the subleading logarithms. In the present paper, we supply both a direct computation of $\bm{\Gamma}^{(2)}$, by extracting it from the divergences from the relevant Feynman diagrams, and an explicit check of the anomalous dimension. Our computation reproduces the result in \cite{Caron-Huot:2015bja}, but we find that the anomalous dimension in this paper does not correspond to the one in the standard minimal subtraction ($\overline{\text{MS}}$) scheme. As a check on our result, we analytically verify that our anomalous dimension correctly reproduces the logarithms in the interjet energy flow at two-loop order, which were computed in \cite{Becher:2016mmh}. 
 
The anomalous dimension matrix $\bm{\Gamma}^{(2)}$ can be extracted by considering double soft limits of hard functions, which are in essence cross sections of $m$ hard partons along fixed directions. The result for $\bm{\Gamma}^{(2)}$ has three distinct entries: i) double real-emission terms $\bm{d}_{m}$, ii) real-virtual terms $\bm{r}_{m}$ and iii) double virtual contributions $\bm{v}_{m}$. An important difficulty in the determination of the anomalous dimension is the presence of collinear singularities in these terms. These divergences cancel in the end result after combining the different contributions and integrating over the directions, but have to be tracked at intermediate stages of the computation. In order to make the cancellation of collinear singularities manifest and to have a form suitable for implementation in a parton shower framework, we write all terms in $\bm{\Gamma}^{(2)}$ as angular integrals. To obtain the angular integrals, we perform the energy integrations in the loop diagrams using the residue theorem. The soft singularities in both the real and virtual terms are then isolated using an upper cutoff on the momenta. At two-loop order there are different ways to impose this cutoff and we find that the presence of intermediate collinear singularities leads to an ambiguity in the extraction of the coefficient of a term $C_A \pi^2 \,\bm{\Gamma}^{(1)}$ in the two-loop anomalous dimension $\bm{\Gamma}^{(2)}$. For hard functions with a small number of external legs, one can avoid the use of a cutoff by computing the divergences with the full kinematics and we fix the missing coefficient by comparing to the explicit results for the hard and soft functions for dijet production computed in \cite{Becher:2016mmh}.

Our paper is organized as follows. In Section \ref{sec:factorization}, we review the factorization theorem for non-global observables, setup the notation and write down the relation between divergences and the anomalous dimension at two-loop order. As a warm up we then discuss the extraction of the one-loop anomalous dimension $\bm{\Gamma}^{(1)}$  in detail in Section \ref{sec:oneloop}. Already in $\bm{\Gamma}^{(1)}$, one encounters collinear singularities and we discuss their subtraction and cancellation in Section \ref{sec:collSing}. We then proceed in Section \ref{sec:Gamma2_result} to the two- loop anomalous dimension and present the result for $\bm{\Gamma}^{(2)}$ as one obtains it from an analysis of the soft limit of two-loop diagrams (the analysis of the diagrams itself is presented later in Section \ref{sec:gamma2_extraction}). This raw form of the result still contains implicit and explicit collinear divergences which we eliminate by shifting collinear terms in the double emission part $\bm{d}_{m}$ to the real-virtual part $\bm{r}_{m}$. After this collinear rearrangement we obtain the anomalous dimension in a form that agrees with the result of \cite{Caron-Huot:2015bja} (in what the author refers to as $\overline{\text{MS}}$ scheme), but this result corresponds to a renormalization scheme in which all angular integrals are kept $d$-dimensional. We then determine the extra terms that are needed to change to the standard $\overline{\text{MS}}$ scheme. With the final result for the anomalous dimension, we verify in Section \ref{sec:finite} that we correctly reproduce all divergences in the two-loop soft function for interjet energy flow. The most technical part of our paper, namely the extraction of the diagrammatic result for $\bm{\Gamma}^{(2)}$ is presented in Section \ref{sec:gamma2_extraction}. We summarize our results in Section \ref{sec:summary} and go over some of the subtleties encountered in the extraction of the anomalous dimension. We conclude in Section \ref{sec:conclusion}.

\section{Factorization of non-global observables\label{sec:factorization}}

The factorization formula for jet production in $e^+e^-$ collisions with a veto on radiation in part of the phase space takes the form \cite{Becher:2015hka,Becher:2016mmh}
\begin{align}\label{eq:fact}
\sigma(Q, Q_0) &=  \sum_{m=m_0}^\infty \big\langle \bm{\mathcal{H}}_m(\{\underline{n}\},Q,\mu) \otimes \bm{\mathcal{S}}_m(\{\underline{n}\},Q_0,\mu) \big\rangle\, ,
\end{align}
where $m_0$ is the number of final-state jets. The hard function $\bm{\mathcal{H}}_m$ describes the production of $m$ partons in the unconstrained region and the soft function $\bm{\mathcal{S}}_m$ is the matrix element squared of the emission from Wilson lines along the $m$ partons of the hard function. Both of these functions depend on the directions of the $m$ partons $\{\underline{n}\}=\{n_{1},\dots,n_m\}$, which we take as massless. The symbol $\otimes$ indicates the angular integration over these directions
\begin{equation}\label{angInt}
\bm{\mathcal{H}}_m(\{\underline{n}\},Q,\mu) \otimes \bm{\mathcal{S}}_m(\{\underline{n}\},Q_0,\mu) = \prod_{i=1}^m \int \frac{d\Omega_i}{4\pi} \bm{\mathcal{H}}_m(\{\underline{n}\},Q,\mu) \bm{\mathcal{S}}_m(\{\underline{n}\},Q_0,\mu) \,.
\end{equation}
The generalization of \eqref{eq:fact} for processes involving heavy quarks was discussed in \cite{Balsiger:2020ogy}. The above factorization theorem holds for jets with large radius, but similar factorization theorems are also available for narrow jets \cite{Becher:2015hka,Becher:2016mmh} and a variety of other non-global observables such as certain event shape variables \cite{Becher:2016omr,Becher:2017nof} or isolation cone cross sections \cite{Balsiger:2018ezi}.

For $Q_0 \ll Q$, the cross section \eqref{eq:fact} will involve large logarithms irrespective of the choice of the renormalization scale $\mu$. These large logarithms can be resummed by solving the RG equation of the hard function and evolving it from its characteristic scale $\mu_h \sim Q$ down to a soft scale  $\mu_s \sim Q_0$, leading to 
\begin{align} \label{eq:crssctEvo}
\sigma(Q, Q_0) &= \sum_{m=m_0}^\infty 
\big\langle \bm{\mathcal{H}}_m(\{\underline{n}^\prime\},Q,\mu_h) 
\otimes \sum_{l\geq m}^\infty \bm{U}_{ml}(\{\underline{n}\},\mu_s,\mu_h)\,\hat{\otimes}\, 
\bm{\mathcal{S}}_l(\{\underline{n}\},Q_0,\mu_s) \big\rangle \,,
\end{align}
where the evolution factor is the path-ordered exponential of the anomalous dimension 
\begin{align}\label{eq:evolmat}
\bm{U}(\{\underline{n}\},\mu_s,\mu_h) = {\rm \bf P} \exp\left[ \int_{\mu_s}^{\mu_h} \frac{d\mu}{\mu} \bm{\Gamma}(\{\underline{v}\} , \{\underline{n}\},\mu) \right], 
\end{align}
which evolves the $m$-parton configuration along the directions $\{\underline{n}^\prime\}=
\{n_{1},\dots,n_m\}$ into an $l$-parton final state along the directions $\{\underline{n}\}=\{n_{1},\dots,n_l\}$. RG evolution generates additional particles and the symbol $\hat{\otimes}$ denotes the integration over their directions before integrating over the directions of the original hard partons. 

The anomalous dimension in \eqref{eq:evolmat} is the main subject of the present paper. It captures the ultraviolet divergences of the soft emission matrix elements, which are in one-to-one correspondence to the infrared region of the hard parton matrix elements $\bm{\mathcal{H}}_m$. This correspondence of the infrared singularities of hard amplitudes $|\mathcal{M}_m(\{\underline{p}\}) \rangle$ to ultraviolet divergences
can be used to obtain stringent factorization constraints on the infrared divergences of scattering amplitudes \cite{Becher:2009cu, Gardi:2009qi, Becher:2009qa,Dixon:2009ur,Ahrens:2012qz}, which now have been worked out up to four loops \cite{Becher:2019avh}. The situation we analyze is more complicated because our bare hard functions 
\begin{align}\label{eq:Hm}
\bm{\mathcal{H}}_m(\{\underline{n}\},Q,\ep) =\frac{1}{2Q^2} \sum_{\rm spins}
\prod_{i=1}^m & \int \! \frac{dE_i \,E_i^{d-3} }{{\tilde{c}}^\ep\,(2\pi)^{2}} \, |\mathcal{M}_m(\{\underline{p}\}) \rangle \langle  \mathcal{M}_m(\{\underline{p}\}) |\nonumber \\
&\times (2\pi)^d \,\delta\Big(Q - \sum_{i=1}^m E_i\Big) \,\delta^{(d-1)}(\vec{p}_{\rm tot})\,{\Theta }_{\rm in}\!\left(\left\{\underline{n}\right\}\right)
\end{align}
consist of the amplitudes squared, integrated over the particle energies at fixed directions $n_i^\mu = p_i^\mu/E_i$. The function ${\Theta }_{\rm in}\!\left(\left\{\underline{n}\right\}\right) = \theta_{\rm in}(n_1)   \theta_{\rm in}(n_2) \dots \theta_{\rm in}(n_m)$ restricts the $m$ hard particles to be inside the jet region, i.e.\ it prevents these particles from entering the veto region. Both the hard and soft function depend on the geometry of the veto region in which the energy is constrained. The hard function depends on it because the hard partons are restricted to the jet region and the soft function because the energy constraint on soft radiation is only applied in the veto region. In the example of cone jet cross sections, the hard and soft functions depend on the cone angle. We do not need the operator definition of the soft function for our computations in the present paper and refer the interested reader to \cite{Becher:2016mmh}. 

In the definition \eqref{eq:Hm} of the bare hard functions in $d=4-2 \ep$ dimensions, we have strategically included a factor $\tilde{c} = e^{\gamma_E}/\pi$ to the power $\ep$ in the denominator of each energy integral, where $\gamma_E$ is the Euler–Mascheroni constant. This factor cancels a corresponding one arising when expressing the bare coupling $\alpha_0 = g_s^2/(4\pi)$ in terms of the $\overline{\rm MS}$ renormalized one through $\alpha_0 = Z_\alpha \alpha_s (\mu^2\,\tilde{c}/4)^{\ep}$. We will include the $\tilde{c}$-factor in the $d$-dimensional angular integrals defined at the end of this section (see Table \ref{tab:angInts}) to avoid a proliferation of $\ln \tilde{c}$ terms in intermediate expressions. The normalization differs from \cite{Becher:2016mmh}, which had factors of $(2\pi)^{d-2}$ instead of ${\tilde{c}}^\ep\,(2\pi)^{2}$ in the denominators of the energy integrals in the hard function. 

There are two types of infrared singularities which arise in $\bm{\mathcal{H}}_m$ and must be absorbed into $\bm{\Gamma}$: Divergences when one or several of the energies $E_i$ go to zero and IR divergences associated with loop corrections to $|\mathcal{M}_m(\{\underline{p}\}) \rangle$. The first type of singularity is associated with real emissions, the second one with virtual corrections. Of course, at higher orders combinations of both effects arise. In the next section, we will analyze the singularities in detail for the one-loop case, but from the above discussion, we can already anticipate the general structure of the anomalous dimension. Expanding it as
\begin{equation}
\bm{\Gamma} =\frac{\alpha_s}{4\pi} \bm{\Gamma}^{(1)} +\left( \frac{\alpha_s}{4\pi}\right)^2 \bm{\Gamma}^{(2)} + \dots \,,
\end{equation}
the one- and two-loop matrices for the dijet case $m_0=2$ take the form
\begin{align}\label{eq:gammaOneTwo}
\bm{\Gamma}^{(1)} &=  \left(
\begin{array}{ccccc}
   \, \bm{V}_{2} &   \bm{R}_{2} &  0 & 0 & \hdots \\
 0 & \bm{V}_{3} & \bm{R}_{3}  & 0 & \hdots \\
0 &0  &  \bm{V}_{4} &  \bm{R}_{4} &   \hdots \\
 0& 0& 0 &  \bm{V}_{5} & \hdots
   \\
 \vdots & \vdots & \vdots & \vdots &
   \ddots \\
\end{array}\right),
&
\bm{\Gamma}^{(2)} &=  \left(
\begin{array}{ccccc}
   \, \bm{v}_{2} &   \bm{r}_{2} &  \bm{d}_{2} & 0 & \hdots \\
 0 & \bm{v}_{3} & \bm{r}_{3}  & \bm{d}_{3} & \hdots \\
0 &0  &  \bm{v}_{4} &  \bm{r}_{4} &   \hdots \\
 0& 0& 0 &  \bm{v}_{5} & \hdots
   \\
 \vdots & \vdots & \vdots & \vdots &
   \ddots \\
\end{array}
\right).
\end{align}
At $\mathcal{O}(\alpha_s)$, we can either have divergences from a one-loop $\bm{V}_{m}$ virtual correction, or from a single real emission $\bm{R}_{m}$, which maps from the $m$ to the $m+1$ parton space. It therefore occupies the row above the diagonal in $\bm{\Gamma}$. At two loops, we can either have a double real emission $\bm{d}_{m}$, a real-virtual correction $\bm{r}_{m}$ or a two-loop virtual term $\bm{v}_{m}$. We see that each additional order fills in one more row above the diagonal in $\bm{\Gamma}$, while there are never any entries below the diagonal.

As usual, the anomalous dimension is related to the $\bm{Z}$-matrix, which absorbs the divergences of the hard and soft functions. At the two-loop level the $\bm{Z}$-matrix is given by
\begin{equation}\label{twoLoopZ}
{\bm Z}_{lm} = \bm{1}- \frac{\alpha_s}{4\pi}\frac{1}{2\epsilon}\bm{\Gamma}_{lm}^{(1)}  +\left( \frac{\alpha_s}{4\pi}\right)^2\left[ \frac{1}{8\epsilon^2}\sum_{k=l}^{m}\bm{\Gamma}_{lk}^{(1)}  \hat{\otimes} \bm{\Gamma}_{km}^{(1)} +  \frac{\beta_0}{4\epsilon^2}\bm{\Gamma}_{lm}^{(1)} -\frac{1}{4\epsilon}\bm{\Gamma}_{lm}^{(2)}  \right] .
\end{equation}
Applying this matrix to the bare soft functions must render them finite
\begin{equation}\label{eq:rencond}
\sum_{m=l}^\infty {\bm Z}_{lm} \hat{\otimes}\, \bm{\mathcal{S}}_m = \bm{\mathcal{S}}_m^{\rm ren} = \text{ finite}\,.
\end{equation}
Writing this out order-by-order leads to the following finiteness conditions
\begin{align}\label{eq:finitenessOne} 
&-\frac{1}{2\epsilon}\left[\bm{V}_{m}  + \bm{R}_{m} \hat{\otimes} \bm{1}\right] + \bm{\mathcal{S}}^{(1)}_m =  \bm{\mathcal{S}}^{{\rm ren}(1)}_m= \text{ finite}\,, \\ \label{eq:finitenessTwo} 
&  + \frac{1}{8\epsilon^2} \left[\bm{R}_{m}\hat{\otimes} \left(\bm{R}_{m+1}\hat{\otimes} \bm{1}+  \bm{V}_{m+1}\right)+\bm{V}_{m}\left ( \bm{R}_{m}\hat{\otimes} \bm{1} +\bm{V}_{m}\right) \right] \nonumber \\
&\quad +  \frac{\beta_0}{4\epsilon^2}\left[\bm{V}_{m}  + \bm{R}_{m} \hat{\otimes} \bm{1}\right]  -  \frac{\beta_0}{\epsilon}  \bm{\mathcal{S}}^{(1)}_{m}  -\frac{1}{2\epsilon}\left[\bm{V}_{m} \bm{\mathcal{S}}^{(1)}_{m} + \bm{R}_{m}\hat{\otimes} \bm{\mathcal{S}}^{(1)}_{m+1}  \right] \nonumber \\
&\quad\quad - \frac{1}{4\epsilon} \left[\bm{v}_{m}  + \bm{r}_{m} \hat{\otimes} \bm{1}+ \bm{d}_{m} \hat{\otimes} \bm{1} \right]+ \bm{\mathcal{S}}^{(2)}_m =  \bm{\mathcal{S}}^{{\rm ren}(2)}_m= \text{ finite} \,,
\end{align}
where we have used that the leading order soft functions are trivial, $ \bm{\mathcal{S}}^{(0)}_m=\bm{1}$. The term $-\frac{\beta_0}{\epsilon} \bm{\mathcal{S}}_m^{(1)}$ in the second line arises because we expand the bare function in the bare coupling $\alpha_0$ and then perform coupling constant renormalization by replacing $\alpha_0$ with the renormalized coupling $\alpha_s$, wich involves a factor $Z_\alpha = 1- \frac{\beta_0 }{\epsilon} \frac{\alpha_s}{4\pi}$.

We can use the one-loop relation to get explicit expressions for all the divergences at the two-loop level. After expressing $ \bm{\mathcal{S}}^{(1)}_m$ through the renormalized function, we have
\begin{align}\label{resDiv}
 \bm{\mathcal{S}}^{{\rm ren}(2)}_m &=  \bm{\mathcal{S}}^{(2)}_m  - \frac{1}{8\epsilon^2} \left[\bm{R}_{m}\hat{\otimes} \left(\bm{R}_{m+1}\hat{\otimes} \bm{1}+ \bm{V}_{m+1}\right)+(\bm{V}_{m} + 2\beta_0)\left ( \bm{R}_{m}\hat{\otimes} \bm{1} +\bm{V}_{m}\right)  \right] \nonumber \\
& - \frac{1}{4\epsilon} \left[\bm{v}_{m}  + \bm{r}_{m} \hat{\otimes} \bm{1}+ \bm{d}_{m} \hat{\otimes} \bm{1} +2 (\bm{V}_{m}+2\beta_0) \bm{\mathcal{S}}^{{\rm ren}(1)}_{m} + 2 \bm{R}_{m}\hat{\otimes}\, \bm{\mathcal{S}}^{{\rm ren}(1)}_{m+1} \right] ,
\end{align}
where the divergences are now fully explicit. The first line is indeed what we found in \cite{Becher:2016mmh} by iterating the anomalous dimension. In Section \ref{sec:finite}, we will use the second line to check our explicit expression for the two-loop anomalous dimension against the computation of $\bm{\mathcal{S}}^{(2)}_m$ in \cite{Becher:2016mmh}. The relevant result was given in (B.13) of that reference.

\begin{table}
\begin{align}
& \text{Convolution symbol} & & \hspace{-1.8cm} \text{Associated angular integrals} \nonumber  \\  
& \hspace{1.5cm} \otimes &\Dqqq{i} &= \tilde{c}^\ep \int \hspace{-0.10cm} \frac{d^{d-2}\Omega_i}{2 (2\pi)^{d-3}} \nonumber \\
& \hspace{1.5cm} \otimes_2 &   \DQQQ{i}  &= \int \hspace{-0.08cm} \frac{d^2\Omega_i}{4\pi} \nonumber \\
& \hspace{1.5cm} \otimes_\ep &  \otimes &= \otimes_2 + 2\ep\, \otimes_\ep \nonumber
\end{align}
\caption{Angular convolution symbols and associated angular integrals for the individual partons $i=1 \dots m$. In the final result, the factors $\tilde{c} = e^{\gamma_E}/\pi$ cancel against the ones in the energy integrals in \eqref{eq:Hm}. We will use the same notation also for the integrations over additional partons indicated by $\hat{\otimes}$.}\label{tab:angInts}
\end{table}

The $\bm{Z}$-matrix in \eqref{twoLoopZ} and the expressions \eqref{eq:finitenessOne}, \eqref{eq:finitenessTwo} and \eqref{resDiv} involve bare functions and therefore angular integrals in $d$ dimensions. At the end of the day, we will express everything through angular integrations in $d=4$ as in \eqref{angInt} but at intermediate stages it is crucial to keep track of the $\ep$ dependence of the integrals. To do so, we will use the notation in Table \ref{tab:angInts} to distinguish angular integrals in $d=4-2\ep$ and $d=4$. As shown in Table \ref{tab:angInts}, we add a prefactor $\tilde{c}^\ep= (e^{\gamma_E}/\pi)^\ep$ to the angular integrals in $d$-dimensions
 \begin{equation}
\Dqqq{q} =\tilde{c}^\ep \int \hspace{-0.10cm} \frac{d^{d-2}\Omega_q}{2 (2\pi)^{d-3}} =e^{\ep \gamma_E} \int \hspace{-0.10cm} \frac{d^{d-2}\Omega_q}{(4\pi)^{1-\ep}} 
= 1+2 \ep +\left(4-\frac{\pi ^2}{4}\right) \ep ^2+O\left(\ep ^3\right)
\end{equation}
so that their $\ep$-expansion is free from $\gamma_E$'s and $\ln(4\pi)$'s. Note that the angular integral over the $(d-1)$-dimensional unit sphere $\Omega_{d-1}$ is parameterized by $d-2$ angles.

\section{One-loop anomalous dimension from soft factorization}
\label{sec:oneloop}

The basis for the derivation of the anomalous dimension is the factorization of hard amplitudes in the soft limit. For an amplitude with $m$ hard partons with momenta $\{\underline{p}\} = \{p_1,\dots , p_m\}$ and one soft momentum $q$, we have
\begin{align}\label{eq:oneSoftEmission}
| {\cal M}_{m+1}(\{ \underline{p}, q \}) \rangle = \varepsilon^{*\mu} \bm{J}_{\mu,a}(q)  | {\cal M}_{m} (\{ \underline{p} \}) \rangle 
=g_s \sum_{i=1}^m   \bm{T}_i^a \,\frac{ \varepsilon^*\cdot n_i}{n_i\cdot q} \, | {\cal M}_{m} (\{ \underline{p} \}) \rangle\,,
\end{align}
where $\varepsilon^*_\mu$ is the polarization vector of the outgoing gluon. Throughout our paper, we use the color-space formalism \cite{Catani:1996vz} in which the amplitudes are written as ket vectors in the color space of the partons in the amplitude. The color-space notation suppresses the color indices and works with color generators $\bm{T}_i^a$ for the $i$-th external line, where $(\bm{T}_i^a)_{\alpha\beta}=t_{\alpha\beta}^a$ for a final-state quark or initial-state anti-quark, $(\bm{T}_i^a)_{\alpha\beta}=-t_{\beta\alpha}^a$ for a final-state anti-quark or initial-state quark, and $(\bm{T}_i^a)_{bc}=-if^{abc}$ for a gluon. We have written the soft current $\bm{J}_{\mu,a}(q)$ at leading order in the coupling constant, but also the one \cite{Catani:2000pi}
 and two-loop results \cite{Duhr:2013msa,Dixon:2019lnw} for this quantity are available. Note that the soft current is transverse when applied to the amplitude
\begin{equation}
q^\mu\, \bm{J}_{\mu,a}(q) |{\cal M}_{m} ( \{ \underline{p} \}) \rangle  = g_s \sum_{i=1}^m   \bm{T}_i^a  | {\cal M}_{m} (\{ \underline{p} \}) \rangle = 0
\end{equation}
due to color conservation. Below, we will use soft factorization to also extract the two-loop anomalous dimension. 

To get the leading-order real emission $\bm{R}_{m}$ entry in the anomalous dimension, we start with the hard function for $m+1$ partons, 
\begin{align}\label{eq:HmP1}
\bm{\mathcal{H}}_{m+1}(\{\underline{n},n_q\},Q,\ep) =\frac{1}{2Q^2}& \sum_{\rm spins}
 \prod_{i=1}^{m+1}  \int \! \frac{dE_i \,E_i^{d-3} }{{\tilde{c}}^\ep\,(2\pi)^{2}} \, |\mathcal{M}_{m+1}(\{\underline{p},q\}) \rangle \langle  \mathcal{M}_{m+1}(\{\underline{p},q\}) |\nonumber \\
&\times (2\pi)^d \,\delta\Big(Q - \sum_{i=1}^{m+1} E_i\Big) \,\delta^{(d-1)}(\vec{p}_{\rm tot}+\vec{q})\,{\Theta }_{\rm in}\!\left(\left\{\underline{p},q\right\}\right), 
\end{align}
and assume that the momentum $q$ becomes soft. To isolate the associated infrared divergence, we put a cutoff $\Lambda$ on the energy $E_q$. This cutoff is a technical tool to simplify our computation. As is obvious from \eqref{eq:HmP1}, even without a cutoff the energy is constrained in the hard function and for small $m$ we can simply compute $\bm{\mathcal{H}}_{m+1}$ in dimensional regularization to obtain its divergences. However, the cutoff allows us to isolate the divergence associated with $q$ becoming soft and to use the associated soft factorization formula \eqref{eq:oneSoftEmission}. In the presence of the cutoff, we can also expand the soft momentum $q$ out of the phase-space constraints and get results for arbitrary $m$. Introducing the cutoff and expanding in the soft momentum yields
\begin{align}\label{eq:realSoft}
\bm{\mathcal{H}}_{m+1}(\{\underline{n},n_q \},Q,\ep) = -g_s^2 \int_0^{\Lambda} \frac{dE_q \,E_q^{d-5} }{\tilde{c}^\ep\,(2\pi)^{2}}\,{\theta }_{\rm in}\!\left(q\right)  \sum_{(ij)} \frac{n_i\cdot n_j}{n_i\cdot n_q\, n_j \cdot n_q}  \bm{T}^a_i\, \bm{\mathcal{H}}_{m}(\{\underline{n}\},Q,\ep)\, \bm{T}_j^{\tilde{a}}\, .
\end{align}
The minus sign arises from the sum over polarizations and the notation $(ij)$ refers to unordered pairs of indices. The color index of the soft gluon in the amplitude is $a$, the one in the conjugate amplitude $\tilde{a}$. The color indices of the hard function are contracted with the ones of the soft function in the factorization theorem \eqref{eq:fact}. To evaluate the color trace in \eqref{eq:fact} with the trivial leading-order soft function, $ \bm{\mathcal{S}}^{(0)}_m=\bm{1}$, one will set $a=\tilde{a}$ and sum over all values of $a$, but in general we must distinguish the color indices of the amplitude and the conjugate amplitude. Having explicit color indices is against the spirit of the color-space formalism in which indices are suppressed, but the left-hand side of \eqref{eq:realSoft} lives in the color space of $m+1$ particles, while $\bm{\mathcal{H}}_{m}$ on the right-hand side lives in the space of $m$ partons so that we have to make the extra color index explicit.

Expressing the bare coupling $g_s$ through the $\overline{\text{MS}}$ one and performing the energy integral, we then obtain 
\begin{align}
\bm{\mathcal{H}}_{m+1}(\{\underline{n},n_q \},Q,\ep) =\frac{2}{\ep} \frac{\as}{4\pi}\left(\frac{\mu}{2\Lambda}\right)^{2\ep}  \theta_{\rm in}(n_q) \sum_{(ij)} W_{ij}^q\,  \bm{T}^a_i\, \bm{\mathcal{H}}_{m}(\{\underline{n}\},Q,\ep)\,  \bm{T}_j^{\tilde{a}} \, .
\end{align}
Note that the color generator multiplying the amplitude appears on the left, while the one multiplying the conjugate amplitude appears on the right. In order to indicate where they act, we will denote the generators by $\bm{T}^a_{i,L}$ and $\bm{T}^{\tilde{a}}_{i,R}$ respectively. In this notation the corresponding entry in the anomalous dimension matrix then reads
\begin{align}\label{eq:oneLoopR}
 \bm{R}_m & = -4\,\sum_{(ij)}\,\bm{T}_{i,L}^a \bm{T}_{j,R}^{\tilde{a}}  \,W_{ij}^{q}\,  \theta_{\rm in}(n_q)\,,
\end{align}
since the anomalous dimension is $(-2)$ times the coefficient of the divergence. The anomalous dimension $\bm{R}_m$ maps a hard function with $m$ partons onto a hard function with $m+1$ partons.

There is an important comment we need to make at this point. It is obvious that integrals over the anomalous dimension \eqref{eq:oneLoopR} suffer from collinear divergences when $n_q$ becomes collinear to $n_i$ or $n_j$. These collinear divergences will cancel against collinear divergences in the virtual corrections when the anomalous dimension is applied to the soft functions.  This cancellation of divergences can be made manifest on the level of the integrand by writing also the virtual corrections as angular integrals, which is what we will do in the following. The cancellation of collinear divergences will be discussed in detail in the next section. While the collinear divergences cancel in the end, their presence means that \eqref{eq:oneLoopR} is not well defined on its own. To make it well defined, one could introduce an intermediate collinear regulator when computing the individual pieces. This is done e.g.\ in parton showers, which put a small cone around each hard parton and then compute both the real and virtual contributions with a cutoff.  However, with the implicit divergences present, \eqref{eq:oneLoopR} is not the anomalous dimension in standard dimensional regularization: for this to be true, we would need to regularize and extract all divergences dimensionally. We will come back to this issue in the next section.

Applying the soft factorization \eqref{eq:oneSoftEmission} at the one-loop-level, the virtual correction takes the form
\begin{multline}\label{eq:oneLoopVirtual}
\bm{\mathcal{H}}_{m}(\{\underline{n}\},Q,\ep) = \frac{g_s^2}{2} \sum_{(ij)} \int\! \frac{d^d q}{(2\pi)^d} \frac{-i }{q^2+i0} \frac{n_i\cdot n_j}{[n_i\cdot q+i 0]\, [-n_j \cdot q+ i 0]}\\
\times \bm{T}_{i}\cdot\bm{T}_{j} \, \bm{\mathcal{H}}_{m}(\{\underline{n}\},Q,\ep) + \text{h.c.} \,,
\end{multline}
where the factor two corrects for the fact that the unordered sum $(ij)$ includes every term twice and $\bm{T}_{i}\cdot\bm{T}_{j}=\sum_a \bm{T}_{i}^a \bm{T}_{j}^a$. We have written out the loop correction to the amplitude, the hermitian conjugate adds the correction from the loop in the conjugate amplitude, which has the opposite $i0$ prescription and has the color matrices act on the right-hand side of the hard function. The $i0$ prescriptions in the above integral refer to outgoing lines $i$ and $j$. The result with incoming lines is obtained by substituting $n_k \to -n_k$ for $k=i$ or $k=j$, which switches the $i0$ on the corresponding line. 

To write the infrared divergence as an angular integral, we first perform the integration over $q^0$ using the residue theorem and then extract the IR divergence of the $|\mathbf{q}|$ integral. To simplify the discussion, we use the identity
\begin{equation}
\frac{1}{n_i\cdot q+i 0} =  \frac{-1}{-n_i\cdot q+i 0} - 2\pi i \delta(n_i \cdot q)
\end{equation}
inside the integral \eqref{eq:oneLoopVirtual}. In the part without the $\delta$-function both poles of the light-cone propagators have positive imaginary parts. They can thus be avoided closing the contour in the lower half-plane, which picks up the residue at $q^0=\E{q}-i0$. Doing so and then performing the energy integral with an upper cutoff, as in \eqref{eq:realSoft}, we get
\begin{equation}
\bm{\mathcal{H}}_{m}^{(a)}(\{\underline{n} \},Q,\ep) = - \frac{\alpha_s}{4\pi}\frac{1}{\ep} \left(\frac{\mu}{2\Lambda}\right)^{2\ep}  \sum_{(ij)}\bm{T}_i \cdot \bm{T}_j  \,\Dqqq{q} W_{ij}^q \, \bm{\mathcal{H}}_m(\{\underline{n} \},Q,\ep)  + \text{h.c.} \, ,
\end{equation}
which corresponds to an anomalous dimension entry of
\begin{align}\label{eq:oneLoopV}
 \bm{V}^{(a)}_m &=  2\,\sum_{(ij)}\,(\bm{T}_{i,L}\cdot  \bm{T}_{j,L}+\bm{T}_{i,R}\cdot  \bm{T}_{j,R})  \Dqqq{q}\, W_{ij}^q\, ,
\end{align}
where we added the contribution of the conjugate amplitude.

For the part with the $\delta$-function we first symmetrize the integrand under $q\to-q$ to get a second delta function 
\begin{equation}
\bm{\mathcal{H}}_{m}^{(b)}(\{\underline{n} \},Q,\ep) = i\pi^2 g_s^2\sum_{(i j)}\bm{T}_i^a \bm{T}_j^a  \Dq{q}\frac{n_i\cdot n_j \,\delta(n_i\cdot q)\delta(n_j\cdot q)}{q^2+i0}\bm{\mathcal{H}}_{m}(\{\underline{n} \},Q,\ep)\,.
\end{equation}
The leading IR pole of this is unambiguous, subleading terms in $\ep$ depend on the details of the UV regulator. The easiest way to obtain the IR pole is the replacement 
\begin{equation}
 \Dq{q}\to\frac{1}{n_i\cdot n_j}\int\frac{d(n_i\cdot q)d(n_j\cdot q)d^{d-2}\mathbf{q}_\perp}{(2\pi)^d}\theta(\Lambda-|\mathbf{q}_\perp|)\,,
\end{equation}
where treating the perpendicular component as purely spatial gives the correct result for the leading pole because such a frame obviously exists. Then the divergent part reads
\begin{equation}
 \bm{\mathcal{H}}_{m}^{(b)}(\{\underline{n} \},Q,\ep) = \frac{\alpha_s}{4\pi}\,\frac{i\pi}{\ep}\sum_{(ij)}\bm{T}_i\cdot \bm{T}_j \bm{\mathcal{H}}_{m}(\{\underline{n} \},Q,\ep) \,,
\end{equation}
which implies that the contribution to the anomalous dimension is
\begin{equation}
\bm{V}^{(b)}_m =  - 2 \sum_{(ij)} \left[\bm{T}_{i,L}\cdot \bm{T}_{j,L} - \bm{T}_{i,R}\cdot \bm{T}_{j,R}\right]\times i\pi \, \Pi_{ij}\,.
\end{equation}
The minus sign for the conjugate amplitude arises because it has the opposite imaginary part. The imaginary part originates from the cut through the two eikonal lines and is often referred to as the Glauber or Coulomb phase. It is not present when one line is incoming and the other outgoing. This is reflected by the factor 
\begin{equation}
 \Pi_{ij} = \begin{cases}
                1,\hspace{1cm}i,j\text{ both incoming or outgoing}\,,\\
                0,\hspace{1cm} \text { otherwise}\,.
               \end{cases}
\end{equation}
For the rest of the paper we will focus on cases where all lines are either incoming or outgoing. For these cases $\Pi_{ij}=1$ for all $i,j$ and we can then use color conservation
\begin{equation}
\sum_i \bm{T}_{i} = 0
\end{equation}
to eliminate the Glauber phase
\begin{equation}
 \sum_{(ij)} \left[\bm{T}_{i,L}\cdot \bm{T}_{j,L} - \bm{T}_{i,R}\cdot \bm{T}_{j,R}\right] = \sum_{i} \left[- \bm{T}_{i,L}\cdot\bm{T}_{i,L}  + \bm{T}_{i,R}\cdot \bm{T}_{i,R} \right] = 0\,,
 \end{equation} 
 since $\bm{T}_{i,L}^2 =  \bm{T}_{i,R}^2 = C_i$, the quadratic Casimir associated with the corresponding line. 

\section{Collinear singularities}
\label{sec:collSing}

We now discuss the collinear singularities in the one-loop anomalous dimension and their cancellation. The one-loop anomalous dimension was obtained by considering a single soft emission and the collinear divergences arise when the emission is along the direction of one of the $m$ hard partons in the function $\bm{\mathcal{H}}_{m}$. As we will explain in this section, these collinear configurations cancel between real and virtual terms in the anomalous dimension. Since they cancel within the anomalous dimension they can be kept implicit and do not need to be subtracted. However, in the determination of the two-loop anomalous dimension we also encounter collinear singularities when two soft emissions become collinear. These cancel against collinear configurations in the soft functions and are part of  the anomalous dimensions, as will be explained in the later sections.

\subsection{Subtraction and cancellation of collinear singularities}
\label{sec:colllinearCancellation}
When adding the real and virtual contributions and applying them to the trivial tree-level soft functions $\bm{\mathcal{S}}_m =\bm{1}$, one finds
\begin{equation}\label{oneloop}
\langle \bm{\mathcal{H}}_{m} \otimes \left(   \bm{V}_{m} \, \bm{1}  + \bm{R}_{m} \hat{\otimes} \bm{1} \right)  \rangle  = 4\,\sum_{(ij)} \Dqqq{q}\, W_{ij}^q  \,\theta_{\rm out}(n_q)   \langle  \bm{\mathcal{H}}_{m}  \otimes  \bm{T}_{i}\cdot  \bm{T}_{j} \rangle \,,
\end{equation}
where we used the cyclicity of the color trace $\langle \dots \rangle$ to move the color generators acting on the left of the hard function to the right. We also defined $\theta_{\rm out}(n_q)= 1- \theta_{\rm in}(n_q)$. Since the gluon is always outside, it can never become collinear to the hard partons along directions $i$ and $j$ so that the angular integration is finite. A useful property of the result is that the soft function is trivial in the fully inclusive limit where no veto is imposed, i.e.when there is no outside region. This property links the virtual corrections directly to the real emission result and is useful at higher orders.

To demonstrate the cancellations of collinear singularities more generally and to have a form of the anomalous dimension that is valid within ordinary dimensional regularization, we now make the collinear divergences in $\bm{V}_m$ and $\bm{R}_m$ explicit. To do so, we introduce an angular $\delta$-function
 \begin{equation}\label{angDelta}
 \Dqqq{q} \delta(n_q-n_i) f(n_q) = f(n_i)\,
\end{equation}
and a collinearily subtracted dipole operator, which acts on angular functions $f(n_q)$ as follows \cite{Forshaw:2019ver,Becher:2021zkk}
\begin{equation}\label{eq:collSub}
 \Dqqq{q}  \overline{W}_{ij}^q\, f(n_q) =  \Dqqq{q}  \left[W_{ij}^q f(n_k)- \frac{1}{n_i\cdot n_q} f(n_i)  - \frac{1}{n_j\cdot n_q} f(n_j)\right]\, .
\end{equation}
In the subtraction terms, we can carry out the $d$-dimensional angular integration using
 \begin{equation} \label{eq:collint1}
c(\ep) = \Dqqq{q}   \frac{1}{n_i\cdot n_q} = \frac{e^{\ep \gamma_E} \Gamma(-\ep)}{2\Gamma(1-2\ep)} = -\frac{1}{2\e}+\mathcal{O}(\e)\,.
  \end{equation}
We can now separate out the collinear pieces from the anomalous dimension  
\begin{align}\label{eq:oneLoopSub}
 \bm{R}_m & =  \overline{\bm{R}}_m +  \bm{R}^c_m  = -4\,\sum_{(ij)}\,\bm{T}_{i,L}\cdot\bm{T}_{j,R}  \,\left[\overline{W}_{ij}^{q}  \theta_{\rm in}(n_q) + c(\ep)\, \delta(n_q-n_i) + c(\ep)\,\delta(n_q-n_j) \right] ,\nonumber\\
\bm{V}_m & =  \overline{\bm{V}}_m +  \bm{V}^c_m  =2\,\sum_{(ij)}\,(\bm{T}_{i,L}\cdot  \bm{T}_{j,L}+\bm{T}_{i,R}\cdot  \bm{T}_{j,R})  \,\left[ \Dqqq{q} \overline{W}_{ij}^{q}  \theta_{\rm in}(n_q) + 2\, c(\ep) \right]\, ,
\end{align}
and we can further simplify the collinear pieces using color conservation
\begin{align}\label{eq:oneLoopColl}
\bm{R}^c_m  &= +8\,c(\ep) \sum_{i=1}^m\,\bm{T}_{i,L}^a\bm{T}_{i,R}^{\tilde{a}} \,\delta(n_q-n_i) \,, \nonumber \\
\bm{V}^c_m  &=-8 \, c(\ep)\,\sum_{i=1}^m\,C_i\, \bm{1}\,.
\end{align}
To establish the cancellation of collinear singularities we should show that when the combination acts
on a soft function we get zero. Applying the real and virtual part, we obtain
\begin{multline}
\left \langle \bm{\mathcal{H}}_{m} \otimes \left( \bm{V}^c_{m} \bm{\mathcal{S}}_m+  \bm{R}^c_{m} \hat{\otimes}\bm{\mathcal{S}}_{m+1}\right) \right \rangle \\
=-8 \, c(\ep)\,\sum_{i=1}^m \,C_i   \, \left \langle \bm{\mathcal{H}}_{m} \otimes \bm{\mathcal{S}}_m \right \rangle  + 8\,c(\ep)\,  \left \langle  \bm{T}_{i}^a  \bm{\mathcal{H}}_{m} \bm{T}^{\tilde{a}}_{i}  \otimes \bm{\mathcal{S}}_{m+1} \right\rangle  \,.  \label{eq:GammacollAction}
\end{multline}
Note that we have performed the trivial angular integration over the extra gluon produced by $\bm{R}^c_m$, so that two of the Wilson lines in $\bm{\mathcal{S}}_{m+1}$ are collinear. For the trivial tree-level soft functions $\bm{\mathcal{S}}_{m} = \bm{1}$ the cancellation of collinear singularities among the real and virtual contributions immediately follows from the cyclicity of the trace
\begin{align}
\left \langle  \bm{T}_{i}^a \bm{\mathcal{H}}_{m} \bm{T}^{\tilde{a}}_{i}  \otimes \bm{1} \right\rangle &= 
\left \langle  \bm{\mathcal{H}}_{m} \bm{T}_{i}^a \bm{T}^{a}_{i}  \otimes \bm{1} \right\rangle = C_i \left \langle  \bm{\mathcal{H}}_{m}  \otimes \bm{1} \right\rangle . \label{eq:collcanc}
\end{align}
However, it also holds for the nontrivial higher-order soft functions $\bm{\mathcal{S}}_m$ as we shall demonstrate now. To see this consider the splitting amplitude for the process in which a parent parton $P$ splits into collinear partons $1$ and $2$:
\begin{equation}
\mbox{\bf Sp}(\{p_1,p_2\})\, |\mathcal{M}_{m}(\{P, p_3, \dots, p_{m+1}\}) \rangle =
 |\mathcal{M}_{m+1}(\{p_1,p_2, p_3 ,\dots , p_{m+1}\}) \rangle \, .
\end{equation}
Charge conservation implies that the splitting amplitude fulfills (see e.g.\ 
\cite{Catani:2003vu,Becher:2009qa})
\begin{equation}\label{eq:splitColor}
\left(\bm{T}_1^a+\bm{T}_2^a\right) \mbox{\bf Sp}(\{p_1,p_2\}) = \mbox{\bf Sp}(\{p_1,p_2\})\, \bm{T}_P^a\,.
\end{equation}
The color state of the partons after the decay corresponds to the color state of the parent parton.
The soft function for the two collinear partons contains Wilson lines for partons $1$ and $2$ along the common direction $P$. Since the generators $1$ and $2$ commute, the Wilson lines combine into a single Wilson line with color $\bm{T}_1^a+\bm{T}_2^a$:
\begin{equation}
\bm{S}_1(n_1) \bm{S}_2(n_2 ) = \bm{S}_1(n_P) \bm{S}_2(n_P) = \bm{S}_{1+2}(n_P)\,.
\end{equation} 
Then one can use the color identity for splitting functions \eqref{eq:splitColor} to exchange the Wilson line and the splitting amplitude
\begin{multline}\label{eq:WilsonSplit}
\bm{S}_{1+2}(n_P) \,\mbox{\bf Sp}(\{p_1,p_2\})\, |\mathcal{M}_{m}(\{P, p_3, \dots, p_{m+1}\}) \rangle \\
=  \mbox{\bf Sp}(\{p_1,p_2\})\, \bm{S}_{P}(n_P) \, |\mathcal{M}_{m}(\{P, p_3, \dots, p_{m+1}\}) \rangle\,.
\end{multline}
In other words, the Wilson lines after the collinear splitting can be combined into the Wilson line of the parent parton. The operator $\bm{R}^c_{m}$ is the soft limit of a sum of splitting functions for the emission of an additional gluon from each leg and we can thus use the identity \eqref{eq:WilsonSplit} to move the splitting after the soft function, i.e.
\begin{equation}\label{eq:splitSoft}
 \bm{\mathcal{H}}_m\, \bm{R}^c_{m} \,\bm{\mathcal{S}}_{m+1} =  \bm{\mathcal{H}}_m \,\bm{\mathcal{S}}_{m} \bm{R}^c_{m} \, ,
\end{equation}
where the soft function $\bm{\mathcal{S}}_{m+1}$ on the left-hand side contains two collinear partons, while the one on the right-hand side  only has the parent parton $\bm{\mathcal{S}}_{m}$. Equation \eqref{eq:splitSoft} expresses soft coherence on the operator level and once we use it to shift the collinear pieces to the end of the emission chain they cancel against the collinear singularities of the virtual diagrams when taking the color trace as in \eqref{eq:collcanc}. With this we establish the cancellation of collinear singularities also for nontrivial soft functions.

Suppressing the multiplicity index and writing the one-loop anomalous dimension as
\begin{equation}
\bm{\Gamma}^{(1)} = \bar{\bm{\Gamma}} + \bm{\Gamma}^c
\end{equation}
one can also show that $\bm{\Gamma}^c\, \bar{\bm{\Gamma}}  =  \bar{\bm{\Gamma}} \,\bm{\Gamma}^c$ \cite{Becher:2021zkk}, using the same arguments as for the soft function, so that we can commute all collinear pieces $ \bm{\Gamma}^c$ to the right in expressions such as
\begin{equation}\label{eq:GammaOneProd}
\langle \bm{\mathcal{H}}\otimes  \bm{\Gamma}^{(1)} \hat{\otimes}  \bm{\Gamma}^{(1)} \dots \hat{\otimes}  \bm{\Gamma}^{(1)} \hat{\otimes} \bm{\mathcal{S}} \rangle \,.
 \end{equation}
 Then one moves the collinear pieces past the soft function using \eqref{eq:splitSoft}
where they vanish as shown in \eqref{eq:collcanc}. We can thus replace $\bm{\Gamma}^{(1)}$ by $\bar{\bm{\Gamma}}$ in \eqref{eq:GammaOneProd}. Due to the cancellation of collinear pieces, the form of the anomalous dimension is not unique, we can always add or subtract collinear pieces of the form \eqref{eq:oneLoopColl} with coefficients that depend on the individual legs. 

Let us add an important side remark: in processes with initial-state partons the Glauber phases do not cancel. The Glauber phases do not commute with the collinear emissions $\bm{\Gamma}^c$ associated with initial state legs, which then leads to the appearance of super-leading logarithms \cite{Forshaw:2006fk,Forshaw:2008cq,Becher:2021zkk}.

\subsection{Collinear anomalous dimensions in dimensional regularization}
\label{sec:collineardimreg}
As discussed above, the form of the infrared singularities of massless scattering amplitudes $|\mathcal{M}_m(\{\underline{p}\}) \rangle$ is by now very well known \cite{Becher:2009cu, Gardi:2009qi, Becher:2009qa,Dixon:2009ur,Ahrens:2012qz,Becher:2019avh}. Up to two loops their structure is especially simple as they follow from an anomalous dimension which has the dipole form
\begin{equation}\label{eq:GammaAmp}
\bm{\Gamma}(\{\underline{s}\},\mu) 
   = \sum_{(ij)}\,\frac{\bm{T}_i\cdot\bm{T}_j}{2}\,\gamma_{\rm cusp}(\alpha_s)\,\ln\frac{\mu^2}{-s_{ij}} 
    + \sum_i\,\gamma^i(\alpha_s)\,\bm{1}\,.
\end{equation}
This anomalous dimension does not immediately translate into a result for $\bm{V}_m$ and $\bm{v}_m$ because the definition of the hard function \eqref{eq:Hm} involves integrals over the energies of the hard partons. However, for small values of $m$ the energy integrals are fixed by the momentum conservation constraints since the directions of the hard partons are fixed. The hard function involves $m-d$ integrations, so that for $m\leq 4$, the energies of the partons can be expressed in terms of directions and the center-of-mass energy $Q$. For $m\leq 4$, equation \eqref{eq:GammaAmp} directly yields the result for the virtual anomalous dimensions in dimensional regularization. At one-loop, we have
\begin{equation}\label{eq:oneLoopVirt}
  \bm{V}^{\rm dim.reg.}_m =- \sum_{(ij)}\,\frac{\Gamma_0}{2}\, \left(\bm{T}_{i,L}\cdot  \bm{T}_{j,L} \ln\frac{\mu^2}{-s_{ij}+i 0} +\bm{T}_{i,R}\cdot  \bm{T}_{j,R} \ln\frac{\mu^2}{-s_{ij}-i 0}\right) 
    - \sum_i\, 2 \gamma^i_0\,\bm{1} \,.
 \end{equation}
 The one-loop coefficients are $\Gamma_0= 4$, and $\gamma^i_0 = \gamma^q_0 = -3 C_F$ for quarks and $\gamma^i_0 = \gamma^g_0 = -\beta_0$ for gluons. The formula for the two-loop result $\bm{v}_m$ has the same form with the corresponding two-loop coefficients. We note that $s_{ij}\equiv 2\sigma_{ij}\,p_i\cdot p_j+i0$, where the sign factor $\sigma_{ij}=+1$ if the momenta $p_i$ and $p_j$ are both incoming or outgoing, and $\sigma_{ij}=-1$ otherwise. Using that
\begin{equation}
\ln\frac{\mu^2}{-s_{ij}+i 0}  = \ln\frac{\mu^2}{2 p_i\cdot p_j} - i \pi\, \Pi_{ij}
\end{equation}
we see that formula indeed produces the Glauber phases, which cancel by color conservation if all particles are outgoing.

According to the discussion above, we should find that this anomalous dimension can be written in terms of the collinear subtracted anomalous dimension plus collinear pieces
\begin{equation}\label{eqref:collStruct}
 \bm{V}^{\rm dim.reg.}_m  = \overline{\bm{V}}_m + \sum_i \bm{V}^{\rm coll.}_i \,.
\end{equation}
To match the two forms, we can use that
\begin{equation}
\DQQQ{q}\, \overline{W}_{ij}^q = \ln\frac{n_i\cdot n_j}{2}\,,
\end{equation}
which allows us to write
\begin{align}
\bm{V}^{\rm dim.reg.}_m  &= \overline{\bm{V}}_m - \sum_{(i,j)}\,\frac{\Gamma_0}{2}\, \left(\bm{T}_{i,L}\cdot  \bm{T}_{j,L}  +\bm{T}_{i,R}\cdot  \bm{T}_{j,R} \right) \ln\frac{ \mu^2}{2 E_i 2 E_j }  - \sum_i\, 2 \gamma^i_0\,\bm{1}  \\
&=  \overline{\bm{V}}_m + \sum_{i} 2 \left(C_i \,\Gamma_0\, \ln\frac{ \mu}{2 E_i} -\gamma^i_0 \right)\bm{1} \,,
\end{align}
where we have used color conservation in the second line. The result clearly displays the structure anticipated in \eqref{eqref:collStruct}.

\section{\boldmath Two-loop anomalous dimension $\Gamma^{(2)}$\label{sec:Gamma2_result}}

In this section, we now turn to the two-loop anomalous dimension. As in the one-loop case discussed in detail in Section \ref{sec:oneloop}, the anomalous dimension can be extracted by considering the soft limit of amplitudes and extracting the infrared singularities associated with this limit. To get the contribution associated with the two-loop anomalous dimension, one has to first remove the contributions which arise from iterating the one-loop anomalous dimension. We will discuss this in detail in Section~\ref{sec:gamma2_extraction}, but want to first present the result of this extraction and discuss subtleties associated with collinear divergences and with the choice of the renormalization scheme. 

\subsection{Diagrammatic result for the anomalous dimension\label{sec:Gamma2_result_diag}}

As indicated in \eqref{eq:gammaOneTwo}, the two-loop anomalous dimension has three different types of contributions. The entries $\bm{d}_m$ are associated with double real emissions, $\bm{r}_m $ describe real-virtual terms and $\bm{v}_m$ contains purely virtual divergences. The part of the anomalous dimension, which describes the emission of two particles along the directions $n_q$ and $n_r$ is extracted by considering the soft current for the emission of two particles, identifying the associated divergences and removing the strongly ordered contribution which corresponds to the iterated one-loop result, as was done in \cite{Caron-Huot:2015bja} and will be detailed in Section \ref{sec:gamma2_extraction}. The result for the double real part of the anomalous dimension is given by 
\begin{align}
 \bm{d}_m ={}& \sum_{(i j)}\sum_k i f^{abc}\left(\bm{T}_{i,L}^a \bm{T}_{j,L}^b \bm{T}_{k,R}^c - \bm{T}_{i,R}^a\bm{T}_{j,R}^b \bm{T}_{k,L}^c\right) K_{ijk;qr} \theta_\text{in}(n_q)\theta_\text{in}(n_r) \nonumber\\
                 & - 2 \sum_{(ij)}\bm{T}_{i,L}^a \bm{T}_{j,R}^a\left ( K_{ij;qr} \theta_\text{in}(n_q) \theta_\text{in}(n_r) 
                  +  \Gamma_{\rm coll} W_{ij}^q\, \theta_\text{in}(n_q) \delta(n_q-n_r) \right) .
 \label{eq:dm0}
\end{align}
As in the one-loop result \eqref{eq:oneLoopR}, we should distinguish the color indices of the emitted particles in the amplitude from the ones in the conjugate amplitude.  To be precise, we should for example replace
\begin{equation}\label{eq:rec}
i f^{abc}\bm{T}_{i,L}^a \bm{T}_{j,L}^b \bm{T}_{k,R}^c \to \bm{T}_{i,L}^a \bm{T}_{j,L}^b i f^{\tilde{a}\tilde{b}c} \bm{T}_{k,R}^c
\end{equation}
in the first term in $\bm{d}_m$ to properly indicate that two new gluons are produced with color indices $a$, $b$ in the amplitude and $\tilde{a}$, $\tilde{b}$ in the conjugate amplitude. To keep the notation compact, we will write the color indices in contracted form as in \eqref{eq:dm0} throughout the main text. Similarly to the example \eqref{eq:rec} it is easy to reconstruct the full result with open color indices for the other terms. We will do so in Section \ref{sec:summary} when we present the final result for the anomalous dimension.

The anomalous dimension $\bm{d}_m$ in \eqref{eq:dm0} consists of three pieces. The first is a term involving three color generators, proportional to the angular function \cite{Caron-Huot:2015bja}
\begin{align}\label{eq:Kijkqr}
 K_{ijk;qr} ={}& 8\left(W_{ik}^q W_{jk}^r - W_{ik}^q W_{jq}^r - W_{ir}^q W_{jk}^r  + W_{ij}^qW_{jq}^r\right) \ln\left(\frac{n_{kq}}{n_{kr}}\right) ,
\end{align}
where we introduced the abbreviation $n_{ab} = n_a \cdot n_b$. This function has the property that it vanishes in all collinear limits. The notation $(ij)$ in \eqref{eq:dm0} refers to a sum over unordered pairs of legs, which includes both a term with $i=1$, $j=2$ and one with $i=2$, $j=1$. Below we will also encounter sums over unordered triplets $(ijk)$ of legs. As it stands, the first term in \eqref{eq:dm0} also generates contributions involving only two legs, since the sum over $k$ is unconstrained and includes $k=i$ and $k=j$, however due to $K_{iji;qr}=K_{ijj;qr} = 0$ these terms vanish. Following Caron-Huot \cite{Caron-Huot:2015bja}, we split the two-particle term as follows
\begin{equation}\label{eq:Kijqr}
 K_{ij;qr} = C_A K_{ij;qr}^{(a)} + \left[ n_F T_F-2C_A \right] K_{ij;qr}^{(b)} + \left[C_A -2n_FT_F+n_ST_S \right] K_{ij;qr}^{(c)}\,,
 \end{equation}
where $n_F$ and $n_S$ are the number of fermions and scalars included in the theory and $T_F$ and $T_S$ the traces of the associated generators. The individual functions are given by  \cite{Caron-Huot:2015bja}
\begin{align} \label{eq:KijqrFuns}
K_{ij;qr}^{(a)} &= \frac{4n_{ij}}{n_{iq}n_{qr}n_{jr}}\left[1+\frac{n_{ij}n_{qr}}{n_{iq}n_{jr}-n_{ir}n_{jq}}\right]\ln\frac{n_{iq}n_{jr}}{n_{ir}n_{jq}} \, , \nonumber\\
K_{ij;qr}^{(b)} &= \frac{8 n_{ij}}{n_{qr}(n_{iq}n_{jr}-n_{ir}n_{jq})}\ln\frac{n_{iq}n_{jr}}{n_{ir}n_{jq}} \, ,\\
K_{ij;qr}^{(c)} & =\frac{4}{n_{qr}^2}\,\left( \frac{n_{iq}n_{jr}+n_{ir}n_{jq}}{n_{iq}n_{jr}-n_{ir}n_{jq}}\ln\frac{n_{iq}n_{jr}}{n_{ir}n_{jq}} -2 \right)  .  \nonumber
\end{align}
These functions are finite when $q$ or $r$ become collinear the legs $i$ and $j$, but there are collinear divergences when $q$ becomes collinear to $r$. In \cite{Caron-Huot:2015bja} the motivation for splitting the two-particle term in the above form was that only the first term is present in $\mathcal{N}=4$ supersymmetric Yang-Mills theory. We are primarily interested in QCD, but allowing for $n_S$ scalars is also useful when adopting renormalization schemes such as dimensional reduction (DRED) which involve $\ep$-scalars, see e.g. \cite{Gnendiger:2017pys}.

In addition to the implicit collinear divergences present in $K_{ij;qr}$, the final term in the anomalous dimension  involves a purely collinear term proportional to  
\begin{equation}\label{eq:Gammacoll}
\Gamma_{\rm coll} = \gamma_1^\text{cusp} - \frac{2C_A}{3}(2-\CDRflag+2\pi^2) + \frac{4}{3}(C_A-2n_FT_F+n_ST_S)\,,
\end{equation}
which is proportional to the angular $\delta$-distribution introduced in \eqref{angDelta}. The value of the coefficient $\Gamma_{\rm coll}$ depends on the treatment of spins in $d$-dimensions. To be able to consider different renormalization schemes, we have introduced a variable $\CDRflag$ which tracks $\ep$-terms in gluon spin sums. For the result in conventional dimensional regularization (CDR) we need to set $\CDRflag = 1$. The collinear terms involve the two-loop cusp anomalous dimension
\begin{equation}
\gamma_1^\text{cusp} = 4 \left(\left(\frac{67}{9}-\frac{\pi ^2}{3}\right) C_A-\frac{20}{9}  n_F T_F -\frac{8}{9}  n_S T_S\right) .
\end{equation}
Note that, throughout our paper, the symbol $\gamma_1^\text{cusp}$  refers to the value of this anomalous dimension in conventional dimensional regularization (CDR). The terms proportional to $\Gamma_{\rm coll}$ are different from the rest of the terms in $\bm{d}_m$ in that it is the angular integration, which produces the divergence, while the soft integral itself is finite.  The terms can be extracted by considering the collinear limit of the real emissions and extracting the associated divergence, see Section \ref{sec:dcomp}. As mentioned earlier the $q\parallel r $ collinear divergences cancel soft function collinear divergences and are part of the anomalous dimension. To demonstrate this we extract the collinear divergence of the soft function in Appendix \ref{app:soft}, and show that it indeed matches $\Gamma_{\rm coll}$.

The real-virtual part extracted from soft limits of diagrams takes the form\footnote{There is an ambiguity in the extraction of the terms $\propto C_A \pi^2$ in our computation of $\bm{r}_m$ and $\bm{v}_m$, see \eqref{eq:pi2problem} in Section \ref{sec:gamma2_extraction}. In the results presented here, the coefficient of the $C_A \pi^2$ terms in the last line of \eqref{eq:rm0} and third line of \eqref{eq:vm0} were adjusted to be consistent with the checks performed in Section \ref {sec:finite}.} 
\begin{align}\label{eq:rm0}
 \bm{r}_m ={}& -2 \sum_i\sum_{(j k)}if^{abc}(\bm{T}_{i,L}^a \bm{T}_{j,R}^b \bm{T}_{k,R}^c - \bm{T}_{i,R}^a \bm{T}_{j,L}^b \bm{T}_{k,L}^c) \Dqqq{r} K_{ijk;qr} \theta_\text{in}(n_q) \nonumber\\
                 & +8i\pi \sum_i\sum_{(j k)} if^{abc}\left( \bm{T}_{i,L}^a \bm{T}_{j,R}^b \bm{T}_{k,R}^c + \bm{T}_{i,R}^a \bm{T}_{j,L}^b \bm{T}_{k,L}^c  \right)
                   W_{ij}^q\ln W_{jk}^q \,\theta_\text{in}(n_q) \nonumber\\
                 & + \left(\frac{4\beta_0}{\ep} -\frac{8 \pi^2 C_A}{3} \right) \sum_{(ij)} \bm{T}_{i,L}^a \bm{T}_{j,R}^a W_{ij}^q \theta_\text{in}(n_q) \,,
\end{align}
where we have taken the real part from the exclusive calculation, performing the energy integrals using the residue theorem, and the imaginary part from the inclusive calculation with the one-loop soft current. We want to have the real emission part in the form of an angular integral in order to make cancellations of collinear divergences manifest, while it is sufficient to obtain the imaginary part inclusively. Let us note that in contrast to $\bm{d}_m$ the sums involving the angular function $K_{ijk;qr}$ in  $\bm{r}_m$ do produce two-leg contributions from the terms where $i=j$ since $K_{iik;qr}$ is non-zero.

We have extracted soft divergences associated with energy integrals but have kept all angular integrals $d$-dimensional. The entire anomalous dimension should thus be integrated over angles in $d$ dimensions and this result is clearly not the anomalous dimension in the standard $\overline{\rm MS}$ scheme. This is particularly important for the divergent term in the last line, where $O(\ep)$ terms from the angular integral can contribute finite pieces in the anomalous dimension. The divergence in the last line also makes it clear that the anomalous dimension $\bm{r}_m$ in \eqref{eq:rm0} is not yet in a useful form since it contains an explicit collinear divergence which must cancel against collinear divergences from other terms. A related problem affects $\bm{d}_m$: the angular integrals over $K_{ij;qr}$ produce a collinear divergence when $q$ becomes collinear to $r$ which must be regularized dimensionally. After giving also the result for the fully virtual piece, we will rearrange the anomalous dimension in such a way that the collinear singularities are manifestly cancelled when adding real and virtual contributions. After this we change the scheme to the standard $\overline{\rm MS}$ scheme in which angular integrations in the anomalous dimension will be performed in $d=4$.

For the double-virtual contribution the calculation in Section \ref{sec:gamma2_extraction} yields
\begin{align}
 \bm{v}_m ={}& \sum_{(i j k)} if^{abc}\left(\bm{T}_{i,L}^a \bm{T}_{j,L}^b \bm{T}_{k,L}^c - \bm{T}_{i,R}^a \bm{T}_{j,R}^b \bm{T}_{k,R}^c\right) 
                   \Dqqq{q}\Dqqq{r} K_{ijk;qr}  \nonumber\\
                 & + \sum_{(i j)}\left(\bm{T}_{i,L}^a \bm{T}_{j,L}^a+\bm{T}_{i,R}^a \bm{T}_{j,R}^a\right) \Bigg\{ \Dqqq{q}\Dqqq{r} K_{ij;qr}   \nonumber \\
                 & +\left( - \frac{2\beta_0}{\ep} + \Gamma_\text{coll}+ \frac{4 \pi^2 C_A}{3} \right) \Dqqq{q} W_{ij}^q \Bigg\} \nonumber\\
                 & - \sum_{(i j)}\left(\bm{T}_{i,L}^a \bm{T}_{j,L}^a-\bm{T}_{i,R}^a \bm{T}_{j,R}^a\right)\gamma_1^\text{cusp}\frac{i\pi \Pi_{ij}}{2} \,.
\label{eq:vm0}
\end{align}
The real part of the result was obtained by performing the energy integrals using the residue theorem. The extraction of the imaginary part is delicate since it is associated with Glauber phases, but the above form is consistent with the general result for the IR divergences shown in \eqref{eq:GammaAmp}. All angular integrations in \eqref{eq:vm0} have to be performed in $d$ dimensions because of collinear divergences, which are present both  in implicit form in $K_{ij;qr}$ and explicitly in the term proportional to $\beta_0$. 

\subsection{Collinear rearrangement}

 The anomalous dimension presented above suffers from both implicit and explicit collinear divergences. In order to have these regularized, it is necessary to keep all angular integrals in $d$ dimensions. Such a renormalization scheme is unconventional and not suited for implementation into a parton shower framework. The key to solving this problem is the identity
\begin{align}
\Dqqq{r} K_{ij;qr} ={}& 2W_{ij}^q \Bigg[
       \beta_0\left(\frac{1}{\ep}+\ln(2W_{ij}^q)\right) + \frac{1}{4} \, \gamma_1^{\rm cusp} \nonumber\\
     & \hspace{1cm} - \frac{\Gamma_{\rm coll}}{2} -\frac{2 \pi^2 C_A}{3} +\frac{c_R-1}{3} C_A \Bigg]\,,
 \label{eq:cuspRel}
\end{align}
which is derived in Appendix \ref{app:KijInt}. We first use this identity to add zero to the real-virtual part in such a way that the explicit divergence proportional to $\beta_0$ is converted into the angular integral on the left-hand side \eqref{eq:cuspRel}. This yields
\begin{align}
 \bm{r}_m ={}& -2 \sum_i\sum_{(jk)}if^{abc}(\bm{T}_{i,L}^a\bm{T}_{j,R}^b\bm{T}_{k,R}^c - \bm{T}_{i,R}^a\bm{T}_{j,L}^b\bm{T}_{k,L}^c) \Dqqq{r} K_{ijk;qr} \theta_\text{in}(n_q) \nonumber\\                
                   & + 8i\pi \sum_i\sum_{(j k)} if^{abc}\left( \bm{T}_{i,L}^a \bm{T}_{j,R}^b \bm{T}_{k,R}^c + \bm{T}_{i,R}^a \bm{T}_{j,L}^b \bm{T}_{k,L}^c  \right)  W_{ij}^q\ln W_{jk}^q \,\theta_\text{in}(n_q)  \nonumber\\
                 & + 2 \sum_{(ij)} \bm{T}_{i,L}^a \bm{T}_{j,R}^a \Dqqq{r} K_{ij;qr} \theta_\text{in}(n_q) 
                   - \sum_{(ij)}\bm{T}_{i,L}^a\bm{T}_{j,R}^a  W_{ij}^q \Bigg[4 \beta_0 \ln(2W_{ij}^q) \nonumber\\
                 &  + \gamma_1^{\rm cusp} -2 \Gamma_{\rm coll}  +\frac{4\left(c_R-1 \right) C_A}{3} \Bigg] \theta_\text{in}(n_q) \, .
\label{eq:rm1}
\end{align} 
Then, we observe that the collinear contribution proportional to $\delta(n_q-n_r)$ in the double-real part is physically indistinguishable from the real-virtual part. This allows us to move this term from $\bm{d}_m$ to the real-virtual part. Adding it cancels the $\Gamma_{\rm coll}$ term and yields 
\begin{align}
 \bm{r}_m ={}& -2 \sum_i\sum_{(jk)}if^{abc}(\bm{T}_{i,L}^a\bm{T}_{j,R}^b\bm{T}_{k,R}^c - \bm{T}_{i,R}^a\bm{T}_{j,L}^b\bm{T}_{k,L}^c) \Dqqq{r} K_{ijk;qr} \theta_\text{in}(n_q) \nonumber\\
                 & + 8i\pi \sum_i\sum_{(j k)} if^{abc}\left( \bm{T}_{i,L}^a \bm{T}_{j,R}^b \bm{T}_{k,R}^c + \bm{T}_{i,R}^a \bm{T}_{j,L}^b \bm{T}_{k,L}^c  \right)  W_{ij}^q\ln W_{jk}^q \,\theta_\text{in}(n_q)  \nonumber\\
                 & + 2 \sum_{(ij)} \bm{T}_{i,L}^a \bm{T}_{j,R}^a \Dqqq{r} K_{ij;qr} \theta_\text{in}(n_q) \nonumber\\
                 & - \sum_{(ij)}\bm{T}_{i,L}^a\bm{T}_{j,R}^a  W_{ij}^q \left[4 \beta_0 \ln(2W_{ij}^q) + \gamma_1^{\rm cusp} + \frac{4(\CDRflag-1)C_A}{3} \right] \theta_\text{in}(n_q)\,, 
\label{eq:rm}
\end{align}
while the double-real part simplifies to 
\begin{align}
 \bm{d}_m ={}& \sum_{(i j)}\sum_k i f^{abc}\left(\bm{T}_{i,L}^a \bm{T}_{j,L}^b \bm{T}_{k,R}^c - \bm{T}_{i,R}^a\bm{T}_{j,R}^b \bm{T}_{k,L}^c\right) K_{ijk;qr} \theta_\text{in}(n_q)\theta_\text{in}(n_r) \nonumber\\
                 & - 2 \sum_{(ij)}\bm{T}_{i,L}^a \bm{T}_{j,R}^a K_{ij;qr} \theta_\text{in}(n_q) \theta_\text{in}(n_r) \,.
 \label{eq:dm}
\end{align}

Next, we use the identity \eqref{eq:cuspRel} to carry out one of the angular integrals in the $K_{ij;qr}$ term in the purely virtual part which then becomes 
\begin{align}
 \bm{v}_m ={}& \sum_{(i j k)} if^{abc}\left(\bm{T}_{i,L}^a \bm{T}_{j,L}^b \bm{T}_{k,L}^c - \bm{T}_{i,R}^a \bm{T}_{j,R}^b \bm{T}_{k,R}^c\right) 
                   \Dqqq{q}\Dqqq{r} K_{ijk;qr}  \nonumber\\
                 & + \sum_{(i j)}\frac{\bm{T}_{i,L}^a \bm{T}_{j,L}^a+\bm{T}_{i,R}^a \bm{T}_{j,R}^a}{2}
                   \Dqqq{q} W_{ij}^q  \left[4\beta_0 \ln(2W_{ij}^q)  + \gamma_1^{\rm cusp} + \frac{4C_A(\CDRflag-1)}{3}\right]   \nonumber\\
                 & - \sum_{(i j)}\left(\bm{T}_{i,L}^a \bm{T}_{j,L}^a- \bm{T}_{i,R}^a \bm{T}_{j,R}^a\right)\gamma_1^\text{cusp}\frac{i\pi\Pi_{ij}}{2} \,.
\label{eq:vm}
\end{align}
This eliminates the  $\Gamma_{\rm coll}$ term present in \eqref{eq:vm0}. As a final simplification, we can set $c_R=1$ to get the result in CDR.

Even after this rearrangement, the individual pieces of the anomalous dimension contain collinear divergences. These cancel in physical quantities in the same way as for the one-loop anomalous dimension  discussed at length in Section \ref{sec:colllinearCancellation}. For example, the divergences in the last line of \eqref{eq:rm} cancel against the ones in the second line of \eqref{eq:vm}. As in \eqref{eq:collSub}, we could introduce subtracted dipole functions in which the divergence is removed, but equally well we can introduce some intermediate angular cutoff, as is done in a parton-shower framework. After this, the angular integrals in these terms can be carried out in $d=4$. 

We already discussed that $K_{ijk;qr} $ is free of collinear divergences so also these terms can be integrated in $d=4$. The only terms in which implicit collinear singularities are still present are the integrals over $ K_{ij;qr}$ in $\bm{d}_m$ given in \eqref{eq:dm} and $\bm{r}_m$ in \eqref{eq:rm}, but note that in the combination  these terms take the form
\begin{equation}
 \bm{d}_m + \bm{r}_m =  - 2  \sum_{(ij)}\bm{T}_{i,L}^a \bm{T}_{j,R}^a K_{ij;qr} \theta_\text{in}(n_q)  ( \theta_\text{in}(n_r) -1 ) + \dots
\end{equation}
and $1- \theta_\text{in}(n_r)  = \theta_\text{out}(n_r)$. Since the vector $n_q$ is inside the jet and $n_r$ in the veto region, the combination $\bm{d}_m + \bm{r}_m$ is free of collinear divergences so that also for these terms the angular integrals can be carried out in $d=4$.  In the form \eqref{eq:rm} and \eqref{eq:vm}, we can thus take the limit $d=4$ in all the angular integrals associated with the two-loop anomalous dimension and this form is suitable for implementation in a parton shower framework.

\subsection{\boldmath Change to the $\overline{\rm MS}$ scheme}
\label{sec:MSbar}

The renormalization condition \eqref{eq:rencond} and the associated anomalous dimensions involve angular integrals in $d$-dimensions. It would be inconvenient and unconventional to keep these integrations $d$ dimensional and in the previous subsection, we have written the anomalous dimension in such a way that we can make the transition to angular integrals in $d=4$ and perform subtractions in the usual $\overline{\rm MS}$ scheme. Before changing scheme, let us first rewrite the renormalization conditions \eqref{eq:finitenessOne} and \eqref{resDiv}  in compact form
\begin{align}\label{eq:subtra}
 \bm{\mathcal{S}}^{{\rm ren}(1)} &=  \bm{\mathcal{S}}^{(1)} -\frac{1}{2\ep}\bm{\Gamma}^{(1)} \hat{\otimes} \bm{1} \,,\nonumber \\
  \bm{\mathcal{S}}^{{\rm ren}(2)} &=   \bm{\mathcal{S}}^{(2)}  - \frac{1}{8\ep^2} \left[ \bm{\Gamma}^{(1)} \hat{\otimes} \bm{\Gamma}^{(1)} \hat{\otimes} \bm{1} + 2\beta_0 \bm{\Gamma}^{(1)} 
  \hat{\otimes} \bm{1} \right] \nonumber \\
  & \quad - \frac{1}{4\ep} \left[ \bm{\Gamma}^{(2)} \hat{\otimes} \bm{1} +2\bm{\Gamma}^{(1)}\hat{\otimes} \bm{\mathcal{S}}^{{\rm ren}(1)}+4\beta_0  \bm{\mathcal{S}}^{{\rm ren}(1)}  \right] .
\end{align}
For brevity, we have dropped the multiplicity indices on the anomalous dimensions and the soft functions.

To see that the renormalization conditions \eqref{eq:subtra} differ from standard minimal subtraction, let us give the explicit result for the renormalized the one-loop soft function $\bm{\mathcal{S}}_m^{{\rm ren}(1)}$ in this scheme. The bare soft function is
\begin{align}\label{oneloopsoft}
\bm{\mathcal{S}}_m(\{ \underline{n} \}, Q_0) \!=\! - g_s^2  \sum_{(ij)} \bm{T}_{i}\cdot  \bm{T}_{j}\DqqE{q} \frac{1}{\E{q}^2 } W_{ij}^q\,\theta(Q_0 - 2 E_q)  \theta_\text{out}(q) \,
\end{align}
and the one-loop subtraction was given in \eqref{oneloop}
\begin{equation}
-\frac{1}{2\ep} \bm{\Gamma}^{(1)} \hat{\otimes} \bm{1} =-\frac{2}{\ep} \sum_{(ij)} \bm{T}_{i}\cdot  \bm{T}_{j} \Dqqq{q}\, W_{ij}^q  \,\theta_{\rm out}(n_q)\,.
\end{equation}
This form of the subtraction removes all higher-order terms in $\ep$ associated with the angular part of the integral in \eqref{oneloopsoft}. We thus also subtract finite terms so that this is not a minimal subtraction. Expressing the bare quantity $g_s$ in terms of the $\overline{\rm MS}$ coupling $\alpha_s$ and performing the energy integral, we find
\begin{equation}\label{eq:S1renddim}
\bm{\mathcal{S}}^{\rm ren (1)}_m = -2 \sum_{(ij)} \bm{T}_{i}\cdot  \bm{T}_{j}  \ln\frac{Q_0}{\mu} \DQQQ{q}\, W_{ij}^q \theta_{\rm out}(n_q) \,.
\end{equation}
This result is not equal to $\overline{\rm MS}$ renormalization for the soft function as we will see explicitly in the next section, when we compute $\bm{\mathcal{S}}^{\rm ren (1)}_2$ for the case of the two-jet cross section.

To change to the standard  $\overline{\rm MS}$ subtraction scheme, we should perform the subtractions in \eqref{eq:subtra} using angular integrals in $d=4$. To do so, we now rewrite angular integrals as integrals in $d=4$ plus a remainder. Using the notation defined in Table \ref{tab:angInts}, we write
\begin{equation}
\bm{\Gamma}^{(1)} \hat{\otimes} \bm{1} = \bm{\Gamma}^{(1)} \hat{\otimes}_2 \bm{1} + 2\e\, \bm{\Gamma}^{(1)} \hat{\otimes}_\ep \bm{1}\,,
\end{equation}
where we denoted the two-dimensional angular integral in $d=4$ by the symbol $\otimes_2$ and the remainder using the symbol  $\otimes_\ep$. In the remainder, we have factored out a prefactor $2\e$ for convenience. Using angular integrations in $d=4$, the one-loop renormalization reads
\begin{equation}\label{eq:onelooprenMS}
\bar{\bm{\mathcal{S}}}^{{\rm ren}(1)} =  \bm{\mathcal{S}}^{(1)} - \frac{1}{2\e} \bm{\Gamma}^{(1)} \otimes_2  \bm{1}\,,
\end{equation}
where $\bar{\bm{\mathcal{S}}}^{{\rm ren}(1)} $ is the one-loop soft function renormalized in the $\overline{\rm MS}$ scheme. Compared to the earlier prescription, this corresponds to a finite shift 
\begin{equation}\label{oneLoopRel}
\bm{\mathcal{S}}^{{\rm ren}(1)} = \bar{\bm{\mathcal{S}}}^{{\rm ren}(1)} -\bm{\Gamma}^{(1)} \otimes_\ep \bm{1} \,.
\end{equation}
Converting to the $\overline{\rm MS}$ scheme not only induces a shift in the renormalized function but also changes the expression for the two-loop anomalous dimension
\begin{align}\label{eq:renMSbar}
  \bar{\bm{\mathcal{S}}}^{{\rm ren}(2)} &=   \bm{\mathcal{S}}^{(2)}  - \frac{1}{8\ep^2} \left[ \bm{\Gamma}^{(1)} \otimes_2 \bm{\Gamma}^{(1)} \otimes_2 \bm{1} + 2\beta_0 \bm{\Gamma}^{(1)} 
  \otimes_2 \bm{1} \right] \nonumber\\
  & \quad - \frac{1}{4\ep} \left[ \bar{\bm{\Gamma}}^{(2)} \otimes_2 \bm{1} +2\bm{\Gamma}^{(1)}\otimes_2 \bar{\bm{\mathcal{S}}}^{{\rm ren}(1)}+4\beta_0  \bar{\bm{\mathcal{S}}}^{{\rm ren}(1)}  \right] ,
\end{align}
where the two-loop anomalous dimension in the $\overline{\rm MS}$ is given by
\begin{align}\label{eq:GammaExtra}
 \bar{\bm{\Gamma}}^{(2)} \otimes_2 \bm{1} = \bm{\Gamma}^{(2)} \otimes_2 \bm{1} -2 \beta_0\, \bm{\Gamma}^{(1)}\otimes_\ep \bm{1} - \left( \bm{\Gamma}^{(1)} \otimes_2 \bm{\Gamma}^{(1)}\otimes_\ep \bm{1} - \bm{\Gamma}^{(1)}\otimes_\ep \bm{\Gamma}^{(1)} \otimes_2 \bm{1}\right) \,.
 \end{align}
 To derive the additional terms in the anomalous dimension, one starts with \eqref{eq:subtra}, rewrites the $d$-dimensional integrals in terms of $4$-dimensional ones and uses \eqref{oneLoopRel} to express $\bm{\mathcal{S}}^{{\rm ren}(1)}$ in terms of the $\overline{\rm MS}$ one. Note that the second term has the form of a commutator of angular integrations associated with the two emissions. For independent emissions, this part vanishes because of the symmetry among the emissions. However, for the two-loop non-global piece we get a non-zero contribution, because this part involves one gluon outside and one gluon inside the jet region and the $\mathcal{O}(\e)$ pieces of the two angular integrals are not identical.
 
The result for the two-loop anomalous dimension in \eqref{eq:GammaExtra}, together with the two-loop ingredients $\bm{\Gamma}_m^{(2)} = \bm{d}_m+ \bm{r}_m+ \bm{v}_m$, given in equations \eqref{eq:dm}, \eqref{eq:rm} and \eqref{eq:vm} and evaluated with angular integrals in $d=4$, is the final result for the anomalous dimension in the $\overline{\rm MS}$ scheme, suitable for numerical implementation. 
 
\section{Finiteness check for the two-jet cross section}
\label{sec:finite}

Given the subtleties in the extraction of the anomalous dimension, it is important to have a consistency check to verify that it indeed correctly renormalizes the divergences in the hard and soft functions in the factorization theorem \eqref{eq:crssctEvo}. We will now verify finiteness for the case of the Sterman-Weinberg two-jet cross section \cite{Sterman:1977wj}, for which the hard and soft functions were explicitly given in \cite{Becher:2016mmh}. To simplify the two-loop computations \cite{Becher:2016mmh} used the thrust axis as the jet axis. As in the original definition, back-to-back cones with a half-opening angle $\alpha$ are put around the jet axis and one then restricts the energy outside the jets to $2E_{\rm out} \leq Q_0 = Q\beta$. We define the abbreviations $r=\delta^2=\tan^2(\alpha/2)$ and $\Delta=\cos\alpha=(1-r)/(1+r)$. While we assume that $\beta$ is small, we consider a situation where $\delta\sim 1$ so that we do not encounter large logarithms of the opening angle.

The results for the bare soft functions can be most compactly written in terms of the harmonic polylogarithms $H_{a_1, \dots, a_n}(r)$ introduced in \cite{Remiddi:1999ew} (see \cite{Maitre:2005uu,Maitre:2007kp} for a {\sc Mathematica} implementation). In terms of these functions, the bare one-loop soft function reads 
\begin{equation} \label{eq:S21}
 \bm{\mathcal{S}}_2^{(1)} = 4 C_F \left(\frac{\mu^2}{Q_0^2}\right)^{\ep} \left[\frac{H_0(r)}{\ep} + 2H_{-2}(r) - H_{0, 0}(r) -\frac{\pi^2}{6} +\mathcal{O}(\ep)\right]  \,.
\end{equation}
The poles of the bare two-loop soft function are
\begin{align}\label{eq:S22}
 \bm{\mathcal{S}}_2^{(2)} ={}& \frac{1}{2} \left( \bm{\mathcal{S}}_2^{(1)}\right)^2 + C_F \left(\frac{\mu^2}{Q_0^2}\right)^{2\ep}\Bigg[\frac{2\beta_0 H_0(r) + C_A\left(8H_2(r)-8H_{-2}(r)-\frac{2 \pi^2}{3}\right)}{\ep^2}\nonumber\\
                       & + \frac{1}{\ep} \Bigg(\frac{\gamma_1^\text{cusp}}{2}H_0(r) + 4\beta_0 \Big [3 H_{-2}(r)-H_2(r)-H_{-1,0}(r)-H_{0,0}(r)+H_{1,0}(r) \Big] \nonumber\\
                       & -8C_A\Big [6 H_{-2,-1}(r)-5 H_{-2,0}(r)-2 H_{-2,1}(r)-2 H_{2,-1}(r)+3 H_{2,0}(r)-2 H_{2,1}(r) \nonumber\\
                       &\hspace{1.4cm} -\frac{\pi^2}{6} H_0(r)+\zeta_3 \Big]
                         -2(C_A-2 n_F T_F) \frac{1-r^4+4 r^2 H_0(r)}{3 (1-r^2)^2}\Bigg) + \mathcal{O}(\ep^0) \Bigg]\,.
\end{align}
We have not written out the $C_F^2$ terms, which follow trivially from exponentiation.

Let us start with the one-loop counter term. Applying the anomalous dimension to the tree-level soft function $ \bm{\mathcal{S}}_m^{(0)}=\bm{1}$ and using color conservation $\bm{T}_1^a +\bm{T}_2^a =0$, we get
\begin{align}\label{eq:oneLoopCT}
 \bm{R}_2 \otimes_2 \bm{\mathcal{S}}_3^{(0)} +\bm{V}_2 \bm{\mathcal{S}}_2^{(0)} ={}& 
  2\sum_{(ij)}\DQQQ{q} W_{ij}^q\left(\bm{T}_{i}\cdot \bm{T}_{j}\, \bm{\mathcal{S}}_2^{(0)}+ \bm{\mathcal{S}}_2^{(0)} \bm{T}_{i} \cdot \bm{T}_{j}-2 \,\bm{T}_{i}^a \bm{\mathcal{S}}_3^{(0)} \bm{T}^a_{j} \, \Theta_\text{in}(q)\right) \nonumber\\
                                         ={}& -8C_F\DQQQout{q}W_{12}^q
                                         ={} -8C_F\int_{-\Delta}^{+\Delta}\frac{dc}{2}\frac{2}{1-c^2}\nonumber\\
                                         ={}& 8C_F\ln(r)
                                         ={}  8C_FH_0(r)\,,
\end{align}
where we used the short-hand notation
\begin{equation}\label{eq:ang}
\DQQQout{q} W_{12}^q = \DQQQ{q} \theta_\text{out}(q) W_{12}^q =\DQQQ{q} \left[1-\theta_\text{in}(q)\right] W_{12}^q \,.
\end{equation}
Inserting this result into the renormalization condition  \eqref{eq:onelooprenMS} and comparing with \eqref{eq:S21}, we see that we indeed obtain the renormalized function in the $\overline{\rm MS}$ scheme. 

The  angular integral  \eqref{eq:ang} also arises when computing the renormalized function in the renormalization scheme with $d$-dimensional angular integrals in the subtractions. Inserting it into expression \eqref{eq:S1renddim}, we get
\begin{equation}\label{eq:S2rend}
\bm{\mathcal{S}}^{\rm ren (1)}_2 = - 8 C_F  \ln\frac{Q_0}{\mu} H_0(r) \,,
\end{equation}
which explicitly demonstrates that this scheme does not correspond to the usual $\overline{\rm MS}$ subtraction.

Now we proceed to the two-loop results. The leading divergence in \eqref{eq:renMSbar} involves the square of the one-loop anomalous dimension
\begin{equation}
 \bm{\Gamma}^{(1)} \otimes_2 \bm{\Gamma}^{(1)} \otimes_2 \bm{1} =  \bm{R}_2\otimes_2\left(\bm{R}_3\otimes_21+\bm{V}_3\right) + \bm{V}_2\left(\bm{R}_2\otimes_21+\bm{V}_2\right) ,
\end{equation}
whose individual terms are given by
\begin{align}\label{eq:GaGa1}
 \bm{R}_2\otimes_2 (\bm{R}_3\otimes_2\bm{1}+ \bm{V}_3) ={}& 32C_F \DQQQin{q}\DQQQout{r}\Bigg[-2C_F W_{12}^qW_{12}^r \nonumber\\
    & + C_A\left(W_{12}^qW_{12}^r -W_{12}^{qr}-W_{12}^{rq} \right)\Bigg] \,,\\
 \bm{V}_2\left( \bm{R}_2\otimes_2\bm{1}+ \bm{V}_2\right) ={}& -64C_F^2 \DQQQ{q}\DQQQout{r}W_{12}^qW_{12}^r \,, \nonumber
 \end{align}
 where we introduced the abbreviation
 \begin{equation}
 W_{12}^{qr} = W_{12}^q W_{q2}^r = W_{12}^r W_{r1}^q = \frac{n_{12}}{n_{1q} n_{qr} n_{r2} }\,.
 \end{equation}
 Combining the real and virtual contribution has cancelled the collinear divergences in the angular integral over $r$ and adding the two pieces in \eqref{eq:GaGa1} also removes the ones associated with the $q$ integral so that the result
\begin{align}\label{eq:GaGa2}
   \bm{\Gamma}^{(1)} \otimes_2 \bm{\Gamma}^{(1)} \otimes_2 \bm{1} ={}& 64C_F^2\DQQQout{q}\DQQQout{r}W_{12}^qW_{12}^r \nonumber\\
    & + 32C_FC_A\DQQQin{q}\DQQQout{r}\left(W_{12}^qW_{12}^r-W_{12}^{qr}-W_{12}^{rq}  \right) \nonumber\\
    ={}& 64C_F^2 \left[ H_0(r) \right]^2 + 64 C_F C_A \left[ H_2(r)-H_{-2}(r)-\frac{\pi^2}{12} \right]
\end{align}
is finite. Looking at the result and inserting it into \eqref{eq:renMSbar}, we immediately see that these terms indeed cancel the $1/\ep^2$ divergences in the two-loop soft function \eqref{eq:S21}.

As discussed in Section \ref{sec:MSbar}, we also need to compute the $\mathcal{O}(\ep)$ terms in the angular integrals in \eqref{eq:GaGa2} since these arise when transitioning to the $\overline{\rm MS}$, see \eqref{eq:GammaExtra}. Specifically, we need the term
\begin{multline}
\bm{\Gamma}^{(1)} \otimes_2 \bm{\Gamma}^{(1)}\otimes_\ep \bm{1} - \bm{\Gamma}^{(1)}\otimes_\ep \bm{\Gamma}^{(1)} \otimes_2 \bm{1} = 16 C_F C_A \big[-2 H_{-3}(r) +2 H_{-2,0}(r)-4 H_{-2,1}(r)   \\
-4 H_{2,-1}(r)+4 H_{2,1}(r)+4 H_{3}(r)-\zeta_3 \big] \,,
\end{multline}
whose computation is detailed in Appendix \ref{app:GaGa}. The other term in \eqref{eq:GammaExtra} is $2 \beta_0\, \bm{\Gamma}^{(1)}\otimes_\ep \bm{1}$, determined by the $\mathcal{O}(\ep)$ terms in the one-loop angular integral \eqref{eq:oneLoopCT}
\begin{equation}\label{eq:Ga1b0}
\bm{\Gamma}^{(1)} \hat{\otimes} \bm{1} = -8 C_F \Dqqqout{q}W_{12}^q = 8C_F\left[H_0(r) + \ep \left(2H_{-2}(r) - H_{0, 0}(r) -\frac{\pi^2}{6}\right) \right] .
\end{equation}
Since the dipole $W_{12}^q$ consists of back-to-back vectors in the jet direction, the angular integral can also be written in the form
\begin{equation}
\Dqqqout{q}W_{12}^q = \frac{4^\ep e^{\gamma_E \ep}}{\Gamma(1-\ep)} \int_{-\Delta}^{+\Delta}\! dc  \left(\frac{1}{1-c^2}\right)^{1+\ep}= \frac{e^{\gamma_E \ep}}{\Gamma(1-\ep)} \DQQQout{q} \frac{1}{2}\left(2 W_{12}^q \right)^{1+\ep} \,.
\end{equation}
We can trade the $d$-dimensional angular integral for integral over the logarithm of the dipole in $d=4$. For the two-jet case, we  have
\begin{equation}\label{eq:Ga1Be}
\bm{\Gamma}^{(1)}\otimes_\ep \bm{1}  = -4 C_F \DQQQout{q} W_{12}^q \ln(2 W_{12}^q) = 4C_F \left(2H_{-2}(r) - H_{0, 0}(r) -\frac{\pi^2}{6}\right)\,.
\end{equation} 

Having evaluated the terms associated with the scheme change from $\bm{\Gamma}^{(2)}$ to the $\overline{\rm MS}$ anomalous dimension $\overline{\bm{\Gamma}}^{(2)}$, let us now evaluate the contributions from the anomalous dimension $\bm{\Gamma}^{(2)}$ itself. Evaluating the sums for the two-particle case, where the indices can only take the values $1$ or $2$, and performing the trivial color algebra, we get
\begin{align}\label{eq:Ga2}
\bm{\Gamma}^{(2)} \otimes_2 \bm{1} = {}&  \bm{d}_2\otimes_2 \bm{1} + \bm{r}_2\otimes_2\bm{1} + \bm{v}_2  \nonumber\\
= {}& 2C_F C_A \DQQQin{q}\DQQQ{r}\left(K_{112;qr} + K_{221;qr}\right) \nonumber\\
&  - 2C_F \DQQQin{q}\DQQQout{r}\left( K_{12;qr} + K_{21;qr}\right)\nonumber\\
    & - 8C_F \DQQQout{q}W_{12}^q \left( \beta_0 \ln(2W_{12}^q) +\frac{1}{4}\gamma_1^\text{cusp}\right) .\end{align}
Note that the imaginary parts of the anomalous dimension do not contribute when it acts on the trivial tree-level soft function. Because of this, our consistency check does not test the imaginary parts. Note that the test does involve the three-particle correlations $K_{ijk;qr}$, but here only the real-virtual part $\bm{r}_2$ contributes. The double-real contribution vanishes because $K_{iji;qr}=K_{ijj;qr} = 0$ and the double-virtual part requires three different legs. 

We see that the $\ln(2W_{12}^q)$ term the last line of \eqref {eq:Ga2} cancels against $-2\beta_0 \bm{\Gamma}^{(1)}\otimes_\ep \bm{1}$ given in \eqref{eq:Ga1Be}. We still need to evaluate the integral over the two particle terms, for which we find
\begin{align}
     \DQQQin{q}\DQQQout{r} &\left( K_{12;qr} + K_{21;qr}\right)  \nonumber\\
 ={}& 4(C_A-2n_F T_F)\frac{1-r^4+4r^2 H_0(r)}{3(1-r^2)^2}  + 8C_A[2H_{2,0}(r)-2H_{-2,0}(r)+\zeta_3] \nonumber\\
    & + 8\beta_0[H_2(r)+H_{-1,0}(r)-H_{-2}(r)-H_{1,0}(r)-\frac{\pi^2}{6}]  \,.
\end{align}
Finally, we need the contribution from the three-particle correlations for which we get 
\begin{align}
 \DQQQin{q}& \DQQQ{r} \left(K_{112;qr} + K_{221;qr}\right) \nonumber\\
                      ={}& -16 \DQQQin{q}\DQQQout{r} 
                           \frac{n_{1q} n_{2r}+n_{1r} n_{2q} -n_{12} n_{qr}}{n_{1q} n_{2q}  
                           n_{1r} n_{2r} n_{qr}}
                           \ln\frac{n_{1q} n_{2q}}{n_{1r} n_{2r}}\nonumber\\
                      ={}& 16 [H_{-3}(r)-2 H_{-2,-1}(r)+H_{-2,0}(r)+\zeta_3]\,,
\end{align}
where we have used the antisymmetry of the integrand under $q\leftrightarrow r$ to eliminate the 'in-in' contribution. 

With this we have all two-loop terms associated with the anomalous dimension. The final ingredient to check finiteness in \eqref{eq:renMSbar} is the product
\begin{equation}
\bm{\Gamma}^{(1)}\otimes_2 \bar{\bm{\mathcal{S}}}^{{\rm ren}(1)} =\bm{\Gamma}^{(1)}\otimes_2 \left[ \bm{\mathcal{S}}^{(1)} - \frac{1}{2\ep} \bm{\Gamma}^{(1)}\otimes_2 \bm{1} \right]\,.
\end{equation}
Rewriting it in the form shown on the right hand side is convenient because we can use the result \eqref{eq:GaGa2} together with the results for the angular integrals in Appendix \ref{app:GaGa}, which leads to
\begin{align} \label{eq:GaSm}
\bm{\Gamma}^{(1)}\otimes_2\, \bar{\bm{\mathcal{S}}}^{{\rm ren}(1)} =& 
32 C_F^2 H_0(r) \left(H_0(r) \ln \!\left(\frac{\mu ^2}{Q_0^2}\right)-H_{0,0}(r)+2
   H_{-2}(r)-\frac{\pi ^2}{6}\right) \nonumber \\
  & +32  C_F C_A \Bigg[
   \left(-H_{-2}(r)+H_2(r)-\frac{\pi ^2}{12}\right)  \ln\!\left(\frac{\mu ^2}{Q_0^2}\right) \nonumber \\
&\hspace{2.1cm}-H_{-3}(r)-2H_{-2,-1}(r)+2 H_{-2,0}(r)- H_{2,0}(r)\nonumber \\
&\hspace{2.1cm}+2 H_{2,1}(r)+H_3(r)+\frac{\pi^2}{12}H_{0}(r)-\zeta_3  \Bigg]\,.
\end{align}
The detailed derivation can be found in Appendix \ref{app:GaGa}.

With this we have evaluated all the individual pieces entering \eqref{eq:renMSbar}. Plugging in and adding up, we find that the pole terms cancel the ones present in the bare soft function \eqref{eq:S22} so that we end up with a finite renormalized function \eqref{eq:renMSbar}.

\section{\boldmath Extraction of $\Gamma^{(2)}$ from soft limits\label{sec:gamma2_extraction}}

In the following we derive the diagrammatic result for the two-loop anomalous dimension $\bm{\Gamma}^{(2)}$ presented in Section~\ref{sec:Gamma2_result_diag}. As in the one-loop case discussed in Section \ref{sec:oneloop}, we proceed by considering the soft limits of the hard function. Thus, the anomalous dimension does not subtract hard-collinear divergences from the hard function, but those can only appear in the in-region where the energy is unconstrained and thus cancel in the sum of real and virtual contributions by virtue of the KLN theorem. On the other hand, soft-collinear divergences are absorbed into the anomalous dimension. However, contrary to the one-loop case analyzed in Section \ref{sec:collSing}, they no longer completely cancel between the real and virtual contributions at two-loop order. The reason is that the double-virtual configuration where the two-loop momenta are in the out-region and become collinear to each other has no analogue in the double-real and real-virtual contribution. For this reason, we will have to carefully track  divergences arising when two soft momenta become collinear, as will be explained in detail in the calculation presented below. 

\subsection{\boldmath Double-real contribution $d_m$}\label{sec:dcomp}

We determine the double-real emission entry $\bm{d}_m$ of the NLO anomalous dimension by considering the limit where the last two partons in the hard function $\bm{\mathcal{H}}_{m+2}$ become soft. At leading power, the two soft partons can either be two gluons or a quark-antiquark pair and we denote their momenta with $q$ and $r$. In addition we consider the production of a pair of soft scalars in the color-adjoint representation to allow a more detailed comparison with the results of \cite{Caron-Huot:2015bja} in the DRED scheme. In \eqref{eq:Kijqr} we have grouped the two-parton function $K_{ij;qr}$ into SUSY multiplets as \cite{Caron-Huot:2015bja}, but in our explicit computation, we will separately extract the gluonic, fermionic and scalar pieces and split
\begin{equation}\label{eq:KijqrCi}
 K_{ij;qr} = C_A K_{ij;qr}^{A} + n_F T_F K_{ij;qr}^{F} +  n_ST_S K_{ij;qr}^{S}\,.
 \end{equation}
The individual pieces are of course immediately obtained from the functions $K_{ij;qr}^{(a)}$, $K_{ij;qr}^{(b)}$, $K_{ij;qr}^{(c)}$ given in \eqref{eq:KijqrFuns}. We note, however, that the function $K_{ij;qr}$ arising in the explicit computation is manifestly symmetric in $q$ and $r$, in contrast to the expressions \eqref{eq:KijqrFuns} which were simplified using the fact that the anti-symmetric part does not contribute to the anomalous dimension result.

In the case of two soft gluons with polarization vectors $\varepsilon_{m+1}^*(q)$ and $\varepsilon_{m+2}^*(r)$, the amplitude takes the form 
\begin{align}\label{eq:two_gluons}
| {\cal M}_{m+2}^{gg}(\{ \underline{p}, q,r \}) \rangle = 
   \Bigg[ \frac{1}{2} \Big\{ \bm{J}^{\mu,a}(q),\bm{J}^{\nu,b}(r) \Big\} + 
          g_s^2 i f^{abc} \sum_i \bm{T}_i^c J^{\mu\nu}_i(q,r) 
   \Bigg] \varepsilon_{m+1}^{*\mu} \varepsilon_{m+2}^{*\nu} | {\cal M}_{m} (\{ \underline{p} \}) \rangle\,,
\end{align}
where the color-connected part reads \cite{Catani:1999ss}
\begin{equation}\label{eq:Jmunu}
J^{\mu\nu}_i(q,r) = \frac{1}{2 n_i\cdot (q+r)} \left[  \frac{n_i^\mu n_i^\nu}{n_i\cdot q}-\frac{n_i^\mu n_i^\nu}{n_i\cdot r}+ \frac{g^{\mu\nu} n_i \cdot (r-q) + 2 n_i^\mu q^\nu - 2n_i^\nu r^\mu}{q\cdot r}\right] .
\end{equation}
When integrating the current over the phase space of the two soft gluons, one needs to add a factor of $1/2!$ since they are identical particles. Note that the color generators $\bm{T}_i$ associated with different legs commute since they act in a different space. The anti-commutator in \eqref{eq:two_gluons} is thus only relevant when the two generators involve the same leg, i.e. for the terms $\{\bm{T}_i^a, \bm{T}_i^b\}$. The two terms in \eqref{eq:two_gluons} are not separately transverse but their sum is. As in the one-emission case, the non-transverse pieces vanish due to color conservation. To apply it, one needs to combine the anti-commutator from the first terms with the commutator in the second, as explained in  \cite{Catani:1999ss,Caron-Huot:2013fea}. The above results are valid in a physical gauge, where the sum over polarizations takes the form
\begin{equation}
d_{\mu\nu}(q) = -g_{\mu\nu} + \frac{n_\mu q_\nu+q_\mu n_\nu }{n\cdot q}\,,
\end{equation}
but since the currents are transverse, we can omit the gauge-vector dependent terms and replace $d_{\mu\nu}(q) \to -g_{\mu\nu} $.

Using \eqref{eq:two_gluons}, the hard function takes on a factorized form in the double-soft region 
\begin{align}
\bm{\mathcal{H}}_{m+2}^{gg}
={}& \frac12 \int\frac{dE_q E_q^{d-3}}{\tilde{c}^\ep(2\pi)^2}\int\frac{dE_r E_r^{d-3}}{\tilde{c}^\ep(2\pi)^2}
     \Bigg[\frac14 \{\bm{J}^{\mu,a}(q),\bm{J}^{\nu,b}(r)\} \bm{\mathcal{H}}_{m} \{\bm{J}^{\mu,a}(q),\bm{J}^{\nu,b}(r)\} \nonumber\\
   & + \frac{g_s^2if^{abc}}{2}\sum_k J_k^{\mu\nu}(q,r)\left(\bm{T}_k^c \bm{\mathcal{H}}_{m} \{\bm{J}^{\mu,a}(q),\bm{J}^{\nu,b}(r)\} - 
                                                           \{\bm{J}^{\mu,a}(q),\bm{J}^{\nu,b}(r)\} \bm{\mathcal{H}}_{m} \bm{T}_k^c\right)\nonumber\\
   & + g_s^4C_A\sum_{i,j}J_i^{\mu\nu}(q,r)J_j^{\mu\nu}(q,r) \bm{T}_i^a \bm{\mathcal{H}}_{m} \bm{T}_j^a \Bigg]\theta_\text{in}(n_q) \theta_\text{in}(n_r) \theta(\Lambda-E_q-E_r)\,.
\end{align}
After the energy integration we obtain 
\begin{align}
\bm{\mathcal{H}}_{m+2}^{gg} ={}& \left(\frac{\alpha_s}{4\pi}\right)^2 \left(\frac{\mu}{2\Lambda}\right)^{4\ep}\theta_\text{in}(n_q) \theta_\text{in}(n_r)\Bigg\{
                                 \sum_{i,j,k,l} \{\bm{T}_i^a,\bm{T}_j^b\}\bm{\mathcal{H}}_m\{\bm{T}_k^a,\bm{T}_l^b\} \frac{W_{ik}^qW_{jl}^r}{2}\left(\frac{1}{\ep^2}-\frac{2\pi^2}{3}\right)\nonumber\\
                               & + \sum_{(ij)}\sum_k if^{abc} \left(\bm{T}_k^c \bm{\mathcal{H}}_{m} \{\bm{T}_i^a,\bm{T}_j^b\} - \{\bm{T}_i^a,\bm{T}_j^b\} \bm{\mathcal{H}}_{m} \bm{T}_k^c\right)
                                 \Bigg[\frac{K_{ijk;qr}+K_{ijk;qr}^{(\ep)}\ep}{8\ep}\nonumber\\
                               & + \frac{W_{ik}^qW_{jq}^r-W_{jk}^rW_{ir}^q}{2}\left(\frac{1}{\ep^2}-\frac{2\pi^2}{3}\right) \Bigg]
                                 + C_A\sum_{(ij)} \bm{T}_i^a\bm{\mathcal{H}}_{m} \bm{T}_j^a  \nonumber\\
                               & \times \left[\frac{W_{ij}^qW_{ij}^r - 2W_{ij}^qW_{jq}^r - 2W_{ij}^rW_{jr}^q}{2}
                                 \left(\frac{1}{\ep^2}-\frac{2\pi^2}{3}\right) + \frac{K_{ij;qr}^A+K_{ij;qr}^{A,(\ep)}\ep}{2\ep}\right] + \mathcal{O}(\ep)\Bigg\}\,,
 \label{eq:RRgg_energy_int}
\end{align}
where we have expanded to $\mathcal{O}(\ep^0)$. The finite terms in the energy integration contribute to the anomalous dimension because of implicit $1/\ep$ poles that are generated by the angular integration. However, in the angular integration of the finite terms we only need to consider those $1/\ep$ poles which arise from the configuration where the emitted gluons with momenta $q$ and $r$ become collinear to each other, because the cancellation of singularities from emissions collinear to the hard lines is guaranteed by the KLN theorem, as detailed in Section \ref{sec:collSing}. Thus, we can expand the finite terms in \eqref{eq:RRgg_energy_int} around the limit where both emissions are collinear, i.e. $n_{qr}\to0$. Dropping integrable terms which do not result in $1/\ep$ poles, we obtain
\begin{equation}
 K_{ij;qr}^{A,(\ep)} = \frac{20-6\CDRflag}{9 \, n_{qr}^2}\left(\frac{n_{ir}}{n_{iq}} - \frac{n_{jr}}{n_{jq}}\right)^2 - \frac{64W_{ij}^q}{n_{qr}} + \dots \,,
\end{equation}
whereas $K_{ijk;qr}^{(\ep)}$ vanishes. The collinear expansion of the finite $\pi^2$ terms is straightforward. The fact that the $\mathcal{O}(\ep^0)$ terms in \eqref{eq:RRgg_energy_int} only contribute via the collinear configuration allows us to perform one of the angular integrations once and for all -- independently of the shape of the veto region. We find 
\begin{align}
\bm{\mathcal{H}}_{m+2}^{gg}\hat{\otimes}\bm{1} ={}& \left(\frac{\alpha_s}{4\pi}\right)^2 \left(\frac{\mu}{2\Lambda}\right)^{4\ep} \Bigg\{\Dqqqin{q}\Dqqqin{r}\Bigg[
                                 \sum_{i,j,k,l} \{\bm{T}_i^a,\bm{T}_j^b\}\bm{\mathcal{H}}_m\{\bm{T}_k^a,\bm{T}_l^b\} \frac{W_{ik}^qW_{jl}^r}{2\ep^2}\nonumber\\
                               & + \sum_{(ij)}\sum_k if^{abc} \left(\bm{T}_k^c \bm{\mathcal{H}}_{m} \{\bm{T}_i^a,\bm{T}_j^b\} - \{\bm{T}_i^a,\bm{T}_j^b\} \bm{\mathcal{H}}_{m} \bm{T}_k^c\right)
                                 \left(\frac{W_{ik}^qW_{jq}^r}{\ep^2} + \frac{K_{ijk;qr}}{8\ep}\right)\nonumber\\
                               & + C_A\sum_{(ij)} \bm{T}_i^a\bm{\mathcal{H}}_{m} \bm{T}_j^a  \left(\frac{W_{ij}^qW_{ij}^r - 4W_{ij}^qW_{jq}^r}{2\ep^2}
                                 + \frac{K_{ij;qr}^A}{2\ep}\right)\Bigg] \nonumber\\
                               & + \frac{2C_A}{\ep}\left(\frac{67}{9} + \frac{\CDRflag}{6} - \frac{2\pi^2}{3}\right) 
                                 \sum_{(ij)} \bm{T}_i^a\bm{\mathcal{H}}_{m} \bm{T}_j^a \Dqqqin{q} W_{ij}^q  + \mathcal{O}(\ep^0)\Bigg\}\,.
 \label{eq:RRgg_ang_int}
\end{align}
The renormalization condition states that the poles in this expression must be removed by the counterterms contribution 
\begin{equation}
 \left(\frac{\alpha_s}{4\pi}\right)^2 \left[\frac{-1}{8\ep^2}\bm{\mathcal{H}}_m\hat{\otimes} \bm{R}_m \hat{\otimes} \bm{R}_{m+1} + 
 \frac{1}{2\ep}\bm{\mathcal{H}}_{m+1}^{\text{ren}} \hat{\otimes} \bm{R}_{m+1} + 
 \frac{1}{4\ep}\bm{\mathcal{H}}_m \hat{\otimes} \bm{d}_m^A\right] .
\end{equation}
Here $\bm{\mathcal{H}}_{m+1}^{\text{ren}}$ denotes the hard function in which the divergence associated with the extra emission has been renormalized. This does not arise in our computation since we only consider divergences in the region where both gluon momenta $q$ and $r$ become soft. The first term removes the $1/\ep^2$ poles in \eqref{eq:RRgg_ang_int} which originate from strongly-ordered contributions and we can solve for the two-gluon emission part of the anomalous dimension. We obtain 
\begin{align}
 \bm{d}_m^A={}& \sum_{(ij)}\sum_k if^{abc}\left(\bm{T}_{i,L}^a \bm{T}_{j,L}^b \bm{T}_{k,R}^c - \bm{T}_{i,R}^ a\bm{T}_{j,R}^b \bm{T}_{k,L}^c\right) K_{ijk;qr} \theta_\text{in}(n_q)\theta_\text{in}(n_r) \nonumber\\
                 & - 2C_A\sum_{(ij)}\bm{T}_{i,L}^a \bm{T}_{j,R}^a K_{ij;qr}^A \theta_\text{in}(n_q) \theta_\text{in}(n_r) \nonumber\\
                 & - 8C_A\left(\frac{67}{9} + \frac{\CDRflag}{6} - \frac{2\pi^2}{3}\right) \sum_{(ij)} \bm{T}_{i,L}^a\bm{T}_{j,R}^a W_{ij}^q \theta_\text{in}(n_q) \delta(n_q-n_r)\,.
 \label{eq:dmA}
\end{align}
To extract the anomalous dimension, we can set $\mu=2\Lambda$ in \eqref{eq:RRgg_ang_int}, since the divergent parts involving logarithms of $\mu$ are subtracted by the $\bm{\mathcal{H}}_{m+1}^{\text{ren}} \hat{\otimes} \bm{R}_{m+1}$ contribution.

Next, we turn to the fermionic part of the anomalous dimension. The amplitude with a soft quark-antiquark pair reads 
\begin{align}\label{eq:two_quarks}
| {\cal M}_{m+2}^{\bar{q}q}(\{ \underline{p}, q,r \}) \rangle = 
   g_s\bar{u}(q)\gamma^\mu T^av(r)\,\frac{1}{2q\cdot r}\,\bm{J}_{\mu,a}(q+r)
   | {\cal M}_{m} (\{ \underline{p} \}) \rangle\,,
\end{align}
and we obtain a simplified expression for the hard function in the soft region $E_q +E_r\leq\Lambda$ 
\begin{align}
\bm{\mathcal{H}}_{m+2}^{\bar{q}q} ={}& 
   \frac{g_s^4 n_FT_F}{\tilde{c}^{2\ep}(2\pi)^4} \, \sum_{i,j}\bm{T}_i^a \, \bm{\mathcal{H}}_{m} \bm{T}_j^a \, \theta_\text{in}(n_q) \theta_\text{in}(n_r)\nonumber\\
 & \times\int\frac{dE_q}{E_q^{2\ep}}\frac{dE_r}{E_r^{2\ep}}
   \frac{n_{iq}\, n_{jr} + n_{ir} \, n_{jq} - n_{ij} \, n_{qr}}{(E_q n_{iq} + E_r n_{ir}) \, (E_q n_{jq} + E_r n_{jr}) \, n_{qr}^2} \, \theta(\Lambda-E_q-E_r) 
   \,,
\end{align}
where $n_F$ denotes the number of massless quark flavors and $T_F=1/2$. The energy integration yields 
\begin{equation}
\bm{\mathcal{H}}_{m+2}^{\bar{q}q} = 
   \left(\frac{\alpha_s}{4\pi}\right)^2\left(\frac{\mu}{2\Lambda}\right)^{4\ep} \, n_FT_F \, \sum_{i,j}\bm{T}_i^a \, \bm{\mathcal{H}}_{m} \bm{T}_j^a \, 
   \frac{K_{ij;qr}^F+K_{ij;qr}^{F,(\ep)}\ep}{\ep} \, \theta_\text{in}(n_q) \theta_\text{in}(n_r) \,,
\end{equation}
where we only obtain a single pole because the configuration with a single soft quark or antiquark is power suppressed. Again, we only require the collinear limit of the $\mathcal{O}(\ep^0)$ term which takes the form 
\begin{equation}
 K_{ij;qr}^{F,(\ep)} = -\frac{20}{9 \, n_{qr}^2}\left(\frac{n_{ir}}{n_{iq}} - \frac{n_{jr}}{n_{jq}}\right)^2 + \frac{16W_{ij}^q}{n_{qr}} + \dots\,.
\end{equation}
Performing the angular integration over the finite part, we find the contribution to the anomalous dimension 
\begin{equation}
  \bm{d}_m^F = -4n_FT_F\sum_{(ij)} \bm{T}_{i,L}^a\bm{T}_{j,R}^a \, \left[K_{ij;qr}^F \, \theta_\text{in}(n_r) - \frac{52}{9} \, W_{ij}^q \, \delta(n_q-n_r)\right]\theta_\text{in}(n_q)\,.
\end{equation}

Last and least, we consider the amplitude with a pair of soft scalars in the color-adjoint representation 
\begin{align}\label{eq:two_scalars}
| {\cal M}_{m+2}^{ss}(\{ \underline{p}, q,r \}) \rangle = 
   ig_sf^{abc}\,\frac{r^\mu-q^\mu}{2q\cdot r}\,\bm{J}_{\mu,a}(q+r)
   | {\cal M}_{m} (\{ \underline{p} \}) \rangle\,. 
\end{align}
The hard function in this configuration reads 
\begin{align}
\bm{\mathcal{H}}_{m+2}^{ss} ={}& 
   \frac{g_s^4 n_SC_A}{4 \tilde{c}^{2\ep} (2\pi)^4} \, \sum_{i,j}\bm{T}_i^a \, \bm{\mathcal{H}}_{m} \bm{T}_j^a \, \theta_\text{in}(n_q) \theta_\text{in}(n_r)\nonumber\\
 & \times\int\frac{dE_q}{E_q^{1+2\ep}}\int\frac{dE_r}{E_r^{1+2\ep}}
   \frac{(E_q n_{iq} - E_r n_{ir}) \, (E_q n_{jq} - E_r n_{jr})}{(E_q n_{iq} + E_r n_{ir}) \, (E_q n_{jq} + E_r n_{jr}) \, n_{qr}^2} \, \theta(\Lambda-E_q-E_r)
   \,,
\end{align}
where $n_S$ denotes the number of scalars. The energy integration creates a double pole which vanishes due to color conservation. For the single-pole contribution we find 
\begin{equation}
\bm{\mathcal{H}}_{m+2}^{ss} = 
   \left(\frac{\alpha_s}{4\pi}\right)^2 \left(\frac{\mu}{2\Lambda}\right)^{4\ep}  n_SC_A \, \sum_{(ij)}\bm{T}_i^a \, \bm{\mathcal{H}}_{m} \bm{T}_j^a \, 
   \frac{K_{ij;qr}^S+K_{ij;qr}^{S,(\ep)}\ep}{-2\ep} \, \theta_\text{in}(n_q) \theta_\text{in}(n_r)\nonumber\\
   \,,
\end{equation}
where 
\begin{equation}
 K_{ij;qr}^S = \frac{4(n_{iq}n_{jr}+n_{ir}n_{jq})}{n_{qr}^2(n_{iq}n_{jr}-n_{ir}n_{jq})}\ln\frac{n_{iq}n_{jr}}{n_{ir}n_{jq}} - \frac{8}{n_{qr}^2} \,,
\end{equation}
and the collinear limit of the $\mathcal{O}(\ep^0)$ term is given by 
\begin{equation}
 K_{ij;qr}^{S,(\ep)} = -\frac{20}{9 \, n_{qr}^2}\left(\frac{n_{ir}}{n_{iq}} - \frac{n_{jr}}{n_{jq}}\right)^2 + \dots\,.
\end{equation}
Proceeding as before, we find the following contribution to the anomalous dimension 
\begin{equation}
  \bm{d}_m^S = 2n_SC_A\sum_{(ij)} \bm{T}_{i,L}^a\bm{T}_{j,R}^a \, \left[K_{ij;qr}^S \theta_\text{in}(n_r) + \frac{20}{9} \, W_{ij}^q \, \delta(n_q-n_r) \right] \, \theta_\text{in}(n_q) \,.
\end{equation}
Summing up the contributions from gluons, quarks and scalars, we obtain the result presented in Eq.~\eqref{eq:dm0}.

\subsection{\boldmath Real-virtual contribution $r_m$}

As we have discussed in Section~\ref{sec:oneloop} for the one-loop case, we want to make the cancellation of soft-collinear divergences manifest by also expressing the virtual contributions as angular integrals. Nevertheless, the inclusive results in dimensional regularization will prove useful for the determination of the imaginary parts of the two-loop anomalous dimension where no cancellations between virtual corrections and real emissions can occur. Thus, making use of known results for the one-loop soft emission, we first consider the inclusive version of the real-virtual corrections where the entire loop integration has been performed. The matrix element with $L+1$ loops and $m+1$ legs takes the form \cite{Catani:2000pi}
\begin{align}\label{eq:oneloopsoftEx}
| {\cal M}_{m+1}^{(L+1)}(\{ \underline{p}, q \}) \rangle ={}& \varepsilon_{m+1}^{*\mu} \left[\bm{J}_{\mu,a}^{(1)}(q) | {\cal M}_{m}^{(L)} (\{ \underline{p} \}) \rangle + 
                                                              \bm{J}_{\mu,a}^{(0)}(q) | {\cal M}_{m}^{(L+1)} (\{ \underline{p} \}) \rangle \right] \\
 ={}& \varepsilon_{m+1}^{*\mu} \left[\bm{J}_{\mu,a}^{(1)}(q) + 
      \bm{J}_{\mu,a}^{(0)}(q) \, \frac12 \int\frac{d^dr}{(2\pi)^d}\,\frac{i}{r^2+i0} \bm{J}_{\nu,b}^{(0)\dagger}(r)\bm{J}_{\nu,b}^{(0)}(r)\right] | {\cal M}_{m}^{(L)} (\{ \underline{p} \}) \rangle \nonumber
\end{align}
in the region where the loop $L+1$ with the associated loop momentum $r$ routed through a gluon line\footnote{Quark loops are power-suppressed in the soft limit.} and the leg $m+1$ with momentum $q$ both become soft, while all other internal momenta are considered hard. In this expression, we have expanded the soft current as
\begin{equation}
  \bm{J}_{\mu,a}(q) =   g_s\left[ \bm{J}_{\mu,a}^{(0)}(q)  + \frac{g_s^2}{16\pi^2} \bm{J}_{\mu,a}^{(1)}(q) \right] .
\end{equation}
The term with the integrated one-loop soft current 
\begin{align}\label{eq:oneLoopCurrent}
 \bm{J}_{\mu,a}^{(1)}(q) ={}& -\frac{\Gamma^3(1-\ep)\,\Gamma^2(1+\ep)}{\ep^2\,\Gamma(1-2\ep)} 
                         if^{abc} \, \sum_{(jk)}\bm{T}_i^b \bm{T}_j^c \left(\frac{n_{j\mu}}{n_j\cdot q}-\frac{n_{k\mu}}{n_k\cdot q}\right)
                         \left[\frac{4\pi \, W_{jk}^q \, e^{-i\pi\Pi_{jk}}}{2 E_q^2 \, e^{-i\pi\Pi_{jq}}e^{-i\pi\Pi_{kq}}}\right]^\ep
\end{align}
is purely non-Abelian while the second term in \eqref{eq:oneloopsoftEx} is Abelian and its contribution will be fully removed by the $\bm{\Gamma}^{(1)}\hat\otimes\bm{\Gamma}^{(1)}$ counterterm. 

In the limit where the leg $m+1$ becomes soft the $(L+1)$-loop hard function takes the form 
\begin{align}
\bm{\mathcal{H}}_{m+1}^{(L+1)} ={}& \frac{ g_s^4\, \theta_\text{in}(n_q)}{4\ep \, (2\pi)^{d-2}}
   \left[\bm{J}_{\mu,a}^{(1)}(n_q) + 
       \frac{\bm{J}_{\mu,a}^{(0)}(n_q)}{2} \int\frac{d^dr}{(2\pi)^d}\,\frac{i\bm{J}_{\nu,b}^{(0)\dagger}(r)\bm{J}_{\nu,b}^{(0)}(r)}{r^2+i0} \right] \, 
      \bm{\mathcal{H}}_{m}^{(L)} \, \bm{J}_{\mu,a}^{(0)\dagger}(n_q) + \text{h.c.}\,,
\end{align}
where we have carried out the trivial energy integration. We obtain 
\begin{align}
\label{eq:RV_integrated_Abelian}
 \left.\bm{\mathcal{H}}_{m+1}^{(L+1)}\hat{\otimes}\bm{1}\right|_\text{Abelian}     ={}&  \left(\frac{\alpha_s}{4\pi}\right)^2 \left(\frac{\mu}{2\Lambda}\right)^{4\ep} \frac{1}{4\ep^2} \bm{\mathcal{H}}_{m}^{(L)} \bm{V}_m \hat{\otimes} \bm{R}_m \,, \\
 \left.\bm{\mathcal{H}}_{m+1}^{(L+1)}\hat{\otimes}\bm{1}\right|_\text{Non-Abelian} ={}& -\left(\frac{\alpha_s}{4\pi}\right)^2 \left(\frac{\mu}{2\Lambda}\right)^{4\ep}\frac{\Gamma^3(1-\ep)\,\Gamma^2(1+\ep)}{\ep^3\,\Gamma(1-2\ep)} 
    \sum_{(jk)}\sum_i  if^{abc} \, \bm{T}_j^a \bm{T}_k^b \bm{\mathcal{H}}_{m}^{(L)} \bm{T}_i^c \nonumber\\
 & \times e^{\ep \gamma_E}\Dqqqin{q} (W_{ij}^q-W_{ik}^q) \left[\frac{2 W_{jk}^q \, e^{-i\pi\Pi_{jk}}}{e^{-i\pi\Pi_{jq}}e^{-i\pi\Pi_{kq}}}\right]^\ep \,.
\label{eq:RV_integrated_NonAbelian}
\end{align}
The poles in this expression have to be removed by the counterterm 
\begin{align}
    & \left(\frac{\alpha_s}{4\pi}\right)^2 
      \Bigg[\frac{-1}{8\ep^2}\bm{\mathcal{H}}_m^{(L)} \hat{\otimes} \bm{R}_m (\bm{V}_{m+1} + 2\beta_0)
          - \frac{1}{8\ep^2} \bm{\mathcal{H}}_m^{(L)} \bm{V}_m \hat{\otimes} \bm{R}_{m}
          + \frac{1}{2\ep} \bm{\mathcal{H}}_{m}^{\text{ren},(L+1)} \hat{\otimes} \bm{R}_{m}\nonumber\\
    &     + \frac{1}{2\ep} \bm{\mathcal{H}}_{m+1}^{\text{ren},(L)} \left(\bm{V}_{m+1}-2\beta_0\right)
          + \frac{1}{4\ep} \bm{\mathcal{H}}_m^{(L)} \hat{\otimes} \bm{r}_m \Bigg]\,,
 \label{eq:RV_ct}
\end{align}
where the contributions from the third and fourth terms vanish because the renormalized hard function vanishes in the soft region. Summing up the contribution we find 
\begin{align}\label{eq:RV_intsum}
 \bm{\mathcal{H}}_{m+1}^{\text{ren},(L+1)}\hat{\otimes}\bm{1} ={}& \left.\bm{\mathcal{H}}_{m+1}^{(L+1)}\hat{\otimes}\bm{1}\right|_\text{Non-Abelian} + \left(\frac{\alpha_s}{4\pi}\right)^2  \Bigg\{ 
     \frac{2}{\ep^2} \sum_{(jk)}\sum_i if^{abc} \, \bm{T}_j^a \bm{T}_k^b \bm{\mathcal{H}}_{m}^{(L)} \bm{T}_i^c \nonumber\\
   &\hspace{-0.7cm} \times \Dqqqin{q} W_{ij}^q \left[\frac{\Gamma(-\ep) e^{\ep \gamma_E}}{\Gamma(1-2\ep)}\left[\left(\frac{n_{jk}}{2}\right)^{-\ep}-\left(\frac{n_{kq}}{2}\right)^{-\ep}\right]
     + i\pi\left(\Pi_{kq} - \Pi_{jk}\right) \right] + \text{h.c.}\nonumber\\
   &\hspace{-0.7cm} - \frac{\beta_0}{\ep^2} \sum_{(ij)} \bm{T}_i^a \bm{\mathcal{H}}_{m}^{(L)} \bm{T}_j^a \, \Dqqqin{q} W_{ij}^q
     + \frac{1}{4\ep} \bm{\mathcal{H}}_m^{(L)} \hat{\otimes} \bm{r}_m \Bigg\} \,,
\end{align}
where we have evaluated the contribution of the extra gluon in $\bm{V}_{m+1}$ using the integral
\begin{equation}
\Dqqq{r} W_{ij}^r = \frac{\Gamma(-\ep)e^{\ep \gamma_E}}{\Gamma(1-2\ep)}\left(\frac{n_{ij}}{2}\right)^{-\ep} + \mathcal{O}(\ep)\,
 \label{eq:ang_int_dipole}
\end{equation}
in order to match the current \eqref{eq:oneLoopCurrent} which also is given in integrated form. We note that in the case where the directions $i$ and $j$ are back to back, the higher-order corrections in $\ep$ vanish and formula \eqref{eq:ang_int_dipole} becomes exact. 

From the result \eqref{eq:RV_intsum} we can now read off the imaginary part of the anomalous dimension as 
\begin{align}
 \text{Im}\left[\bm{r}_m \right] ={}& - 8 \pi \sum_{(jk)}\sum_i if^{abc} \, \bm{T}_j^a \bm{T}_k^b \bm{\mathcal{H}}_{m}^{(L)} \bm{T}_i^c \, W_{ij}^q 
               \left(\Pi_{jk}-\Pi_{jq}-\Pi_{kq}\right)\ln\!\left(2W_{jk}^q\right) \theta_\text{in}(n_q)  - \text{h.c.} \, ,
 \label{eq:RV_integrated}
\end{align}
where we have dropped terms that vanish due to color conservation. Since we divide by $i$ to obtain the imaginary part, we need to subtract the hermitian conjugate in \eqref{eq:RV_integrated}. The factor in the imaginary part simplifies to $\Pi_{jk}-\Pi_{jq}-\Pi_{kq}=1$ when $j$ and $k$ are both incoming and $\Pi_{jk}-\Pi_{jq}-\Pi_{kq}=-1$ otherwise, because $q$ is outgoing and this has been exploited to simplify a term proportional to $(\Pi_{jk}-\Pi_{jq}-\Pi_{kq})^2=1$. This implies that the imaginary part takes the simple form 
\begin{align}
 \text{Im}\left[\bm{r}_m\right] ={}& 8\pi \sum_{(jk)}\sum_i if^{abc} \, \bm{T}_j^a \bm{T}_k^b \bm{\mathcal{H}}_{m}^{(L)} \bm{T}_i^c \, W_{ij}^q \ln\!\left(W_{jk}^q\right)  \theta_\text{in}(n_q)  - \text{h.c.} 
 \label{eq:RV_im}
\end{align}
when the hard function involves at most one incoming leg, and we will exploit the observation that the contribution from legs $i$, $j$ and $k$ is the same when $j$ and $k$ are both outgoing or when one is incoming and the other is outgoing below.

\begin{figure}
\begin{center}
\includegraphics[width=0.9\textwidth]{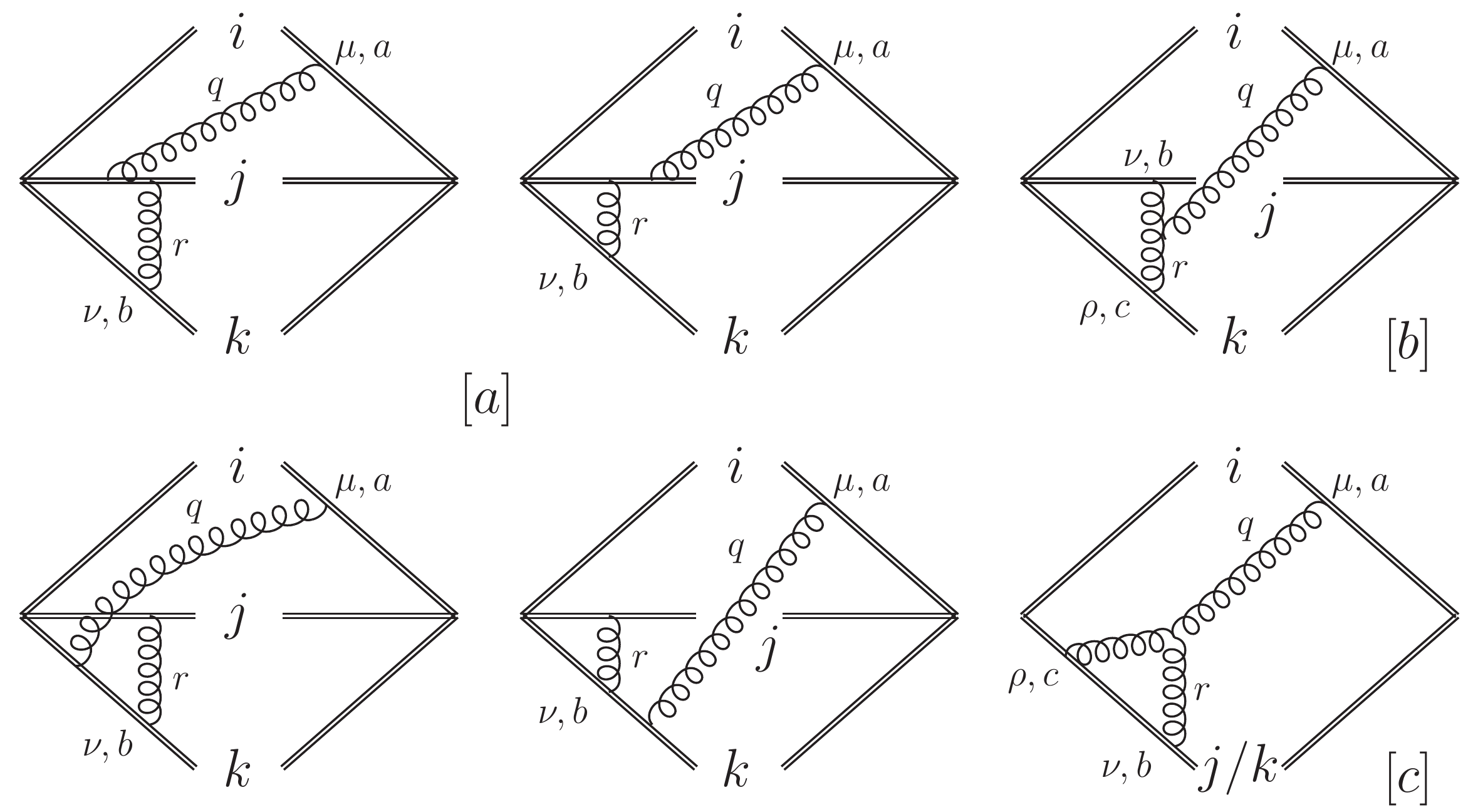}
\end{center}
\caption{Real-virtual diagrams. We denote the contribution from the left four diagrams as part $[a])$ while the upper and lower diagram on the right are parts $[b]$ and $[c]$, respectively. The symmetry of part $[a]$ under exchange of $j$ and $k$ is broken when these lines have opposite directionalities as considered here. In this case there is a second diagram in category $[c]$, where we attach to $k$ instead of $j$. \label{fig:RV_diagrams}}
\end{figure}

As stressed above, the integrated form \eqref{eq:oneLoopCurrent} of the soft current is not suitable to obtain the real part of the anomalous dimension $\bm{r}_m$, since we want to work in a scheme where all energy integrals are evaluated in the presence of a UV cutoff and the angular dependence on the loop momentum $r$ is kept explicit. To determine the real part, we now extract the real-virtual result from the diagrams shown in Figure~\ref{fig:RV_diagrams}. We write the contributions with gluon attachments to the lines $i$, $j$ and $k$ in the form
\begin{align}
\bm{\mathcal{H}}_{m+1}^{(L+1)[\alpha]} \otimes \bm{1}&= \sum_i\sum_{(jk)} h_{ijk}^{[\alpha]} + \text{h.c.}\,, 
 \label{eq:Rv_h_def}
\end{align}
for $\alpha=a,b$. Using color conservation, also the contribution $[c]$ can be written in this way. In this work we are interested in processes where all eikonal lines are outgoing. However, as pointed out above, the configuration where $j$ is outgoing and $k$ is incoming yields the same result as the case where all lines are outgoing. This allows us to perform the calculation of the $h_{ijk}^{[\alpha]}$ for the configuration where the line $j$ is outgoing, the line $k$ is incoming and the directionality of the line $i$ is arbitrary. This has the advantage that no Glauber phases occur which considerably simplifies the calculation. While we adopt this kinematical configuration for the calculation itself, we then use that the result is the same if all lines are outgoing when we carry out the sums over all hard partons pairs $(jk)$ in \eqref{eq:Rv_h_def}. We note that the light-like vectors $n_k$ are always defined with $n_k^0=+1$ and thus $0\leq n_{ik}\leq2$. The contribution from the Abelian diagrams in part $[a]$ of Figure~\ref{fig:RV_diagrams} takes the form 
\begin{align}\label{eq:hijk}
 h_{ijk}^{[a]} ={}&- \frac{g_s^4}{2} \Dqin{q}\Dq{r}\frac{2\pi i\delta_+(q^2)\theta(\Lambda-E_q)}{[r^2+i0]\,n_{i} \cdot q}\\
                {}& \times\Bigg\{\frac{n_{ij}\,n_{jk}}{[-n_k\cdot r+i0][n_j\cdot(q-r)+i0]}
                    \left[\frac{\bm{T}_j^a\bm{T}_j^b\bm{T}_k^b\bm{\mathcal{H}}_m^{(L)} \bm{T}_i^a}{-n_j\cdot r+i0} + \frac{\bm{T}_j^b\bm{T}_j^a\bm{T}_k^b\bm{\mathcal{H}}_m^{(L)} \bm{T}_i^a}{n_j\cdot q}\right] \nonumber\\
                {}& - \frac{n_{ik}\,n_{jk}}{[-n_j\cdot r+i0][-n_k\cdot(q+r)+i0]}
                    \left[\frac{\bm{T}_j^b\bm{T}_k^a\bm{T}_k^b\bm{\mathcal{H}}_m^{(L)} \bm{T}_i^a}{-n_k\cdot r+i0} + \frac{\bm{T}_j^b\bm{T}_k^b\bm{T}_k^a\bm{\mathcal{H}}_m^{(L)} \bm{T}_i^a}{-n_k\cdot q}\right]\Bigg\}\,,\nonumber
\end{align}
where we have routed the loop momentum such that the only pole in the lower half of the complex $r^0$ plane is at $r^0 = E_r - i0$. 

To obtain the anomalous dimension, we now want to separate out the soft region in the energy integrals by putting an upper cutoff on the energy integrations. For the real momentum $q$, we have already imposed a cutoff $\theta(\Lambda-E_q)$ in \eqref{eq:hijk}, and after closing the contour it seems natural to put an upper limit $\theta(\Lambda-E_r)$ also on the energy integration associated with the loop integral. We would then have a cutoff $\theta(\Lambda-E_q) \theta(\Lambda-E_r)$ in the real-virtual diagrams and $\theta(\Lambda-E_q-E_r)$ in double real diagrams. The energy integrations with the two cutoffs differ at $\mathcal{O}(\epsilon^0)$,
\begin{align}\label{eq:pi2problem}
&\int dE_q d E_r  \frac{\theta(\Lambda-E_q) \theta(\Lambda-E_r)}{ E_q^{-2 \ep-1} E_r^{-2 \ep-1} } = \frac{\Lambda^{-4\ep}}{4} \frac{1}{\ep^2} \,  , \nonumber 
\\
&\int dE_q d E_r  \frac{\theta(\Lambda-E_q-E_r) }{ E_q^{-2 \ep-1} E_r^{-2 \ep-1} } = \frac{\Lambda^{-4\ep}}{4} \left(\frac{1}{\ep^2} -\frac{2 \pi^2}{3} \right) +\mathcal{O} (\ep) \, . 
\end{align}
While the difference between the cutoff prescriptions is finite, the $\pi^2$ term becomes relevant when one encounters a collinear singularity in the angular integrals. If this is the case, it produces a contribution $\propto C_A \pi^2$ in the anomalous dimension. While a separate cutoff on the real and virtual contributions seems natural since the observable only constrains the real emission energy, one could also argue the one should impose the same cutoff in both types of diagrams to ensure the real-virtual cancellations predicted by the KLN theorem. On a more basic level, our problem is that we want to extract an anomalous dimension in dimensional regularization by imposing cutoffs in integrals. Unfortunately, for the $\propto C_A \pi^2$ terms, the result depends on the way in which we impose the cutoff. Since we do not see a simple way to resolve this ambiguity, our method does not predict the $\propto C_A \pi^2$ terms. For simplicity, we will impose the cutoff $\theta(\Lambda-E_q-E_r)$ in all computations below, but neither of the two cutoffs prescriptions \eqref{eq:pi2problem} leads to the correct result for the $\pi^2$ terms. However, the correct coefficient of the $\pi^2$ term can be inferred by comparing to the explicit result for soft functions at NNLO. In our results given in Section \eqref{sec:Gamma2_result_diag}, the coefficients of the $C_A \pi^2$ terms in $\bm{r}_m$ and $\bm{v}_m$ were chosen so that the result for $\bm{\Gamma}^{(2)}$ passes the finiteness check in \ref{sec:finite}.

Returning to \eqref{eq:hijk} we now close the countour in the lower half and pick up the residue at $r^0 = E_r - i0$. After performing the energy integration with the cutoff $\theta(\Lambda-E_q-E_r)$,  we obtain the following result for the Abelian diagrams $[a]$.
\begin{align}
 h_{ijk}^{[a]} ={}& \left(\frac{\alpha_s}{4\pi}\right)^{}2 \left(\frac{\mu}{2\Lambda}\right)^{4\ep} \Dqqqin{q}\Dqqq{r}\nonumber\\
                  &\hspace{-0.5cm}\Bigg\{ W_{ij}^qW_{jk}^r\left[ - \left(\frac{1}{\ep^2}- \frac{2 \pi^2}{3} \right) \{\bm{T}_j^a,\bm{T}_j^b\}\bm{T}_k^b\bm{\mathcal{H}}_m^{(L)} \bm{T}_i^a
                    + \frac{2if^{abc}\bm{T}_j^b\bm{T}_k^c\bm{\mathcal{H}}_m \bm{T}_i^a}{\ep}\left[\ln\left(\frac{n_{jq}}{n_{jr}}\right)-i\pi\right]\right]  \nonumber\\
                  &\hspace{-0.5cm}+ W_{ik}^qW_{jk}^r\left[ - \left( \frac{1}{\ep^2}-\frac{2 \pi^2}{3} \right) \bm{T}_j^b\{\bm{T}_k^a,\bm{T}_k^b\}\bm{\mathcal{H}}_m^{(L)} \bm{T}_i^a
                    - \frac{2if^{abc}\bm{T}_j^b\bm{T}_k^c\bm{\mathcal{H}}_m \bm{T}_i^a}{\ep}\ln\left(\frac{n_{kq}}{n_{kr}}\right)\right]\Bigg\}\,. 
 \label{eq:RVa}
\end{align}
The non-Abelian diagram in part $[b]$ takes the form 
\begin{align}
 h_{ijk}^{[b]} ={}& \frac{g_s^4f^{abc}}{2} \bm{T}_j^b\bm{T}_k^c\bm{\mathcal{H}}_m^{(L)} \bm{T}_i^a \Dqin{q}\Dq{r}\frac{2\pi\delta_+(q^2)\theta(\Lambda-E_q)}{[r^2+i0][(q-r)^2+i0]}\nonumber\\
                  & \times\frac{n_{ij}\,n_k\cdot(r-2q) + n_{jk}\,n_i\cdot(q-2r) + n_{ik}\,n_j\cdot(q+r)}{ [n_i \cdot q ] [n_j\cdot (q-r)+i0][-n_k\cdot r+i0]}\,,
\end{align}
and after closing the contour in the lower half to pick up the residues of the two gluon propagators and performing the energy integration we obtain 
\begin{align}
 h_{ijk}^{[b]} ={}& \left(\frac{\alpha_s}{4\pi}\right)^2 \left(\frac{\mu}{2\Lambda}\right)^{4\ep} if^{abc}\bm{T}_j^b\bm{T}_k^c\bm{\mathcal{H}}_m^{(L)} \bm{T}_i^a
                               \Dqqqin{q}\Dqqq{r}\Bigg\{\nonumber\\
                  & \left(2W_{ij}^q W_{kq}^r - 2W_{ik}^q W_{jq}^r +   W_{ij}^r W_{ir}^q-W_{ik}^r W_{ir}^q\right) \left(\frac{1}{2\ep^2}- \frac{\pi^2}{3} \right)\nonumber\\
                  & - \left(2W_{ij}^q W_{kq}^r + 2W_{jk}^r W_{ir}^q - 2W_{ik}^q W_{kq}^r - W_{ij}^r W_{ir}^q\right)
                    \frac{\ln\left(\frac{n_{jq}}{n_{jr}}\right)-i\pi}{\ep} \nonumber\\
                  & + \left(2W_{ik}^q W_{jq}^r + 2W_{jk}^r W_{ir}^q - 2W_{ij}^q W_{jq}^r - W_{ik}^r W_{ir}^q\right)
                    \frac{\ln\left(\frac{n_{kq}}{n_{kr}}\right)}{\ep} \nonumber\\
                  & +W_{jk}^r W_{kr}^q  \left[\frac{1}{2\ep^2}- \frac{\pi^2}{3} - \frac{\ln\left(\frac{n_{kq}}{n_{kr}}\right)}{\ep}\right]
                    -W_{jk}^r W_{jr}^q \left[\frac{1}{2\ep^2} - \frac{\pi^2}{3}- \frac{\ln\left(\frac{n_{jq}}{n_{jr}}\right)-i\pi}{\ep}\right]
                    + \cdots\Bigg\}\,,
 \label{eq:RVb}
\end{align}   
where the terms in the last line are transverse and will vanish in the sum over $i$ because of color conservation. 

 The contribution $[c]$ involves only two legs. In our calculation, we will again consider the case of an outgoing leg $j$ and an incoming leg $k$, but at the end extend the sum over all $j$ and $k$. In order to avoid double counting, we therefore compute the average
\begin{equation}
 \bm{\mathcal{H}}_{m+1}^{(L+1)[c]} \otimes \bm{1} = \frac{1}{2} \left[\sum_{i,j} h_{ij}^{[c]} + \sum_{i,k} h_{ik}^{[c]}\right] + \text{h.c.}\,.
 \label{eq:RVc}
\end{equation}
For the two kinematic configurations, we have 
\begin{align}
 h_{ij}^{[c]} ={}& -\frac{ig_s^4C_A}{2} \bm{T}_j^a\bm{\mathcal{H}}_m^{(L)} \bm{T}_i^a \Dqin{q}\Dq{r}
                      \frac{2\pi\delta_+(q^2)\theta(\Lambda-E_q)W_{ij}^q\, n_j\cdot(q+2r)}{E_q^2[r^2+i0][(q+r)^2+i0][-n_j\cdot r+i0]} \nonumber\\
              ={}& \left(\frac{\alpha_s}{4\pi}\right)^2 \left(\frac{\mu}{2\Lambda}\right)^{4\ep} C_A \bm{T}_j^a\bm{\mathcal{H}}_m^{(L)} \bm{T}_i^a \Dqqqin{q}\Dqqq{r} W_{ij}^r W_{ir}^q
                     \Bigg[\frac{1}{2\ep^2} - \frac{\ln\frac{n_{jq}}{n_{jr}} - i\pi}{\ep} - \pi^2\nonumber\\
                 &   - \left(\ln\frac{n_{jq}}{n_{jr}} - i\pi\right)^2 + 4\text{Li}_2\left(-\frac{n_{jr}}{n_{jq}}\right)
                     - 4\left(\ln\frac{n_{jq}}{n_{jr}} - i\pi\right) \ln\left(1+\frac{n_{jr}}{n_{jq}}\right) + \dots\Bigg]\,,
 \label{eq:RVc_ij}
\end{align}
and 
\begin{align}
 h_{ik}^{[c]} ={}& \frac{ig_s^4C_A}{2} \bm{T}_k^a\bm{\mathcal{H}}_m^{(L)} \bm{T}_i^a \Dqin{q}\Dq{r}
                     \frac{2\pi\delta_+(q^2)\theta(\Lambda-E_q)W_{ik}^q \, n_k\cdot(q-2r)}{E_q^2[r^2+i0][(q-r)^2+i0][-n_k\cdot r+i0]} \nonumber\\
              ={}& \left(\frac{\alpha_s}{4\pi}\right)^2 \left(\frac{\mu}{2\Lambda}\right)^{4\ep} C_A \bm{T}_k^a\bm{\mathcal{H}}_m^{(L)} \bm{T}_i^a \Dqqqin{q}\Dqqq{r} W_{ik}^r W_{ir}^q  
                     \Bigg[\frac{1}{2\ep^2} - \frac{\ln\left(\frac{n_{kq}}{n_{kr}}\right)}{\ep} \nonumber\\
                 &         - \ln^2\frac{n_{kq}}{n_{kr}} - 4\text{Li}_2\left(1-\frac{n_{kr}}{n_{kq}}\right) - \frac{\pi^2}{3} + \dots \Bigg] \,.
 \label{eq:RVc_ik}
\end{align}
In part $[c]$ we have kept the finite terms of order $\ep^0$, because there is an implicit divergence in the angular integration when $q$ and $r$ become collinear and consequently a contribution to the anomalous dimension as discussed above for the double-real contribution. We now rewrite the contribution \eqref{eq:RVc} in the form \eqref{eq:Rv_h_def} by applying the color identity 
\begin{equation}
 \frac{C_A}{2} \bm{T}_j^a\bm{\mathcal{H}}_m^{(L)} \bm{T}_i^a = \sum_{k\neq j} if^{abc}\bm{T}_j^b\bm{T}_k^c\bm{\mathcal{H}}_m^{(L)} \bm{T}_i^a \,
\end{equation}
to \eqref{eq:RVc_ij} and a similar identity to \eqref{eq:RVc_ik}. Averaging the two terms as in \eqref{eq:RVc} yields
\begin{align}
 h_{ijk}^{[c]} ={}& \left(\frac{\alpha_s}{4\pi}\right)^2 \left(\frac{\mu}{2\Lambda}\right)^{4\ep} if^{abc} \bm{T}_j^b\bm{T}_k^c\bm{\mathcal{H}}_m^{(L)} \bm{T}_i^a \Dqqqin{q}\Dqqq{r} \Bigg\{
                        W_{ij}^r W_{ir}^q  \Bigg[\frac{1}{2\ep^2} - \frac{\ln\left(\frac{n_{jq}}{n_{jr}}\right) - i\pi}{\ep} \nonumber\\
                  &  + 4i\pi \ln(2) - \frac{\pi^2}{3}\Bigg] - W_{ik}^r W_{ir}^q \left[\frac{1}{2\ep^2} - \frac{\ln\left(\frac{n_{kq}}{n_{kr}}\right)}{\ep} - \frac{\pi^2}{3}\right] + \dots\Bigg\}\,,
 \label{eq:RVc_ijk}
\end{align}
where we have only kept terms that are singular as $q$ and $r$ become collinear in the term of order $\ep^0$. 
Summing up the contributions, we observe that the logarithmic terms in \eqref{eq:RVc_ijk} cancel against terms in \eqref{eq:RVb}. We find 
\begin{align}
 h_{ijk} ={}& \left(\frac{\alpha_s}{4\pi}\right)^2 \left(\frac{\mu}{2\Lambda}\right)^{4\ep} \Dqqqin{q}\Dqqq{r}\nonumber \\
 & \Bigg\{
             - \{\bm{T}_j^a,\bm{T}_j^b\}\bm{T}_k^b\bm{\mathcal{H}}_m^{(L)} \bm{T}_i^a \frac{W_{ij}^qW_{jk}^r}{\ep^2} - 
              \bm{T}_j^b\{\bm{T}_k^a,\bm{T}_k^b\}\bm{\mathcal{H}}_m^{(L)} \bm{T}_i^a \frac{W_{ik}^qW_{jk}^r}{\ep^2} \nonumber\\
            & + if^{abc}\bm{T}_j^b\bm{T}_k^c\bm{\mathcal{H}}_m^{(L)} \bm{T}_i^a \Bigg[
              \frac{W_{ij}^qW_{kq}^r - W_{ik}^qW_{jq}^r + W_{ij}^r W_{ir}^q- W_{ik}^r W_{ir}^q }{\ep^2} \nonumber\\
            & + \frac{K_{ijk;qr} - K_{ikj;qr}+8i\pi(W_{ij}^qW_{jk}^r+W_{ik}^qW_{kq}^r-W_{ij}^qW_{kq}^r-W_{jk}^rW_{ir}^q)}{-4\ep} \nonumber\\
            &+\frac{W_{ij}^q-W_{ik}^q}{n_{qr}}\left( 2 i\pi \ln(2) - \frac{4 \pi^2}{3}\right) \Bigg] + \dots \Bigg\}\,,
 \label{eq:RV_hijk}
\end{align}
which is renormalized by adding the contribution of the lines $i$, $j$ and $k$ to the counterterm \eqref{eq:RV_ct}. Assuming a process where all color-charged lines are outgoing, the final result for the renormalized hard function then follows by using color conservation to remove terms that are proportional to $\sum_i T_i^a = 0$, using symmetry under exchange of the lines $j$ and $k$ for simplifications, and performing the angular integral over $\Omega_r$ for the imaginary part. We note that the imaginary part in the last line of \eqref{eq:RV_hijk} cancels against the corresponding contribution in the Hermitian conjugate. We have explicitly checked that the terms of order $\ep^0$ in parts $[a]$ and $[b]$, which are not given above, do not contribute to the final result because they are either finite in the collinear limit or also cancel with the Hermitian conjugate. We finally obtain the result 
\begin{align}
 \bm{\mathcal{H}}_{m+1}^{\text{ren}(L+1)} \otimes \bm{1} ={}& \left(\frac{\alpha_s}{4\pi}\right)^2  \Bigg\{ 
                                          \sum_i\sum_{(jk)} if^{abc}\bm{T}_j^b\bm{T}_k^c\bm{\mathcal{H}}_m^{(L)} \bm{T}_i^a \Dqqqin{q} \Bigg[
                                          \Dqqq{r}\frac{2K_{ijk;qr}}{-4\ep} \nonumber\\
                                      & + \frac{8i\pi W_{ij}^q\ln\left(W_{jk}^q\right)}{-4\ep}\Bigg]
                                        + \text{h.c.} \nonumber\\
                                      & - \left(\frac{\beta_0}{\ep^2} - \frac{4 C_A \pi^2}{3\ep} \right) \sum_{(ij)} \bm{T}_i^a \bm{\mathcal{H}}_{m}^{(L)} \bm{T}_j^a \, \Dqqqin{q} W_{ij}^q
                                        + \frac{1}{4\ep} \bm{\mathcal{H}}_m^{(L)} \hat{\otimes} \bm{r}_m \Bigg\} \,,
 \label{eq:RV_Hijk}
\end{align}
where we have set $\mu=2\Lambda$ for simplicity. The imaginary part is in agreement with the expression \eqref{eq:RV_im} obtained in the inclusive calculation.

From these results, we read off the anomalous dimension
\begin{align}\label{eq:rmp}
 \bm{r}^\prime_m ={}& -2 \sum_i\sum_{(j k)}if^{abc}(\bm{T}_{i,L}^a \bm{T}_{j,R}^b \bm{T}_{k,R}^c - \bm{T}_{i,R}^a \bm{T}_{j,L}^b \bm{T}_{k,L}^c) \Dqqq{r} K_{ijk;qr} \theta_\text{in}(n_q) \nonumber\\
                 & + 8i\pi \sum_{(i j)}\sum_k if^{abc}\left( \bm{T}_{i,L}^a \bm{T}_{j,L}^b \bm{T}_{k,R}^c + \bm{T}_{i,R}^a \bm{T}_{j,R}^b \bm{T}_{k,L}^c\right)
                   W_{ik}^q\ln W_{ij}^q \,\theta_\text{in}(n_q) \nonumber\\
                 & + \left(\frac{4\beta_0}{\ep} - \frac{16C_A \pi^2}{3} \right)\sum_{(ij)} \bm{T}_{i,L}^a \bm{T}_{j,R}^a W_{ij}^q \theta_\text{in}(n_q) \,.
\end{align}
The prime in $\bm{r}^\prime_m$ indicates that this result was obtained with the energy cutoff $\theta(\Lambda-E_q-E_r )$. For a cutoff $\theta(\Lambda-E_q) \theta(\Lambda-E_r )$, the $\pi^2$ terms in the third line of \eqref{eq:rmp} would be absent. In the result for $ \bm{r}_m$ given in \eqref{eq:rm0}, the coefficient of the $\pi^2$ term has been adjusted by hand such that the result passes the finiteness check in Section \ref{sec:finite}.

\subsection{\boldmath Double-virtual contribution $v_m$}

Based on \eqref{eq:GammaAmp}, we argue that the imaginary part is given by 
\begin{equation}\label{eq:twoLoopGlauber}
  \bm{v}_m^{\i\pi} = - \sum_{(i j)}\left(\bm{T}_{i,L}^a \bm{T}_{j,L}^a-\bm{T}_{i,R}^a \bm{T}_{j,R}^a\right)\gamma_1^\text{cusp}\frac{i\pi \Pi_{ij}}{2} \,. 
\end{equation}
As in the RV part, this allows us to compute at least the 2-particle contribution in a configuration without picking up residues of Glauber poles and to obtain the full result via appropriate generalization.

\begin{figure}
\centering
\includegraphics[width=0.9\textwidth]{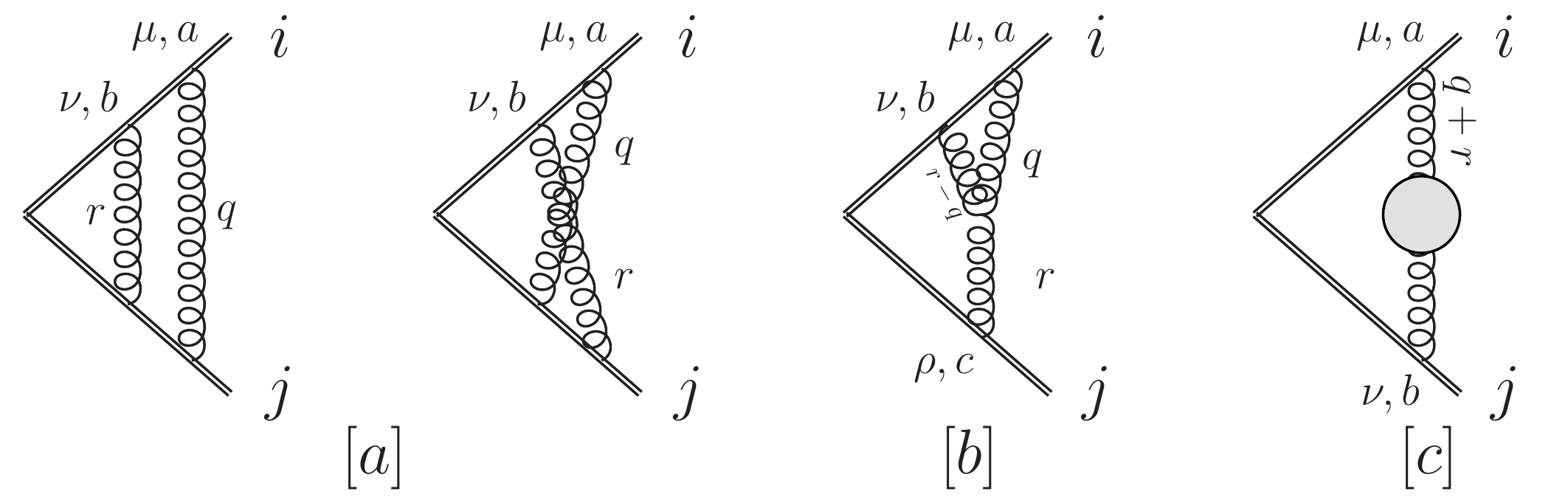}
\caption{Double-virtual diagrams involving two eikonal lines $i$ and $j$. We denote the contribution from the two Abelian diagrams as part $[a])$, the contribution from the diagram with a three-gluon vertex as part $[b]$ and the self-energy contribution as part $[c]$.\label{fig:VV2}}
\end{figure}
We first consider the diagrams involving only two eikonal lines $i$ and $j$ shown in Figure~\ref{fig:VV2}, where we distinguish between three parts $[a]$, $[b]$ and $[c]$. Similar to the real virtual part, we express the full contribution in the form 
\begin{equation}
 \bm{\mathcal{H}}_{m}^{(L+2)[\alpha]} = \sum_{(ij)} h_{ij}^{[\alpha]} + \text{h.c.}\,,
 \label{eq:VV2_h_def}
\end{equation}
and compute the $h_{ij}^{[\alpha]}$ with $\alpha=a,b,c$ for the configuration where the line $i$ is incoming and the line $j$ is outgoing such that no Glauber phases appear. For part $[a]$, we have 
\begin{align}
 h_{ij}^{[a]} ={}& -\frac{g_s^4}{2}\Dq{q}\Dq{r}\frac{n_{ij}^2}{[-n_i\cdot q+i0][-n_i\cdot(q+r)+i0][-n_j\cdot(q+r)+i0]}\nonumber\\
                 & \frac{1}{[q^2+i0][r^2+i0]}\left[\frac{\bm{T}_i^a\bm{T}_i^b\bm{T}_j^a\bm{T}_j^b\bm{\mathcal{H}}_m}{-n_j\cdot q+i0}
                   + \frac{\bm{T}_i^a\bm{T}_i^b\bm{T}_j^b\bm{T}_j^a\bm{\mathcal{H}}_m}{-n_j\cdot r+i0}\right] ,
\end{align}
where the momentum has been routed such that the only poles in the variables $q^0$ and $r^0$ are in the lower half of the complex plane. After closing the contour and picking up these poles we perform the energy integration with the UV cutoff $\theta(\Lambda-E_q-E_r)$ to isolate the IR divergences and obtain
\begin{align}
 h_{ij}^{[a]} ={}& \left(\frac{\alpha_s}{4\pi}\right)^2 \left(\frac{\mu}{2\Lambda}\right)^{4\ep}\Dqqq{q}\Dqqq{r} W_{ij}^q W_{ij}^r \Bigg\{ 
                   \bm{T}_i^a\bm{T}_i^b\{\bm{T}_j^a,\bm{T}_j^b\}\bm{\mathcal{H}}_m \left(\frac{1}{2\ep^2} - \frac{\pi^2}{3}\right) \nonumber\\
                 & - \frac{C_A}{2}\, \bm{T}_i^a\bm{T}_j^a\bm{\mathcal{H}}_m \Bigg[ \left(\frac{1}{2\ep^2} - \frac{\pi^2}{3}\right) + \frac{n_{iq}n_{jr}+n_{ir}n_{jq}}{n_{iq}n_{jr}-n_{ir}n_{jq}} 
                   \Bigg(\frac{\ln\frac{n_{iq}n_{jr}}{n_{ir}n_{jq}}}{\ep} \nonumber\\
                 &  + \ln\frac{n_{iq}n_{jr}}{n_{ir}n_{jq}}\ln\frac{n_{iq}n_{jq}}{n_{ir}n_{jr}}
                     + 4 \text{Li}_2\left(1-\frac{n_{ir}}{n_{iq}}\right) - 4 \text{Li}_2\left(1-\frac{n_{jr}}{n_{jq}}\right)\Bigg)\Bigg]\Bigg\}\,.
\end{align}

For part $[b]$ we consider the average of the diagram shown in Figure~\ref{fig:VV2} and its counterpart with legs $i$ and $j$ interchanged. After some manipulation and momentum shifts, we obtain 
\begin{align}
 h_{ij}^{[b]} ={}& \frac{g_s^4C_A}{4} \Dq{q}\Dq{r} \frac{n_{ij}\bm{T}_i^a\bm{T}_j^a\bm{\mathcal{H}}_m}{[q^2+i0][r^2+i0][(q-r)^2+i0]} \Bigg[\frac{4}{[-n_i\cdot r+i0][-n_j\cdot r+i0]}\nonumber\\
                 & - \frac{1}{[-n_i\cdot q+i0][-n_j\cdot r+i0]} - \frac{1}{[-n_i\cdot r+i0][-n_j\cdot q+i0]}\Bigg] ,
\end{align}
where it is now straightforward to evaluate the $q^0$ and $r^0$ integrals in the spirit of the Feynman tree theorem \cite{Feynman:1963ax,Catani:2008xa}. We use the identity 
\begin{equation}
 \frac{1}{q^2+i0} = \frac{1}{(q^0-i0)^2-E_q^2} - \frac{2\pi i \delta(q^0-E_q)}{2E_q}\,,
\end{equation}
after which we can proceed with the second term by simply closing the $r^0$ contour in the lower half of the complex plane which yields the sum of the residues of the two remaining gluon propagators. In the first term we shift $q\to q+r$ and then apply the same identity after which one can also close the $r^0$ contour to pick up the residue of the third gluon propagator. Overall, this is equivalent to the replacement
\begin{multline}\label{eq:replace}
 \frac{1}{[q^2+i0][r^2+i0][(q-r)^2+i0]}\\
  \to -4\pi^2\Bigg[\frac{\delta_+(q^2)\delta_+(r^2)}{[(q-r)^2+i0]} + 
                                            \frac{\delta_+(q^2)\delta_+((q-r)^2)}{[r^2+i0]} + \frac{\delta_+(r^2)\delta_+((q-r)^2)}{[q^2+i0]}\Bigg]\,,
\end{multline}
i.e. to taking the sum over all possible ways to cut two of the gluon lines. Evaluating the energy integration with the UV cutoff $\theta(\Lambda-E_q-E_r)$, one obtains 
\begin{align}
 h_{ij}^{[b]} ={}& \left(\frac{\alpha_s}{4\pi}\right)^2 \left(\frac{\mu}{2\Lambda}\right)^{4\ep} \frac{C_A}{2} \bm{T}_i^a\bm{T}_j^a\bm{\mathcal{H}}_m \Dqqq{q}\Dqqq{r} \nonumber\\
                &    \left(\frac{n_{ij}}{n_{iq}n_{jr}n_{qr}} + \frac{n_{ij}}{n_{ir}n_{jq}n_{qr}}\right) \frac{n_{iq}n_{jr}+n_{ir}n_{jq}}{n_{iq}n_{jr}-n_{ir}n_{jq}}  \, \Bigg[\frac{\ln\frac{n_{iq}n_{jr}}{n_{ir}n_{jq}}}{\ep} \nonumber\\
                 &  \quad\quad   + \ln\frac{n_{iq}n_{jr}}{n_{ir}n_{jq}}\ln\frac{n_{iq}n_{jq}}{n_{ir}n_{jr}}
                     + 4 \text{Li}_2\left(1-\frac{n_{ir}}{n_{iq}}\right) - 4 \text{Li}_2\left(1-\frac{n_{jr}}{n_{jq}}\right)\Bigg]\,.
\end{align}

The self-energy contribution denoted as part $[c]$ is of the form 
\begin{align}
 h_{ij}^{[c]} ={}& \frac{g_s^4}{2} \bm{T}_i^a\bm{T}_j^a\bm{\mathcal{H}}_m \Dq{q}\Dq{r} \frac{n_i^\mu n_j^\nu}{[-n_i\cdot (q+r)+i0][-n_j\cdot (q+r)+i0][(q+r)^2+i0]^2}\nonumber\\
                 &   \Bigg\{4n_FT_F \frac{g_{\mu\nu}q\cdot r - q_\mu r_\nu - q_\nu r_\mu}{[q^2+i0][r^2+i0]}             - n_ST_S \frac{(q-r)_\mu (q-r)_\nu}{[q^2+i0][r^2+i0]} 
                   + C_A \Bigg[ - \frac{(3-2c_R\ep)g_{\mu\nu}}{[r^2+i0]} \nonumber\\
                 & + \frac{(5q^2+8q\cdot r+5r^2)g_{\mu\nu} - 2(3-c_R\ep)(q_\mu r_\nu+q_\nu r_\mu) - 2(1+c_R\ep)(q_\mu q_\nu+r_\mu r_\nu)}{2[q^2+i0][r^2+i0]} \Bigg]\Bigg\}\,.
\end{align}
We now apply the UV cutoff $\theta(\Lambda-E_q-E_r)$ to isolate the IR divergence and then perform the $q^0$ and $r^0$ integrations similar to part $[b]$. We note that the contributions from the double pole in the squared propagator $1/[(q+r)^2+i0]^2$ involve the self energy and its derivative at vanishing external momenta which are scaleless. Thus, this step effectively reduces to the replacement $1/([q^2+i0][r^2+i0])\to-4\pi^2\delta_+(q^2)\delta_+(r^2)$. Performing the energy integrations, we find 
\begin{align}
 h_{ij}^{[c]} ={}& \left(\frac{\alpha_s}{4\pi}\right)^2 \left(\frac{\mu}{2\Lambda}\right)^{4\ep} \bm{T}_i^a\bm{T}_j^a\bm{\mathcal{H}}_m \Dqqq{q}\Dqqq{r} \Bigg\{
                    \frac{n_ST_S}{n_{qr^2}}\left(\frac{1}{2\ep^2}-\frac{\pi^2}{3}\right) \nonumber\\
                 & + \frac{n_ST_S\left(K_{ij;qr}^S - \frac{8}{n_{qr}^2} + \tilde{K}_{ij;qr}^{S,(\ep)}\ep\right) - 2n_FT_F\left(K_{ij;qr}^F - \frac{8}{n_{qr}^2} + \tilde{K}_{ij;qr}^{F,(\ep)}\ep\right)}{-4\ep} \nonumber\\
                 & + C_A \Bigg[\frac{1}{n_{qr}^2}\left(\frac{1}{2\ep^2}+\frac{c_R}{2\ep}-\frac{\pi^2}{3}\right)
                   + \frac{2n_{ij}n_{qr} - (1-c_R \ep)(n_{iq} n_{jr}+n_{ir} n_{jq})}{(n_{iq} n_{jr} - n_{ir} n_{jq}) n_{qr}^2} \nonumber\\
                 & \times \left[\frac{\ln\frac{n_{iq}n_{jr}}{n_{ir}n_{jq}}}{\ep} 
                    + \ln\frac{n_{iq}n_{jr}}{n_{ir}n_{jq}}\ln\frac{n_{iq}n_{jq}}{n_{ir}n_{jr}}
                     + 4 \text{Li}_2\left(1-\frac{n_{ir}}{n_{iq}}\right) - 4 \text{Li}_2\left(1-\frac{n_{jr}}{n_{jq}}\right)\right]\Bigg]\Bigg\}\,,
\end{align}
where at order $\ep^0$ we again only require the terms that are singular in the configuration where $q$ and $r$ become collinear. These terms 
\begin{align}
 \tilde{K}_{ij;qr}^{S,(\ep)} ={}& - \frac{32}{n_{qr}^2} 
                                  - \frac{32}{9n_{qr}^2}\left(\frac{n_{ir}}{n_{iq}}-\frac{n_{jr}}{n_{jq}}\right)^2 
                                  - \frac{8}{3n_{qr}^2}\left(1-\frac{n_{ir}}{n_{iq}}\right)\left(1-\frac{n_{jr}}{n_{jq}}\right) + \dots\,,\\
 \tilde{K}_{ij;qr}^{F,(\ep)} ={}& \tilde{K}_{ij;qr}^{S,(\ep)}
                                  + \frac{16 W_{ij}^q}{n_{qr}} + \dots\,,
\end{align}
differ from their counterparts in the double-real part by terms that only depend on at most one of the two eikonal lines $i$ and $j$. We now renormalize the sum of the parts $[a]$, $[b]$ and $[c]$ with the two-particle part of the counterterm
\begin{equation}
  \left(\frac{\alpha_s}{4\pi}\right)^2\left[   \frac{-1}{8\ep^2}\bm{\mathcal{H}}_m (\bm{V}_m+2\beta_0) \bm{V}_m - \frac{1}{2\ep}\bm{\mathcal{H}}_{m}^{\text{ren},(1)} (\bm{V}_{m}+2\beta_0) 
      + \frac{1}{4\ep}\bm{\mathcal{H}}_m\bm{v}_m \right]\,,
\end{equation}
and after setting $\mu=2\Lambda$, we obtain 
\begin{align}
 h_{ij}^{[a+b+c+\text{ren}]}={}& \left(\frac{\alpha_s}{4\pi}\right)^2 \bm{T}_i^a\bm{T}_j^a\bm{\mathcal{H}}_m \Dqqq{q}\Bigg[   \frac{(C_A-2n_FT_F+n_ST_S)}{3\ep} \left( \frac{1}{n_{iq}} + \frac{1}{n_{jq}}-1\right) \nonumber\\
                    & \hspace{-1cm}- \frac{\beta_0}{4\ep^2}\,W_{ij}^q - \frac{W_{ij}^q}{\ep}\left(\frac{67C_A}{9}+\frac{C_Ac_R}{6} - \frac{26n_FT_F}{9} - \frac{10 n_ST_S}{9}\right) \nonumber\\
                    & \hspace{-1cm} + \Dqqq{r} \frac{C_A \left(K_{ij;qr}^A + \frac{8-2c_R}{n_{qr}^2}\right)   - 2n_FT_F \left(K_{ij;qr}^F - \frac{8}{n_{qr}^2}\right)
                        + n_ST_S \left(K_{ij;qr}^S - \frac{8}{n_{qr}^2}\right)}{-4\ep}\Bigg]\,,
\end{align}
where the terms in the first line vanish due to color conservation, because they at most depend on a single direction $i$ or $j$. 

\begin{figure}
\centering
\includegraphics[width=0.8\textwidth]{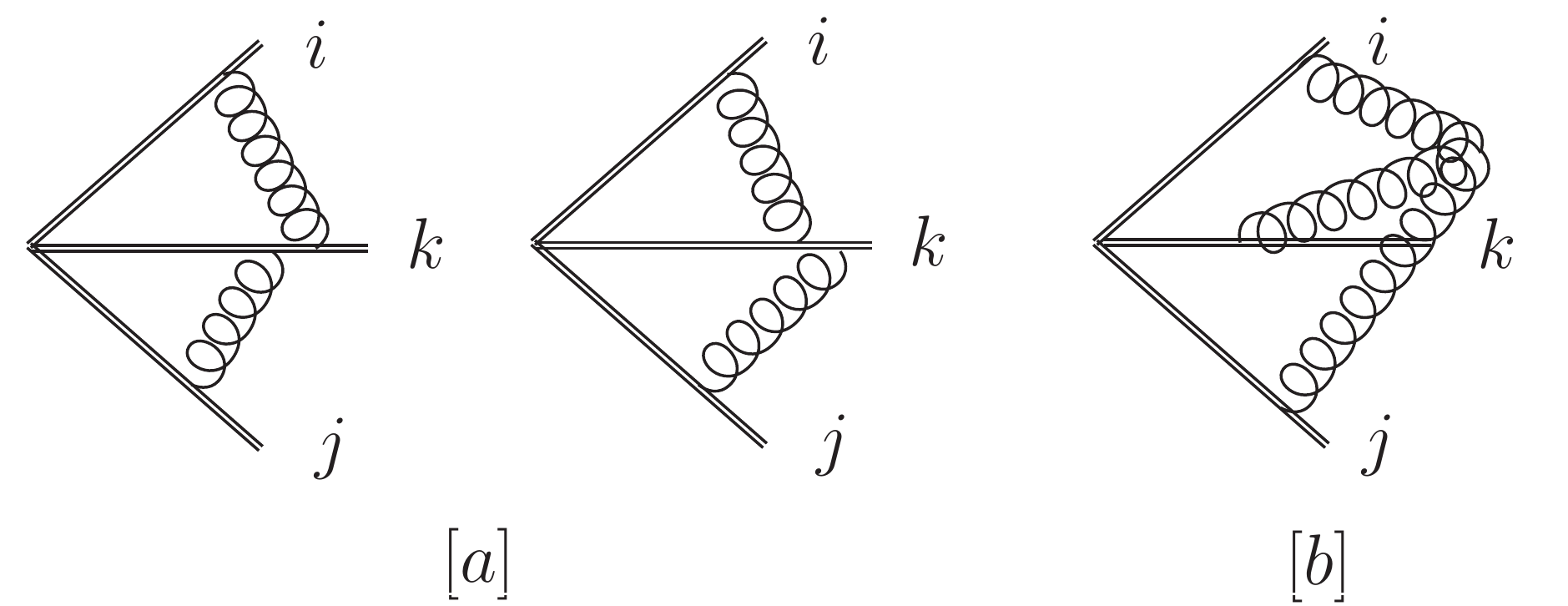}
\caption{Double-virtual diagrams involving three eikonal lines $i$ and $j$.\label{fig:VV3}}
\end{figure}

For the three-particle diagrams part $[a]$, we have
\begin{align}\label{eq:theeVVa}
 h_{ijk}^{[a]} ={}& -\frac{g_s^4}{2}\Dq{q}\Dq{r}\frac{n_{ik} n_{jk}}{[-n_i\cdot r+i0][n_k\cdot(q+r)+i0][-n_j\cdot q+i0]}\nonumber\\
                 & \frac{1}{[q^2+i0][r^2+i0]}\left[\frac{\bm{T}_i^a\bm{T}_j^b\bm{T}_k^b\bm{T}_k^a\bm{\mathcal{H}}_m}{n_k\cdot q+i0}
                   + \frac{\bm{T}_i^a\bm{T}_j^b\bm{T}_k^a\bm{T}_k^b\bm{\mathcal{H}}_m}{n_k\cdot r+i0}\right] .
\end{align}

According to \eqref{eq:twoLoopGlauber} the imaginary part in the double virtual contribution arises from two-particle diagrams and there is no Glauber phase in the three-particle contributions. We can thus ignore Glauber poles and replace the gluon propagators in \eqref{eq:theeVVa} using the Cutkosky rule
\begin{equation}
\frac{1}{[q^2+i0][r^2+i0]}\to -4\pi^2 \delta_+(q^2)\delta_+(r^2) \,
\end{equation}
 to pick up the residues. We observe that there are no singularities when $q$ is collinear to $r$ and will therefore not include the terms at the order $\ep^0$. Evaluating the energy integral with the cutoff $\theta(\Lambda-E_q-E_r )$, we obtain
\begin{align}\label{eq:threeparticlea}
 h_{ijk}^{[a]} = {}& \left(\frac{\alpha_s}{4\pi}\right)^2 \left(\frac{\mu}{2\Lambda}\right)^{4\ep}\Dqqq{q}\Dqqq{r} W_{jk}^q W_{ik}^r \Bigg\{ 
                   \bm{T}_i^a\bm{T}_j^b\{\bm{T}_k^a,\bm{T}_k^b\}\bm{\mathcal{H}}_m \frac{1}{\ep^2}  \nonumber\\
                   &+i f^{abc}\bm{T}_i^a\bm{T}_j^b\bm{T}_k^c\bm{\mathcal{H}}_m \frac{2}{\ep} \ln \frac{n_{kr}}{n_{kq}} \Bigg\} \, .
 \end{align}

For part $[b]$, the integrals are written as
\begin{align}
 h_{ijk}^{[b]} ={}& \frac{g_s^4}{6}\Dq{q}\Dq{r}i f^{abc}\bm{T}_i^a\bm{T}_j^b\bm{T}_k^c\bm{\mathcal{H}}_m
 \nonumber\\
 		    & \frac{n_{ij} \left(n_k \cdot r-n_k \cdot q\right)+n_{ik} \left(-n_j \cdot q-2 n_j \cdot r\right)+n_{jk} \left(2 n_i\cdot q+n_i \cdot r\right)}{[-n_i\cdot r+i0][n_k\cdot(q+r)+i0][-n_j\cdot q+i0][q^2+i0][r^2+i0] [(q+r)^2+i0]} \, .
\end{align}
With the same argument, we replace the gluon propagators by
\begin{multline}
 \frac{1}{[q^2+i0][r^2+i0][(q+r)^2+i0]}\\
  \to -4\pi^2\Bigg[\frac{\delta_+(q^2)\delta_+(r^2)}{[(q+r)^2+i0]} + 
                                            \frac{\delta_+(q^2)\delta_+((q+r)^2)}{[r^2+i0]} + \frac{\delta_+(r^2)\delta_+((q+r)^2)}{[q^2+i0]}\Bigg]\,.
\end{multline}
The integrals contain $q\parallel r$ collinear singularities at order $\ep^0$ but the collinearly divergent terms only depend on two legs and vanish due to color conservation. Evaluating the integral with the cutoff $\theta(\Lambda-E_q-E_r )$, we have
\begin{align}\label{eq:threeparticleb}
h_{ijk}^{[b]}=&\left(\frac{\alpha_s}{4\pi}\right)^2 \left(\frac{\mu}{2\Lambda}\right)^{4\ep}\Dqqq{q}\Dqqq{r}i f^{abc}\bm{T}_i^a\bm{T}_j^b\bm{T}_k^c\bm{\mathcal{H}}_m \Bigg\{ \nonumber \\
		&\frac{1}{2 \ep^2} \left( 2 W^r_{i k} W^q_{j r}-2 W^q_{j k} W^r_{i q}+W^q_{i k} W^r_{i q}- W^q_{j k}W^r_{k q} \right) +  \nonumber \\
		& \frac{1}{\ep} \left( 2 W^r_{i k} W^q_{j r}+2 W^r_{i q} W^q_{j k}-2 W^q_{i j} W^r_{i q} -W^q_{i k} W^r_{i q}-W^q_{j k} W^r_{k q} \right) \ln \frac{n_{kr}}{n_{kq}} \Bigg \} \, .
\end{align}
 The first two $1/\ep^2$ terms of this result have kinematical functions which are symmetric under $i \leftrightarrow k$ and $j \leftrightarrow k$, respectively, but multiply anti-symmetric color structures and therefore vanish when summed over. The last two terms at both $1/\ep^2$ and $1/\ep$ only depend on two legs and vanish by color conservation.  The $1/\ep^2$ terms in \eqref{eq:threeparticleb} therefore all vanish and the counterterm $\frac{-1}{8\ep^2}  \bm{V}_m  \bm{V}_m \bm{\mathcal{H}}_m$ removes the abelian divergence in \eqref{eq:threeparticlea}. Combining the remaining non-abelian term with the one in \eqref{eq:threeparticleb}, we find
 \begin{align}
\left. h_{ijk}^{[a+b]} \right |_\text{non-abelian} =-&\left(\frac{\alpha_s}{4\pi}\right)^2 i f^{abc}\bm{T}_i^a\bm{T}_j^b\bm{T}_k^c\bm{\mathcal{H}}_m\Dqqq{q}\Dqqq{r}   \frac{1}{4\ep} K_{ijk,qr} \, .
\end{align}
Adding up the results, we get the double virtual anomalous dimension
\begin{align}
 \bm{v}^\prime_m ={}& \sum_{(i j k)} if^{abc}\left(\bm{T}_{i,L}^a \bm{T}_{j,L}^b \bm{T}_{k,L}^c - \bm{T}_{i,R}^a \bm{T}_{j,R}^b \bm{T}_{k,R}^c\right) 
                   \Dqqq{q}\Dqqq{r} K_{ijk;qr}  \nonumber\\
                 & + \sum_{(i j)}\left(\bm{T}_{i,L}^a \bm{T}_{j,L}^a+\bm{T}_{i,R}^a \bm{T}_{j,R}^a\right) \Bigg\{ \Dqqq{q}\Dqqq{r} K_{ij;qr}   \nonumber \\
                 & +\left( - \frac{2\beta_0}{\ep} + \Gamma_\text{coll}+ \frac{8 \pi^2 C_A}{3} \right) \Dqqq{q} W_{ij}^q \Bigg\} \nonumber\\
                 & - \sum_{(i j)}\left(\bm{T}_{i,L}^a \bm{T}_{j,L}^a-\bm{T}_{i,R}^a \bm{T}_{j,R}^a\right)\gamma_1^\text{cusp}\frac{i\pi \Pi_{ij}}{2} \,.
\label{eq:vmp}
\end{align}
The prime in $\bm{v}^\prime_m$ indicates that this is the result obtained with the energy cutoff $\theta(\Lambda-E_q-E_r )$, see the remarks after \eqref{eq:rmp}. The explicit $\pi^2$ term in the anomalous dimension \eqref{eq:vmp} differs from the one in \eqref{eq:vm0} by a factor 2 and cancels against the implicit one in $ \Gamma_\text{coll}$.

This concludes the diagrammatic extraction of the anomalous dimension. In Section  \ref{sec:Gamma2_result}, we have taken the diagrammatic results for $\bm{d}_m$, $\bm{r}_m$ and $\bm{v}_m$ derived here and rearranged their collinear singularities in such a way that the anomalous dimensions become suitable for numerical evaluation in $d=4$ and implementation in a parton shower.

\section{Summary and discussion}\label{sec:summary}

In the following we present the final result for the anomalous dimension, compare to the literature and then discuss some of the subtleties we encountered in our computation.

\subsection{Result for the anomalous dimension}

The factorization formula for non-global observables splits the cross section into hard functions $\bm{\mathcal{H}}_m$ and soft functions $\bm{\mathcal{S}}_m$. The hard functions $\bm{\mathcal{H}}_m$ consist of squared amplitudes with $m$ hard partons along fixed directions, while the soft functions $\bm{\mathcal{S}}_m$ describe the soft emissions from the hard partons and are given by matrix elements of Wilson lines along the directions of hard partons. The anomalous dimension of $\bm{\mathcal{H}}_m$ is related to soft singularities of scattering amplitudes with $m$ hard partons. At one-loop level the anomalous dimension matrix has two types of entries: $\bm{R}_m$ absorbs soft singularities in real emissions, while $\bm{V}_m$ contains the ones in the loop diagrams. Explicitly, the one-loop anomalous dimension is given by
\begin{align}\label{eq:GammaOneFinal}
 \bm{R}_m  =& -4\,\sum_{(ij)}\,\bm{T}_{i,L}^a\bm{T}_{j,R}^{\tilde{a}}  \,W_{ij}^{q}\,  \theta_{\rm in}(n_q)\, ,\nonumber \\
 \bm{V}_m =&  2\,\sum_{(ij)}\,(\bm{T}_{i,L}\cdot  \bm{T}_{j,L}+\bm{T}_{i,R}\cdot  \bm{T}_{j,R})  \DQQQ{q}\, W_{ij}^q\,  \nonumber \\
 & - i\pi \sum_{(ij)} \frac{1}{2}\left[\bm{T}_{i,L}\cdot \bm{T}_{j,L} - \bm{T}_{i,R}\cdot \bm{T}_{j,R}\right] \, \Pi_{ij}\, \gamma_0^{\rm cusp}\, ,
\end{align}
where the soft dipole is $W_{ij}^q = n_{ij}/(n_{iq} n_{qj})$. The color generators $\bm{T}_{i,L}$ and $\bm{T}_{i,R}$ act on color of the $i$-th parton in the hard amplitude and its complex conjugate, respectively. The entry $\bm{R}_m$ maps $m$-parton hard functions onto $(m+1)$-parton hard functions and the constraint $\theta_{\rm in}(n_q)$ ensures that the additional hard parton along the direction $n_q$ is inside the jet region. The  additional gluon induced by  $\bm{R}_m$  has color index $a$ in the amplitude and index $\tilde{a}$ in the conjugate amplitude. The virtual piece $\bm{V}_m$ keeps the number of partons unchanged. The individual entries $ \bm{R}_m$ and $\bm{V}_m$ contain collinear divergences when the additional parton $q$ becomes collinear to partons $i$ or $j$. As discussed in detail in Section \ref{sec:collSing}, the collinear singularities associated with final-state partons cancel when the anomalous dimension is applied to the soft functions. The quantity $\Pi_{ij} = 1$ if $i$ and $j$ are both incoming or outgoing and $\Pi_{ij} = 0$ otherwise. For $e^+e^-$ collisions we have only outgoing QCD partons so that $\Pi_{ij} = 1$ and the Glauber-phase terms in $\bm{V}_m$ proportional to $\gamma_0^{\rm cusp}=4$ cancel by color conservation. For hadronic collisions this is not the case, and the Glauber phases  spoil the cancellation of the soft-collinear singularities associated with the initial state. To analyze hadronic collisions, the collinear singularities in \eqref{eq:GammaOneFinal} must be made manifest, see \cite{Becher:2021zkk}. Their presence leads to double logarithms known as super-leading logarithms \cite{Forshaw:2006fk,Forshaw:2008cq}. A form of \eqref{eq:GammaOneFinal} suitable for hadronic collisions was given in \cite{Becher:2021zkk} and used to resum these super-leading logs to all orders. In our paper, we restrict ourselves to $e^+e^-$ collisions.

The two-loop anomalous dimension matrix has three different types of entries. The elements $\bm{d}_m$ describe singularities arising in unordered double emissions from the original hard partons, and it maps from the space of $m$ hard partons to the one of $m+2$ hard partons. The entries $\bm{r}_m$ are related to soft singularities in real-virtual corrections and map $m$ parton hard functions into $(m+1)$-parton hard functions. The third and final entries $\bm{v}_m$ relate to the soft singularities in two-loop virtual corrections and leave the number of hard partons unchanged. Our final result for the three entries reads
\begin{align}\label{eq:GammaFinald}
 \bm{d}_m ={}& \sum_{(i j)}\sum_k i f^{abc}\left(  \bm{T}_{i,L}^a \bm{T}_{j,L}^b \bm{T}_{k,R}^c - \bm{T}_{i,R}^a\bm{T}_{j,R}^b \bm{T}_{k,L}^c\right) K_{ijk;qr} \,\theta_\text{in}(n_q)\theta_\text{in}(n_r) \nonumber\\
                 & - 2 \sum_{(ij)}\bm{T}_{i,L}^c \bm{T}_{j,R}^c K_{ij;qr} \theta_\text{in}(n_q) \theta_\text{in}(n_r) \,,
 \\ \label{eq:GammaFinalr}
  \bm{r}_m ={}& -2 \sum_i\sum_{(jk)}if^{abc}(\bm{T}_{i,L}^a\bm{T}_{j,R}^b\bm{T}_{k,R}^c - \bm{T}_{i,R}^a\bm{T}_{j,L}^b\bm{T}_{k,L}^c) \DQQQ{r} K_{ijk;qr} \theta_\text{in}(n_q) \nonumber\\
  & - \sum_{(ij)}\bm{T}_{i,L}^a\bm{T}_{j,R}^a  \left\{ W_{ij}^q \left[4 \beta_0 \ln(2W_{ij}^q) + \gamma_1^{\rm cusp} \right] - 2  \DQQQ{r} K_{ij;qr}  \right\} \theta_\text{in}(n_q) \nonumber\\               
                 & + 8i\pi \sum_i\sum_{(j k)} if^{abc}\left( \bm{T}_{i,L}^a \bm{T}_{j,R}^b \bm{T}_{k,R}^c + \bm{T}_{i,R}^a \bm{T}_{j,L}^b \bm{T}_{k,L}^c  \right)  W_{ij}^q\ln W_{jk}^q \,\theta_\text{in}(n_q)   \,, \\
                 \label{eq:GammaFinalv}
                   \bm{v}_m ={}& \sum_{(i j k)} if^{abc}\left(\bm{T}_{i,L}^a \bm{T}_{j,L}^b \bm{T}_{k,L}^c - \bm{T}_{i,R}^a \bm{T}_{j,R}^b \bm{T}_{k,R}^c\right) 
                   \DQQQ{q}\DQQQ{r} K_{ijk;qr}  \nonumber\\
       & + \sum_{(i j)}\frac{1}{2}\left(\bm{T}_{i,L}^a \bm{T}_{j,L}^a+\bm{T}_{i,R}^a \bm{T}_{j,R}^a\right)
                   \DQQQ{q} W_{ij}^q  \left[4\beta_0 \ln(2W_{ij}^q)  + \gamma_1^{\rm cusp} \right]   \nonumber\\
           & - i\pi \sum_{(i j)}\frac{1}{2}\left(\bm{T}_{i,L}^a \bm{T}_{j,L}^a- \bm{T}_{i,R}^a \bm{T}_{j,R}^a\right)\,\Pi_{ij} \,\gamma_1^\text{cusp} \,.
\end{align}
This expression involves the functions $K_{ij;qr}$ and $K_{ijk;qr}$ introduced by Caron-Huot \cite{Caron-Huot:2015bja}, which are given in \eqref{eq:Kijqr} and \eqref{eq:Kijkqr}. These angular functions depend on the light-cone directions $n_q$ and $n_r$ associated with emissions or loop-momenta, and the directions $n_i$, $n_j$ and $n_k$ of the hard partons. As they stand, the quantities $\bm{d}_m$, $\bm{r}_m$ and $\bm{v}_m$ contain collinear singularities when the emitted partons are aligned with the hard partons  along directions $n_i$ and $n_j$. In addition, the two-loop hard functions also contain singularities when the emitted partons along $n_q$ and $n_r$ become collinear. In our result for $\bm{\Gamma}^{(2)}$ we have rearranged the collinear terms so that the cancellation of collinear singularities in the angular integrals is manifest. To achieve this, we have shifted terms in $\bm{d}_m$ describing collinear configurations to $\bm{r}_m$.

As stressed earlier, we should distinguish the color indices of the emitted partons in $\bm{d}_m$ and $\bm{r}_m$ in the amplitude from the ones in the conjugate amplitude. To keep the notation compact, we write the indices in contracted form in \eqref{eq:GammaFinald} and \eqref{eq:GammaFinalr}, but it is easy to reconstruct the result with open color indices. For the three-parton correlations in $\bm{d}_m$ we should replace
\begin{equation}
i f^{abc}\left(  \bm{T}_{i,L}^a \bm{T}_{j,L}^b \bm{T}_{k,R}^c - \bm{T}_{i,R}^a\bm{T}_{j,R}^b \bm{T}_{k,L}^c\right) \to \bm{T}_{i,L}^a \bm{T}_{j,L}^b\, i f^{\tilde{a}\tilde{b}c} \bm{T}_{k,R}^c -  \bm{T}_{i,R}^{\tilde{a}}\bm{T}_{j,R}^{\tilde{b}} \, i f^{abc} \bm{T}_{k,L}^c
\end{equation}
to properly indicate that two new gluons are produced with color indices $a$, $b$ in the amplitude and $\tilde{a}$, $\tilde{b}$ in the conjugate amplitude. Similarly in $\bm{r}_m$, we should replace
\begin{equation}
if^{abc}(\bm{T}_{i,L}^a\bm{T}_{j,R}^b\bm{T}_{k,R}^c - \bm{T}_{i,R}^a\bm{T}_{j,L}^b\bm{T}_{k,L}^c)  \to \bm{T}_{i,L}^a\, if^{\tilde{a}bc}\bm{T}_{j,R}^b\bm{T}_{k,R}^{c} - 
  \bm{T}_{i,R}^{\tilde{a}}  if^{abc}\bm{T}_{j,L}^b\bm{T}_{k,L}^c
\end{equation}
to indicate that one extra gluon is emitted, with color index $a$ in the amplitude and $\tilde{a}$ in the conjugate amplitude. The only piece where the restoration of color indices is more involved is the two-parton correlations proportional to $K_{ij;qr}$ in $\bm{d}_m$. The color index in $\bm{T}_{i,L}^c \bm{T}_{j,R}^c$ refers to the parent parton which branches into two quarks, scalars or gluons. To, for example, restore the color indices of the emitted quark-anti-quark pair, one should replace
\begin{equation}
\bm{T}_{i,L}^c \bm{T}_{j,R}^c  C_F n_F T_F \,K^F_{ij;qr} \to \bm{T}_{i,L}^c \bm{T}_{j,R}^d \, (t^c)_{\alpha\beta}\,  (t^d)_{\tilde{\beta}\tilde{\alpha}} \,n_F\, K^F_{ij;qr}\,.
\end{equation}

The results \eqref{eq:GammaFinald}, \eqref{eq:GammaFinalr} and \eqref{eq:GammaFinalv} correspond to a renormalization scheme in which all angular integrals are kept $d$-dimensional. Changing to conventional dimensional regularization and using standard minimal subtraction ($\overline{\text{MS}}$) induces additional contributions related to $\ep$ terms in the angular integrals of the (iterated) one-loop terms. The full anomalous dimension in the $\overline{\text{MS}}$ scheme, integrated over directions is given by
\begin{align}\label{eq:GammaBarFinal}
 \bar{\bm{\Gamma}}^{(2)} \otimes_2 \bm{1} = \bm{\Gamma}^{(2)} \otimes_2 \bm{1} -2 \beta_0\, \bm{\Gamma}^{(1)}\otimes_\ep \bm{1} - \left( \bm{\Gamma}^{(1)} \otimes_2 \bm{\Gamma}^{(1)}\otimes_\ep \bm{1} - \bm{\Gamma}^{(1)}\otimes_\ep \bm{\Gamma}^{(1)} \otimes_2 \bm{1}\right)   ,
 \end{align}
 where $\bm{\Gamma}^{(2)}$ contains the elements $\bm{d}_m$, $\bm{r}_m$ and $\bm{v}_m$ given above. The explicit form of \eqref{eq:GammaBarFinal} for the two-jet case was presented in Appendix \ref{app:GaGa}. 
 
 Let us compare our results to those of Caron-Huot \cite{Caron-Huot:2015bja}, who tracks real-emissions with a color density matrix $\bm{U}$.  In his results for the anomalous dimension, terms with two such matrices correspond to our $\bm{d}_m$, terms with a single one to $\bm{r}_m$. He distinguishes color matrices acting on the left and right, but he defines this relative to the color density matrix. His left color matrices therefore act to the right of the amplitudes and to compare to his results we should thus exchange $\bm{T}_{i,L} \leftrightarrow  \bm{T}_{i,R}$ in our results, which changes the sign of the three-particle terms. In addition, there is a relative minus sign between our  definitions of the anomalous dimensions. Accounting for these conventions, our results \eqref{eq:GammaFinald}, \eqref{eq:GammaFinalr} and \eqref{eq:GammaFinalv} agree with (3.21) - (3.23) in his paper \cite{Caron-Huot:2015bja} when using the value of $\gamma_{\rm cusp}$ in CDR. The Glauber terms in $\bm{v}_m$ are not shown in his result but cancel by color conservation in $e^+e^-$. Note however, that the additional terms in \eqref{eq:GammaBarFinal} are not present in his result. What Caron-Huot denotes as his $\overline{\text{MS}}$ result therefore corresponds to a non-minimal scheme where full $d$-dimensional angular integrals are subtracted, see the detailed discussion in Section \ref{sec:MSbar}. 
 
This year a new formalism for the resummation of subleading non-global logarithms was presented in \cite{Banfi:2021owj,Banfi:2021xzn}. It extends the BMS equation \cite{Banfi:2002hw} to subleading logarithm and is valid in the large $N_c$ limit. As we do, the authors verified their result against the two-loop results for the dijet cross section \cite{Becher:2016mmh} so that there is agreement at this level. It will be interesting to compare the resummation in more detail, for which they presented numerical results very recently \cite{Banfi:2021xzn}. Other recent works considering the all-order structure of non-global observables include \cite{Larkoski:2015zka,Larkoski:2016zzc,Neill:2018mmj} but these papers do not claim higher-logarithmic accuracy. The paper \cite{Platzer:2020lbr} has analyzed the color structures in the anomalous dimension in the color-flow basis, which is suitable to analyze them in an expansion around the large-$N_c$ limit \cite{Platzer:2013fha}. In the strict large-$N_c$ limit, the squared amplitudes can be described by dipoles and the anomalous dimension only acts on the dipoles. Genuine three-parton correlations are then color suppressed and the anomalous dimension can be written in terms of two-parton contributions.

\subsection{Subtleties in the computation}

It is gratifying that our explicit computation of the anomalous dimension matches the indirect determination of \cite{Caron-Huot:2015bja}, but there are several aspects of our computation which make it quite delicate and susceptible to mistakes. Fortunately, our comparison against the two-loop results for the dijet cross section  \cite{Becher:2016mmh} provides for a strong check on our final result, but we nevertheless would like to review some of the issues that complicate the extraction of the anomalous dimension. In the calculation,  multiple issues can destructively interfere, but let's discuss them in turn:

\paragraph{Collinear singularities.}  The individual elements of the anomalous dimension contain two types of collinear singularities. First of all, there are singularities when the emissions become collinear to the hard partons. These cancel within the hard anomalous dimension after applying it to the soft functions since the soft functions are regular in the collinear limits: The higher-multiplicity soft functions describing the process after an emission reduce to the lower multiplicity soft functions in collinear limits. Rather than explicitly subtracting these types of collinear singularities in dimensional regularization, we rearranged and rewrote the matrix elements $\bm{d}_m$, $\bm{r}_m$ and $\bm{v}_m$ in such a way that the singularities and their cancellation becomes manifest in angular integrals. In this form the anomalous dimension is suitable for a parton shower implementation, where the collinear singularities are typically regularized with an intermediate angular cutoff. 

A second type of collinear singularity arises when two soft emissions become collinear. This type of collinear singularity is present in the soft functions and its cancellation involves both the soft and hard functions. Since it does not cancel among the hard functions themselves, it must be subtracted and is part of the hard anomalous dimension. These singularities can be extracted by considering soft limits of hard functions, but the collinear divergence in the angular integrals can multiply a finite contribution in an energy integral. It is therefore generally not sufficient to keep only the divergent part of energy integrals when determining the anomalous dimension. In our derivation we therefore also included the terms $K_{ij;qr}^{(\ep)}$, which are $\ep$-suppressed compared to $K_{ij;qr}$ but contain collinear singularities which contribute to the anomalous dimension. The collinear terms in the double emission $\bm{d}_m$ can be absorbed into $\bm{r}_m$ to cancel real-virtual collinear contributions. The treatment of the collinear singularities is different in \cite{Caron-Huot:2015bja}. Rather than computing the collinear terms, Caron-Huot first extracts the anomalous dimension ignoring the collinear singularities in $K_{ij;qr}^{(\ep)}$, but then performs a collinear subtraction in the form of a scheme change to a renormalization scheme where they are absent. While the end result is the same, we believe that the explicit computations in our paper clarify the extraction and subtraction of these contributions.

\paragraph{Energy cutoff.} We isolate the infrared singularities in the hard functions by putting a UV cutoff on the associated energy integrals. If there is a single divergence, this procedure is unambiguous, but in cases with double divergences, the result can depend on the form of the cutoff and it is then unclear how it translates to the anomalous dimension in dimensional regularization. To isolate the two-loop divergence, we can subtract the strongly ordered part of the soft limit so that we are left with a single divergence. However, due to collinear divergences, higher-order terms in $\ep$ in the strongly ordered limit generate $C_A \pi^2$ terms contributing to the cusp anomalous dimension, which depend on the form of the energy cutoff and cannot be unambiguously extracted.
Let us stress that this ambiguity is a limitation of our cutoff-based method to compute the anomalous dimension, not an ambiguity in $\bm{\Gamma}^{(2)}$ or the resummation formalism. We have determined the coefficient of these terms by comparing to the two-loop results for the dijet cross section \cite{Becher:2016mmh}, which were computed in standard dimensional regularization without a cutoff, but a full computation of the two-loop hard functions for an arbitrary number of external legs is not currently feasible. 

We believe that the ambiguity in the $C_A \pi^2$ terms is also inherent in the calculation of \cite{Caron-Huot:2015bja}. The form of the collinear subtractions performed in \cite{Caron-Huot:2015bja} is not unique and different forms would lead to different results. More specifically, with the form of the angular function $f$ that is chosen in Appendix A of \cite{Caron-Huot:2015bja}, the different $C_A \pi^2$ terms in the anomalous dimension nicely combine into the standard cusp anomalous dimension, but with the form of $f$ present in the splitting functions this would not be the case. The finite difference between the two choices can be interpreted as a difference in the renormalization scheme for the hard functions. In our paper we want to extract the anomalous dimension in the $\overline{\text{MS}}$ scheme, so the scheme is fixed.

\paragraph{Residues of light-cone propagators.} Since we want to write the anomalous dimension in terms of angular integrals, we use the residue theorem to carry out the energy integrals and then extract the associated IR divergences. In our analysis, we have collected the residues by hand, but there are ways to algorithmically perform this step using the Feynman tree theorem, see \cite{Feynman:1963ax,Catani:2008xa}. In the soft limit, we encounter Eikonal propagators associated with the energetic hard partons. Of course, in general one then also picks up residues of these linear light-cone propagators and the algorithm of \cite{Feynman:1963ax,Catani:2008xa} can be generalized to this case \cite{Platzer:2020lbr}. However, in dimensional regularization, the results of the energy integrations are angular distributions and we observe that these angular distributions are difficult to interpret and expand in $\ep$ when we pick up poles of light-cone propagators. More specifically, the angular distributions, in particular those associated with imaginary parts, develop unphysical singularities if they are naively expanded in $\ep$. To avoid dealing with these distributions, we pick up the residues in space-like kinematics for which we can pick contours which avoid the light-cone poles. We then reconstruct the imaginary parts from the inclusive result for the loop diagrams.

We hope that future work will come up with a more algorithmic and less tedious way to obtain the anomalous dimension. We also note that our explicit check using dijet cross section \cite{Becher:2016mmh} does not check the imaginary part of the anomalous dimension since it is not contributing at this order.

\section{Conclusion and outlook}\label{sec:conclusion}

In our paper, we have computed the two-loop anomalous dimension governing the re\-norm\-alization-group evolution from the hard scale to the soft scale in exclusive jet processes and other non-global observables. This anomalous dimension is the final ingredient needed for the resummation of subleading non-global logarithms in $e^+e^-$ cross sections. Our direct computation confirms the result of Caron-Huot \cite{Caron-Huot:2015bja}, but also yields an additional term that is needed if the soft functions are renormalized in the standard $\overline{\text{MS}}$ scheme. We have checked our  result for the anomalous dimension by using it to reconstruct the known two-loop results for the dijet cross section \cite{Becher:2016mmh}.

To perform the resummation, the two-loop anomalous dimension should be implemented into a parton shower framework and combined with the one-loop corrections of the hard and soft functions. The latter two corrections were implemented and computed in \cite{Balsiger:2019tne} so that the implementation of the two-loop anomalous dimension is the only missing piece to achieve next-to-leading logarithmic accuracy in our framework. Both the real and virtual parts of our result for the anomalous dimension are given in terms of angular integrals, which makes them suitable for this task. Our result is valid for finite $N_c$ and there are currently several groups developing methods to go beyond the large-$N_c$ limit in parton shower simulations, see e.g.\ \cite{Nagy:2019pjp,Hoche:2020pxj,Hamilton:2020rcu,DeAngelis:2020rvq}. The paper \cite{Platzer:2020lbr} has analyzed the color structures which arise at two loops and has rewritten them in the color-flow basis, suitable for implementation in the framework \cite{Platzer:2013fha,DeAngelis:2020rvq}. Nevertheless, it is natural to first implement the anomalous dimension in the large-$N_c$  limit. In this limit, the color structure becomes trivial and the amplitudes can be viewed as a products of color dipoles. The anomalous dimension acts on these dipoles and genuine 3-particle correlations are suppressed.
In the absence of 3-particle correlations, the two-loop anomalous dimension has a similar structure as the one-loop result and the implementation can proceed along similar lines, except for three differences: i.) Instead of the simple dipole angular structure $W_{ij}^q$, we encounter more involved angular functions, which must be sampled in an efficient way. ii.) At the two-loop level we have two-emission terms, real-virtual and purely virtual terms and these terms involve double angular integrals so that we must simultaneously generate two vectors to sample these. iii.) Since the contribution of $\bm{\Gamma}^{(2)}$ is single logarithmic but proportional to $\alpha_s^2$, it is suppressed by $\alpha_s$, even when the logarithms are  large $L\sim 1/\alpha_s$. It is therefore sufficient to insert the two-loop anomalous dimension once during the evolution to achieve next-to-leading logarithmic accuracy. 

The first resummation of non-global observables beyond the leading-logarithmic accuracy was obtained very recently using a different formalism \cite{Banfi:2021owj,Banfi:2021xzn}. We look forward to presenting resummed results in our effective theory framework and to comparing to these results.

\begin{acknowledgments}        
The authors thank Ding Yu Shao for collaboration in the early stages of the project and Ze Long Liu, Rudi Rahn and Nicolas Schalch for discussions. The research of T.B.\ and X.X. is supported by the Swiss National Science Foundation (SNF) under grant 200020\_182038.
\end{acknowledgments}  

\begin{appendix}

\section{\boldmath Integral over the two-particle function $K_{ij;qr}$}
\label{app:KijInt}

The goal of this appendix is to establish the relation
\begin{align}
  \Dqqq{r} K_{ij;qr} ={}& 2W_{ij}^q \Bigg[
       \beta_0\left(\frac{1}{\ep}+\ln(2W_{ij}^q)\right) - \frac{1}{4} \, \gamma_1^{\rm cusp} \nonumber\\
     & + \frac{C_A}{3} - \frac{2}{3} (C_A-2n_FT_F+n_ST_S) \Bigg]\,
 \label{eq:KijqrInt}
\end{align}
used to rearrange the anomalous dimension in the main text. The divergence originates from the configuration where the direction $n_r$ is collinear to the direction $n_q$. We proceed by subtracting the divergences in the integrand which allows us to use different parametrizations for the divergent subtraction term and the finite remainder. For the divergent part we use a coordinate system in which 
\begin{align}
 n_q &= (1,\vec{0},0,1)\,, & n_i&=(1,\vec{0},s_i,c_i)\,, & n_j&=(1,\vec{0},s_j,c_j)\,, &n_r&=(1,\hat{n}_rs_1s_2,s_1c_2,c_1)\,,
 \label{eq:KOSY_div}
\end{align}
where $\vec{0}$ and $\hat{n}_r$ denote $(1-2\ep)$-dimensional zero and unit vectors, respectively, and $s_a^2+c_a^2 = 1$. The finite part can be computed in four dimensions and we use coordinates where $i$ and $j$ are back-to-back: 
\begin{align}
 n_i&=(1,0,0,1)\,, & n_j&=(1,0,0,-1)\,, & n_q&=(1,0,s,c)\,, & n_r&=(1,s_1s_2,s_1c_2,c_1)\,.
 \label{eq:KOSY_fin}
\end{align}
Below we outline the integration of the three contributions \eqref{eq:KijqrFuns} that make up the function $K_{ij;qr}$. Part $(a)$ is collinear finite and we obtain 
\begin{equation}
   \Dqqq{r} K_{ij;qr}^{(a)} 
 = \frac{1}{2\pi} \int_{-1}^{1} \frac{dc_1dc_2}{\sqrt{1-c_2^2}} \, K_{ij;qr}^{(a)} 
 = \frac{4 \pi^2}{3(1 - c^2)} 
 = \frac{2\pi^2}{3} \, W_{ij}^q \,,
\end{equation}
where the integral was performed in the coordinates \eqref{eq:KOSY_fin} and we have dropped higher orders in the dimensional regulator. The collinear divergence in part $(b)$ can be subtracted with the function $K_{ij;qr}^{(b),\text{sub}} = 8W_{ij}^q/{n_{qr}}$ and we obtain 
\begin{equation}
   \Dqqq{r} K_{ij;qr}^{(b),\text{sub}} 
 = 8W_{ij}^q  \Dqqq{r} \frac{1}{n_{qr}}
 = - \frac{4W_{ij}^q}{\ep} + \mathcal{O(\ep)} \,,
\end{equation}
with the integral given in \eqref{eq:collint1} for the divergent part and 
\begin{equation}
  \Dqqq{r} \left[K_{ij;qr}^{(b)} - K_{ij;qr}^{(b),\text{sub}} \right]
 = \frac{8 \left(2 + \ln\frac{1 - c^2}{4}\right)}{1 - c^2} 
 = 4W_{ij}^q \left[2 - \ln(2W_{ij}^q)\right] ,
\end{equation}
for the finite part. The presence of two powers of $n_{qr}$ in the denominator makes the calculation for part $(c)$ more complicated. Following \cite{Caron-Huot:2015bja}, we choose the following convenient form 
\begin{equation}
 K_{ij;qr}^{(c),\text{sub}} = \frac{2\left[n_{ir}(n_{jq}-n_{qr}) - n_{jr}(n_{iq}-n_{qr})\right]^2}{3 n_{iq}n_{jq}n_{ir}n_{jr}n_{qr}^2}
\end{equation}
of the subtraction function. The angular integrations yield 
\begin{equation}
   \Dqqq{r} K_{ij;qr}^{(c),\text{sub}} = \frac{2W_{ij}^q}{3} \left[ -\frac{1}{\ep} - 4 + \ln(2W_{ij}^q) + \mathcal{O(\ep)}\right] 
\end{equation}
and 
\begin{equation}
  \Dqqq{r} \left[K_{ij;qr}^{(c)} - K_{ij;qr}^{(c),\text{sub}} \right]
  = \frac{4W_{ij}^q}{9} \left[7 - 3\ln(2W_{ij}^q) + \mathcal{O}(\ep)\right]  .
\end{equation}
The full result for part $(c)$ is 
\begin{equation}
   \Dqqq{r} K_{ij;qr}^{(c)} = W_{ij}^q \left[ -\frac{2}{3\ep} + \frac{4}{9} - \frac{2}{3} \ln(2W_{ij}^q) + \mathcal{O(\ep)}\right]  
\end{equation}
and we obtain \eqref{eq:KijqrInt} by combining the results with the coefficients as given in \eqref{eq:Kijqr}.

\section{Iterated one-loop anomalous dimension} 
\label{app:GaGa}

In this appendix, we compute the $C_F C_A$ part of $\bm{\Gamma}^{(1)} \hat{\otimes} \bm{\Gamma}^{(1)} \hat{\otimes} \bm{1}$ in \eqref{eq:GaGa2} in $d$ dimensions to evaluate the extra term \eqref{eq:GammaExtra} in the anomalous dimension. To be able to separately keep track of the $\ep$ terms in the two angular integrals, we set $d=4-2\ep_q$ in the $\Omega_q$ angular integral and $d=4-2\ep_r$ in the $\Omega_r$ angular integral. We consider the same back-to-back two-jet configuration as in Section \ref{sec:finite} and parameterize the vectors as in \eqref{eq:KOSY_fin}. The angular integral then takes the form
 \begin{align}
  J(\ep_q,\ep_r) ={}   & \Dqqqin{q}\Dqqqout{r} \left(W_{12}^q W_{12}^r -W_{12}^{qr}-W_{12}^{rq}\right)\nonumber\\
 ={}& \frac{4^{-1+\ep _q+\ep _r} e^{\gamma  \left(\ep _q+\ep _r\right)}}{\sqrt{\pi }\, \Gamma \left(1-\ep _q\right) \Gamma \left(\frac{1}{2}-\ep _r\right)}   \left(\int_{-1}^{-\Delta}+\int_\Delta^1\right)\frac{dc_q}{(s_q^2)^{\ep_q}}
      \int_{-\Delta}^\Delta\frac{dc_1}{(s_1^2)^{\ep_r}}\int_{-1}^1\frac{dc_2}{(s_2^2)^{\frac12+\ep_r}}\Bigg[\nonumber\\
    & \frac{4}{s_q^2 s_1^2} -  \frac{2}{(1-c_q)(1+c_1)(1-s_qs_1c_2-c_qc_1)} - \frac{2}{(1+c_q)(1-c_1)(1-s_qs_1c_2-c_qc_1)}\Bigg]\nonumber\\
    ={}& \frac{2 \sqrt{\pi} 4^{\ep _q+\ep _r} e^{\gamma  \left(\ep _q+\ep _r\right)}}{\Gamma \left(1-\ep _q\right) \Gamma \left(\frac{1}{2}-\ep _r\right)}      \left( \int_{-1}^{-\Delta}+\int_\Delta^1\right)\frac{dc_q}{(s_q^2)^{\ep_q}}  \int_{-\Delta}^\Delta\frac{dc_1}{(s_1^2)^{ \ep_r}  }  \Bigg[ \nonumber \\ 
 &\;\; \frac{1}{\left(c_1-1\right) \left(c_1-c_q\right) \left(c_q+1\right)}-\frac{2 \ep_r  \left(c_1 c_q-1\right) \ln \left(\frac{c_q-c_1 }{\left(1-c_1\right) \left(c_q+1\right)}\right)}{s_1^2 \left(c_1-c_q\right) s_q^2}+\frac{2 \ep_r  \ln 2}{s_1^2 s_q^2} \Bigg] \nonumber \\
 ={}& \frac{2 \sqrt{\pi} 4^{\ep _q+\ep _r} e^{\gamma  \left(\ep _q+\ep _r\right)}}{\Gamma \left(1-\ep _q\right) \Gamma \left(\frac{1}{2}-\ep _r\right)}      \int_\Delta^1 dc_q\int_{-\Delta}^\Delta dc_1  \Bigg[ \frac{1}{\left(c_1-1\right) \left(c_1-c_q\right) \left(c_q+1\right)} \nonumber \\ 
 &\;\; -\frac{2 \ep_r  \left(c_1 c_q-1\right) \ln \left(\frac{c_q-c_1 }{\left(1-c_1\right) \left(c_q+1\right)}\right)}{s_1^2 \left(c_1-c_q\right) s_q^2}+\frac{2 \ep_r  \ln 2}{s_1^2 s_q^2} + \ep_q  \ln(s_q^2) +\ep_r \ln (s_1^2)\Bigg] \nonumber
 \\
=&-2 H_{-2}(r)+2 H_2(r)-\frac{\pi ^2}{6} + \ep_q \bigg(\frac{\pi ^2}{6} H_0(r) +2 H_{-2,0}(r) -2 H_{2,0}(r) \nonumber \\ 
 & \hspace{2cm}  -4 H_{-2,-1}(r)+4 H_{-2,1}(r) +4 H_{2,-1}(r)-2 H_3(r)-\zeta_3\bigg) \nonumber\\
&+ \ep_r\bigg(\frac{\pi ^2}{6}H_0(r)+4 H_{-2,0}(r)-2 H_{2,0}(r)  +4 H_{2,1}(r)-4 H_{-2,-1}(r) \nonumber\\
 &\hspace{2cm} -2 H_{-3}(r)+2 H_3(r) -2 \zeta_3 \bigg) \,,
 \end{align}
 where the $\ep_q$ and $\ep_r$ terms are from the $q$ integrals $r$ integrals, respectively. Using this general result, the commutator term in \eqref{eq:GammaExtra} is obtained as
\begin{align}
\bm{\Gamma}^{(1)} \otimes_2 \bm{\Gamma}^{(1)}\otimes_\ep \bm{1} - \bm{\Gamma}^{(1)}\otimes_\ep \bm{\Gamma}^{(1)} \otimes_2 \bm{1} ={}& 32 C_A C_F \lim_{\ep\to 0} \frac{1}{2\ep} \big[ J(0,\ep) -J(\ep,0)  \big]\,.
\end{align}

Note that there is a close relationship between the angular integrals and the one-loop soft functions given in \eqref{oneloopsoft}. Performing the energy integral and rewriting the bare coupling in terms of the $\overline{\rm MS}$ one, we have
\begin{equation}
\bm{\mathcal{S}}_m(\{ \underline{n} \}, Q_0) = \frac{\alpha_s}{4\pi} \left(\frac{\mu^2}{Q_0^2}\right)^{\ep}  \frac{1}{2\ep} \,\bm{\Gamma}^{(1)} \hat{\otimes} \bm{1}  \, , 
 \end{equation}
 where $\bm{\Gamma}^{(1)} \hat{\otimes} \bm{1} = \bm{R}_m \hat{\otimes} \bm{1} + \bm{V}_m$. With this form of the soft function, we can use the result for the integral $J(\ep_q,\ep_r)$ to reconstruct the $C_F C_A$ part of 
 \begin{equation}
 \bm{\Gamma}^{(1)} \otimes_2 \bm{S}^{(1)}  =  \bm{R}_2\otimes_2  \bm{S}_3^{(1)} + \bm{V}_2\,  \bm{S}_2^{(1)} \,.
\end{equation}
For the $C_F^2$ part of this, we need the one-loop angular integral 
 \begin{equation}\label{eq:Ga1b0Alt}
I(\ep)  = \Dqqqout{q}W_{12}^q= -H_0(r) + \ep \left(-2H_{-2}(r) + H_{0, 0}(r) +\frac{\pi^2}{6}\right) ,
\end{equation}
which was given in \eqref{eq:Ga1b0}. Adding the color factors, we have
\begin{equation}
 \bm{\Gamma}^{(1)} \otimes_2 \bm{S}^{(1)} = \frac{1}{2\ep} \left(\frac{\mu^2}{Q_0^2}\right)^{\ep} \left( 64 C_F^2\, I(0)\, I(\ep) + 32 C_F C_A\, J(0,\ep)\right) .
 \end{equation}
Subtracting the divergence, we get the renormalized result for $ \bm{\Gamma}^{(1)} \otimes_2 \bm{S}^{{\rm ren}(1)}$ shown in equation \eqref{eq:GaSm} in the main text.

\section{Extracting the collinear divergence of the soft function}
\label{app:soft}
In the main text we have extracted the divergence which arises when two soft partons become collinear as this divergence is not cancelled among the hard functions themselves. These soft-collinear divergences must cancel against divergences in the soft function and it is therefore instructive to extract the collinear singularities in the soft function itself. 

The situation is especially simple for the $n_F$ and $n_S$ pieces, which in dimensional  regularization only arise in the part of the soft function with two real emissions in the outside region: the double virtual pieces are scaleless and the real-virtual diagrams only have the color structure $C_F C_A$ and the in-out contribution is collinear finite. We consider the two-jet cross section as in Section \ref{sec:finite} and analyze the part of the soft function $S_2^{(2)}$  which involves two soft particles in the veto region. The relevant matrix element can be found in Appendix C of \cite{Becher:2012qc}. Extracting the divergence when the two emissions become collinear, performing the energy integrals and setting $\mu=Q_0$, we find the result
\begin{multline}\label{eq:S2colli1i2}
\left.  S_{2}^{(2)} \right |_{\rm coll} = \frac{1}{2\ep}  \bm{\Gamma}^{(1)} \otimes_2 \bm{1} \left[ \left( 2 C_A -n_F T_F \right) \left(\frac{1}{\ep} +4\right) i_1 - C_A \frac{2\pi^2}{3} i_1   \right. \\
\left. +\left( C_A \left(1- c_R\, \ep\right)  -2n_F T_F + n_S T_S \right) \left( -\frac{1}{6\ep} - \frac{5}{9} \right) i_2 \right] ,
\end{multline}
where the coefficients $i_1$ and $i_2$ are extracted from the angular integrals
\begin{align} \label{eq:i1}
I_1 &= -\frac{i_1}{2\ep} =  \Dqqq{r}   \frac{1}{n_q\cdot n_r} = -\frac{1}{2\ep}+\mathcal{O}(\e)\,, \\
I_2 &= -\frac{i_2}{\ep} \left(W_{ij}^q\right)^{-1} =  \Dqqq{r}   \frac{\left(n_i\cdot n_q n_j\cdot n_r - n_i\cdot n_q n_j\cdot n_r \right)^2 }{(n_i\cdot n_j)^2 (n_q\cdot n_r)^2} = -\frac{1}{\ep} \left(W_{ij}^q\right)^{-1}+\mathcal{O}(\e^0)\,,
 \end{align}
 and have been normalized to one for $\ep \to 0$. As in the rest of the paper $c_R=1$ tags the $\ep$-terms from the gluon spin sums. Since we are interested only in the collinear divergences, we can drop all finite terms in the  integrals $I_1$ and $I_2$. Setting $i_1=i_2=1$ in expression \eqref{eq:S2colli1i2}, we obtain
\begin{equation}\label{eq:S2coll}
\left.  S_{2}^{(2)} \right |_{\rm coll} = \frac{1}{8\ep} \bm{\Gamma}^{(1)} \otimes_2 \bm{1} \left( \frac{2\beta_0}{\ep} + \Gamma_{\rm coll} \right)\,.
\end{equation}
We see that the $\Gamma_{\rm coll}$ term indeed cancels against the combination
\begin{equation}
-\frac{1}{4\ep}\left(\bm{d}_2 \hat{\otimes} \bm{1}+ \bm{v}_2\right)
\end{equation}
 after inserting \eqref{eq:dm0} and \eqref{eq:vm0} obtained from the hard function and the $\beta_0$-term is removed by the $\beta_0/\ep^2$ in the renormalization condition \eqref{resDiv}.
\end{appendix}

\end{document}